\documentclass[a4paper,12pt]{report}
\usepackage{amsmath}
\usepackage{appendix}
\usepackage[linktocpage]{hyperref}
\usepackage[pdftex]{graphicx}
\usepackage[bottom,perpage]{footmisc}

\newcommand{\captionfonts}{\small}

\makeatletter  
\long\def\@makecaption#1#2{%
  \vskip\abovecaptionskip
  \sbox\@tempboxa{{\captionfonts #1: #2}}%
  \ifdim \wd\@tempboxa >\hsize
    {\captionfonts #1: #2\par}
  \else
    \hbox to\hsize{\hfil\box\@tempboxa\hfil}%
  \fi
  \vskip\belowcaptionskip}
\makeatother   

\renewcommand{\baselinestretch}{1.3}
\usepackage{fancyhdr}
\fancypagestyle{plain}{%
\fancyhead{} 
}

\newcommand{\be}{\begin{equation}}
\newcommand{\ee}{\end{equation}}
\newcommand{\bea}{\begin{eqnarray}}
\newcommand{\eea}{\end{eqnarray}}
\newcommand{\nn}{\nonumber\\}

\newcommand{\ap}{\alpha'}

\title{ \bf Non-perturbative Treatments of the Bosonic String and
the Axion with Cosmological Implications}

\author{\Large{Dylan Tanner}\\~~\\ Physics Department\\ King's College London}
\date{October 2011\\ ~~\\~~\\~~\\~~\\~~\\Submitted in partial fulfillment of the requirements for the degree of
Doctor of Philosophy in Physics}

\begin{document}

\maketitle

\addcontentsline{toc}{chapter}{Abstract}

\setcounter{page}{2}
\chapter*{Abstract}

This thesis is about the use of a novel, exact functional quantization method as applied to two commonly studied actions in theoretical physics.  The functional method in question has its roots in the exact renormalisation group flow techniques pioneered by Wilson, but with the flow parameter not limited to the familiar momentum cutoff.  Finding a configuration satisfying an expression for the exact effective action which does not vary with this parameter provides the basis for finding solutions to the physical actions we study. \\\\Firstly, the method is applied to an expression for the bare action of the pseudo-scalar axion used to explain the strong CP problem in QCD.  When quantized, we find that the effective potential of the axion, when interactions are not considered, is necessarily flattened by spinodal instability effects.  We regard this flattening as representing the very early stage in the development of the axion potential, when the Peccei-Quinn $U(1)$ symmetry is spontaneously broken resulting in a double-well potential.  Using commonly quoted values for the parameters of such a potential, we devise an expression for the energy density of the emerging axion potential and this is compared to dark energy.\\\\We then apply the functional method to the bosonic string with time varying graviton, dilaton and antisymmetric tensor (resulting in the string-axion) background fields.  We achieve a demonstration of conformal invariance in a non-perturbative manner in the beta functions, contrasting with conventional string cosmology where cancellation of a perturbative expansion is performed.  We then offer some hints as to possible cosmological implications of our configuration in terms of optical anisotropy.\\\\The research is closely related to the following three published papers.
\begin{itemize}
\item \textbf{Non-perturbative string backgrounds and axion induced optical activity}, J. Alexandre, N. E. Mavromatos and D. Tanner; published in New Journal of Physics 10, 2008, hep-th/0708.1154
\item \textbf{Antisymmetric-tensor and electromagnetic effects in an $\ap$-non-perturbative four-dimensional string cosmology}, J. Alexandre, N. E. Mavromatos and D. Tanner; published in Phys. Rev. D 78, 2008, hep-th/0804.2353
\item \textbf{Quantization leading to a natural flattening of the axion potential},  J. Alexandre and D. Tanner; published in Phys. Rev. D 82, 2010, hep-th/1003.6049
\end{itemize}

\addcontentsline{toc}{chapter}{Declaration}

\chapter*{Declaration}
I confirm that the following thesis does not exceed the word limit prescribed in the Regulations. I further confirm that the work presented in the thesis is my own with all references cited accordingly.

\addcontentsline{toc}{chapter}{Acknowledgments}
\chapter*{Acknowledgments}

First and foremost I wish to thank my supervisor Dr. Jean Alexandre for patient guidance and creative advice during the long process of understanding varied subject areas covered by this thesis, conducting new research and finally compiling the results into a meaningful document. \\\\ I am also grateful to Prof. Nick E. Mavromatos, my second supervisor, for assistance in understanding various aspects of the field, not least through his numerous papers and excellent published reviews.  I would like thank all at the King's College London Physics Department for the opportunity to spend this time pursuing theoretical physics at this level.  In particular Dr. Malcolm Fairbairn offered numerous insights into aspects of axion physics and cosmology.\\\\ Thanks also to several outstanding physicists and educators at Imperial College where I completed a preparatory M.Sc. prior to starting at King's College London.  Prof. Ray Rivers, Prof. Jerome Gauntlett and, in particular, Dr. Tim Evans, my M.Sc. thesis supervisor, stand out. \\\\ Lastly, I would offer deep thanks to all my family and friends in London and around the world who have offered support and encouragement over the last fours years.

\renewcommand{\baselinestretch}{1}
\tableofcontents

\renewcommand{\baselinestretch}{1.3}
\listoffigures
\chapter{Introduction}

\section{Motivation and Context}
This thesis is made up of two parts, one dealing with the behaviour of the QCD-axion under quantization, and the other dealing with the axion evident in the bosonic string through the antisymmetric tensor field.  The uniting feature of both components of the thesis is the use of a functional method to derive non-perturbative evolution expressions for the effective actions: of the QCD-axion; and the bosonic string.  Deriving expressions which may be compared to experimental results in particle physics typically involves commencing with a suitable bare action describing the theory; and then quantizing this (e.g. using path integral methods) and identifying and managing the divergences using a regularisation and/or renormalisation scheme.  The result is usually expressed in a perturbative expansion with terms defined by increasing powers of a small parameter (such as $\hbar$ or a small coupling).  In a useful effective theory, the higher order expansion terms in the series may be neglected in the low energy regime. However the use of perturbative methods may be limited in certain theories or energy regimes.  For example, it is impossible to describe confinement in QCD with perturbative expansions, which are valid only at high energies.  For this reason, it is useful to investigate the the infrared regime of a theory, using a non-perturbative approach such as the Exact Renormalisation Group (ERG). The so-called Wegner-Houghton equation, derived in ERG theory, is built on Wilsonian renormalisation group methods where the behaviour of the parameters of the effective theory with energy level $k$ is studied, where $k$ is some energy scale less than the high energy cutoff of the theory (termed $\Lambda$ in this thesis, Section (\ref{renormalisation Group Theory})).  The renormalisation group is the set of transformations $k \rightarrow k + \delta k$ in which the full quantum theory is described as $k \rightarrow 0$ and the higher energy field degrees of freedom are integrated out leaving any cutoff dependence in the infrared regime.  The Wegner-Houghton approach takes the limit $\frac{\delta k}{k}\rightarrow 0$ and results in an exact expression for the evolution of the effective action with $k$, at constant field configuration, which we derive in Appendix A.  The method used in this thesis is outlined in Section (\ref{An Alternative, Exact Approach}).  A key difference between our method and the Wegner-Houghton approach is that that the evolution parameter is not the energy scale $k$, but another parameter of the theory which when varied can be used to describe the evolution of the theory from classical (bare) to fully quantized, a feature which is desirable for example in string theory which is assumed to be scale invariant.  We apply this new, non-perturbative method to two well known bare actions, both of which are outside the standard model and both of which may be important at high energy scales, potentially near the Planck scale.\\\\
The QCD-axion was postulated by Peccei-Quinn, \cite{Original Peccei Quinn} and the theory subsequently developed into one which dealt with the CP problem.  As a scalar field which acquires mass, it may also offer a contribution to the theorized existence of cold dark matter in the universe, \cite{axion dark matter} and indeed dark energy, (see for example, \cite{A Quintessential Axion}).  A scalar field which solves the CP problem and also provides a meaningful contribution to the missing mass-energy balance in the observed universe would indeed be an elegant addition to the standard model should the axion be successfully detected.  While much theoretical work has been done on the postulated axion, this elegance provides a motivation, in this thesis, for applying the novel, non-perturbative techniques used in shedding more light on its behaviour.\\\\
String theory is the main candidate for unifying quantum field theory with general relativity.  String cosmology is the application of string-based models to cosmological phenomena.  String cosmology models with $D=4$ spacetime dimensions are particularly useful in application to the observed physical world.  A popular application of string cosmology models is in determining any preferred spacetime direction inherent in the universe, or anisotropy.  Lastly, the energy scales of string theory and string cosmology are necessarily high and there is a question as to whether perturbative approaches are appropriate.  This is a key driver for the use of our non-perturbative, exact functional method.  Motivated by these themes, we apply our non-perturbative techniques to create a $D=4$ model and investigate the role of the string-axion in anisotropic effects. 

\section{Structure of the Thesis}

This thesis is divided into two parts, one dealing with a study of the QCD axion and the other with the bosonic string.  (We note that the scalar field $h$ arising from the anti-symmetric tensor in the bosonic string is often referred to as the string axion (see eq. (\ref{string_axion}).  We also refer to this string-axion in this thesis as it is a key parameter in our results.  We stress however that the QCD-axion and the string-axion in the two different sections of this thesis are not related in our research).\\\\ The two sections are united as the treatment of both is based on techniques originating in, but distinct from, the exact renormalisation group formalism pioneered by Wilson and others in the 1970s.  In Chapter (\ref{chapter Topics in Quantum Field Theory}) a review of some aspects of theoretical physics shared by both sections of the thesis is undertaken.  Effective quantum field theory concepts are introduced along with the effective potential.  This naturally leads to a review of the Wilsonian formalism for renormalisation, the exact renormalisation group methods and in (\ref{An Alternative, Exact Approach}) we outline the precise effective field theory manipulations which are the basis of the non-perturbative approach used in this thesis.  These non-perturbative techniques were made extensive use of in work by my thesis supervisor Dr. Jean Alexandre and my second supervisor Prof. Nick E. Mavromatos in \cite{Alexandre0}, \cite{Alexandre1}, \cite{Alexandre2}, \cite{Alexandre3}, \cite{Alexandre4}, \cite{Alexandre5}.  The classical phenomenon of spontaneous symmetry breaking, which underpins many key particle physics theories, including axion theory, is introduced, as is the spinodal instability.\\\\In Chapter (\ref{chapter axion}) a review of topics specific to QCD-axion physics is presented.  A summary of QCD from the viewpoint of the symmetries of the theory is presented.  The concept of instantons in four dimensional non-Abelian gauge theories is covered, which naturally leads to a description of the charge-parity (CP) problem.  The leading candidate for its resolution is the axion, postulated by Peccei and Quinn in 1977, \cite{Original Peccei Quinn}.  Its development, the variety of potential axion models and axion phenomenology are presented.  We should note here an excellent summary and history of axion physics from Kim, one of the world's foremost experts, \cite{Axion_Kim_2010_Summary}, which was relied on heavily for this section.\\\\Chapter (\ref{Full_Quantization of the Axion}) presents our treatment of the axion.  We initially build a justification for the quantisation of the axion and state the relevant actions and potentials we rely on.  We then provide a detailed account of our computations leading to a non-perturbative evolution equation and show how the spinodal instability in the non-interacting theory leads to our formulation of the flattened effective potential of the axion.  The work conducted in this chapter was conducted by myself and Dr. Jean Alexandre and published in 2010 in \cite{Alexandre_Tanner} while analysis of the $\mu=0$ and $\lambda=0$ cases and comparison with accepted dark energy values in section (\ref{Flattening of the Axion Potential}) was largely my own and is unpublished. \\\\Chapter (\ref{chapter String Cosmology}) provides an overview of bosonic string theory with a focus on conformal invariance and the conditions for proving this.  Two excellent and widely known sources from Polchinski, and from Zwiebach, \cite{polchinski}, \cite{zwiebach} were invaluable in this.  A brief summary of critical string cosmology is presented with mention of non-critical and tachyon string cosmology noted.  A brief section on optical anisotropy is presented as an area where our work on the bosonic string may be relevant.  \\\\Chapter (\ref{chapter non-perturbative Approach to String Cosmology}) presents our work on the bosonic string.  We introduce the action we will be using and derive, using non-perturbative methods outlined in section (\ref{An Alternative, Exact Approach}) and exact equation for the effective action of string.  We then show our solution to this exhibits conformal invariance non-perturbatively (our main result for this section) and offer some insights into its cosmology.  The work in this chapter is based on work led by Dr. Jean Alexandre and published in 2007 and 2008 in \cite{Alexandre4}, \cite{Alexandre5}.  My solo contribution to this part of the thesis consisted of computations related to showing conformal invariance presented in sections (\ref{conformal_anisotropy}) and (\ref{conformal_invariance_2}).

\chapter{Topics in Quantum Field Theory}\label{chapter Topics in Quantum Field Theory}
The two parts of this thesis rely on some common themes and calculation methods within classical and quantum field theory.  These are outlined here.  

\section {The Effective Action and Potential} \label{The Effective Action and Potential}

In applying quantum field theory to a bare action, one of the most common regularization processes applies a high energy cutoff, $\Lambda$, above which the validity of the theory is not known (other methods include dimensional regularization).  The resulting expression may contain divergences which need to be isolated and removed prior to arriving at a useful effective action at some energy scale $k_0$ where phenomenology can occur.  In quantizing a scalar field theory such as $\phi^4$ theory introduced in (\ref{SSB}), we seek a function which when minimized provides an exact expression for the expectation value of the field, $<\phi>$, with quantum effects accounted for.  We also want an expression which agrees with the the classically-derived equilibrium of $\phi$ to lowest perturbative order.  We define the generating functional, $Z[J]$ which defines the full quantum theory up to energy scale $\Lambda$ in path integral form in Euclidean D dimensional spacetime:
\be \label{partitiondefi} Z[J]_{\Lambda}=\int {\cal D} \phi \exp -\left(S[\phi]
+ \int d^Dx J(x) \phi(x)\right ), \ee
Here, the Euclidean action is $S[\phi] = \int d^Dx (\frac{1}{2}(\partial\phi)^2  + U(\phi))$ and $J(x)$ is the source term which interacts with $\phi$.  Using $Z[J]$ we can compute all correlation functions for processes by taking taking functional derivatives of $Z[J]$ with respect to the source $J(x)$ at the required space-time points. We further define a functional $W[J]$ such that (now dropping the $\Lambda$ subscript for brevity):
\be \label{def_of_W} Z[J]=\exp (- W[J]).\ee
where $W$ and $Z$ are both functionals of $J$ the external source. Successive functional derivatives of $W$ with respect to $J$ generate the connected correlation functions (as opposed to $Z$ which returns both connected and disconnected correlation functions).  Thus $W(J)$ is more useful in our motivation for deriving measurable quantities in a quantum field theory.  Taking the derivative of $ W(J)$ (noting $-W(J) = \ln Z(J)$), with respect to the source $J$ and using $\frac{\delta W}{\delta J}=\frac{\delta W}{\delta Z}\times\frac{\delta Z}{\delta J}$ gives:
\bea \label{classicalfield}\frac{\delta W}{\delta
J(x_1)}=\frac{1}{Z}\int {\cal D} \phi \, \phi(x_1)\exp \left
(-S[\phi(x)] + \int d^Dx \ \phi(x)J(x)\right
) \nn = \phi_{cl}(x_1)\equiv  <\phi(x_1)>. \eea
This expression for $\phi_{cl}$ is the expectation value of
the quantum field, $<\phi(x)>$.  
With source set to zero(that is a theory with only self interactions) we can take a second functional derivative of $W[J]$, resulting in
two terms.
\begin {multline}  -\frac{\delta^2W}{\delta J(x_1)\delta
J(x_2)}|_{J=0}=\frac{1}{Z}\int {\cal D} \phi \, \phi(x_1)\, \phi(x_2)\exp
\left (-S[\phi(x)] + \int d^Dx \phi(x)J(x)\right ) \\-
\frac{1}{Z^2}\int {\cal D} \phi \, \phi(x_1) \exp \left (-S[\phi(x)] +
\int d^Dx \phi(x)J(x)\right ) \\ \int {\cal D} \phi \, \phi(x_2) \exp
\left (-S[\phi(x)] + \int d^Dx \phi(x)J(x)\right ) .\end
{multline}\\This gives:
\be \label {secondderivative} -\frac{\delta^2W}{\delta J(x_1)\delta
J(x_2)}|_{J=0}=  <\phi(x_1)\phi(x_2)> -  \,\phi_{cl}(x_1)\phi_{cl}(x_2),\ee\\\\
where it is noted that $<\phi(x_1)\phi(x_2)>$ is the two point
correlation function defined as the amplitude of propagation of an excitation (or particle) between $x_1$ and $x_2$:

\be \label{two_point_correlation} <\phi(x_1)\phi(x_2)> \equiv  \frac{ \int {\cal D} \phi\,\phi(x_1)\phi(x_2) \exp -\left(S[\phi]
+ \int d^Dx J(x) \phi(x)\right )}{\int {\cal D} \phi \exp -\left(S[\phi]
+ \int d^Dx J(x) \phi(x)\right )}. \ee
We consider the contributions to the expression $<\phi(x_1)\phi(x_2)>$.  It is made up of two contributions.  Firstly the sum of connected diagrams corresponding to propagation from $x_1$ to $x_2$.  Added to this are contributions from disconnected diagrams corresponding to $x_1$ multiplied by those to $x_2$.  These latter disconnected contributions to $<\phi(x_1)\phi(x_2)>$ exactly cancel with the term  $\phi_{cl}(x_1)\phi_{cl}(x_2)$ (or alternatively $<\phi(x_1)><\phi(x_2)>$) in Eq (\ref{secondderivative}) resulting in:
\be -\frac{\delta^2W}{\delta J(x_1)\delta
J(x_2)}|_{J=0}=  <\phi(x_1)\phi(x_2)>_{connected}, \ee
which is known as the connected correlator.  As it is a propagation amplitude it is necessarily greater than or equal to zero and hence:
\be \label{why_second_W_non_negative} \frac{\delta^2W}{\delta J(x_1)\delta
J(x_2)}|_{J=0} \leq 0. \ee
One notes from (\ref
{classicalfield}) that the expression for the classical field
$\phi_{cl}[J]$ is now a function of $J$, the source.  Here we introduce the Legendre
effective action $\Gamma [\phi_{cl}]$ in four dimensions, which is defined by the Legendre
transformation.
\be\label{legeffact} \Gamma [\phi_{cl}] =  W[J] - \int d^4x
\phi_{cl}(x) J(x).\ee 
We take the first derivative of $\Gamma [\phi_{cl}]$ with respect to the classical field at a point $x_1$, ($\phi_{cl}(x_1)$).
\be \frac{\delta \Gamma [\phi_{cl}]}{\delta \phi_{cl}(x_1)}=
\frac{\delta W[J]}{\delta \phi_{cl}(x_1)} - 
\int d^4x \, \phi_{cl}(x) \frac{\delta J(x)}{\delta \phi_{cl}(x_1)} -
 J(x_1)\ee and utilizing the chain rule,
\be \label{using_chain_rule} \frac{\delta \Gamma [\phi_{cl}]}{\delta \phi_{cl}(x_1)}=
 \int d^4x \, \frac{\delta J(x)}{\delta \phi_{cl}(x_1)}\frac{\delta
W[J]}{\delta J(x)} -  \int d^4x \, \phi_{cl}(x) \frac{\delta
J(x)}{\delta \phi_{cl}(x_1)} - J(x_1).\ee
Using (\ref{classicalfield}), that is: $\frac{\delta W}{\delta J(x)}
= \phi_{cl}(x)$, in the second term on the right-hand-side, this becomes:
\be \label{legdir}\frac{\delta \Gamma [\phi_{cl}]}{\delta
\phi_{cl}(x_1)} = - J(x_1).\ee
Taking a second derivative of
(\ref {legdir}) with respect to $\phi_{cl}(x_2)$ gives the result:
\be \label {inverse relationship}\frac{\delta^2 \Gamma [\phi_{cl}]}{\delta
\phi_{cl}(x_1)\delta \phi_{cl}(x_2)} = -\frac{\delta
J(x_1)}{\delta \phi_{cl}(x_2)}=  -\left(\frac {\delta
\phi_{cl}(x_2)}{\delta J(x_1)}\right)^{-1}_{x_1x_2},\ee
 where $\left(\frac {\delta \phi_{cl}(x_2)}{\delta
J(x_1)}\right)^{-1}_{x_1x_2}$ is a matrix in $x_1$ and $x_2$
representing the inverse of $\frac{\delta J(x_1)}{\delta
\phi_{cl}(x_2)}$.  Using the relationship $\frac{\delta W}{\delta J(x_2)} =
\phi_{cl}(x_2)$, we now have:
\be \label {inverse relationship1}\frac{\delta^2 \Gamma
[\phi_{cl}]}{\delta \phi_{cl}(x_1)\delta \phi_{cl}(x_2)} =
-\left(\frac {\delta^2W[J]}{\delta J(x_1)\delta
J(x_2)}\right)^{-1}_{x_1x_2}. \ee
We now define the effective potential, $U_{eff}$ in terms of the Legendre effective action, as the part of the effective action not containing any parts of the expansion of the kinetic component of the bare action.
\be \label{eff Potential expansion} \Gamma [\phi_{cl}] =   \int d^4x [U_{eff} +  Z(\phi_{cl})(\partial\phi_{cl})^2 + {\cal O} (\partial\phi_{cl})^4]. \ee
Here $Z(\phi_{cl})(\partial\phi_{cl}^2) + {\cal O} (\partial\phi_{cl})^4$ represents an expansion of the kinetic terms in the bare expression, which vanish with a field configuration constant with $x$.  In such a configuration, when minimizing the effective potential with respect to the classical fields we can find the true vacuum state of a quantum field theory.  In general $U_{eff}$ cannot be computed exactly and approximations are deployed.  With source set to zero and $\phi$ constant in space time, $\Gamma [\phi_{cl}] = W$ from  (\ref{legeffact}).  (\ref{partitiondefi}) reduces to:
\be \label{gamma_potential} \Gamma [\phi_{cl}] =  \int d^4x \, U_{eff} = V \cdot U_{eff}, \ee
where $V$ is the space time volume arising from the integral $\int d^4x$.   One may see the use of the effective potential by considering Equs. (\ref{legdir}) and (\ref{gamma_potential}) with source $J=0$.
\be \label{effective_potential_theory} \frac{\delta U_{eff}}{\delta
\phi_{cl}(x_1)}  = 0,\ee
which provides the condition that the effective action $\Gamma
[\phi_{cl}]$ has an extrema, with the solutions to eq (\ref{effective_potential_theory}) representing the stable quantum states of the theory. 
\subsection {Effective Scalar Theory}
In perturbative renormalisation theory, it is conventional to split the Lagrangian into two parts, one containing the renormalised (physical) fields and couplings ($\mathcal {L}_1$) and the other containing the counter terms ($\delta \mathcal {L}$) which absorb the divergences generated by loop corrections.  We thus have the expression $\phi$:
\be \mathcal {L} = \mathcal {L}_1(\phi) + \delta \mathcal {L}(\phi), \ee
from which the Legendre effective action, valid at one loop only, can be computed (quoted without derivation) in Euclidean space, Section 11.4, \cite{Peskin}):
\be \label{Legendre_perturbative} \Gamma(\phi_{cl}) = \int d^4x \mathcal {L}_1(\phi_{cl}) + \frac{\hbar}{2}\ln \det \left( - \frac{\delta^2\mathcal {L}_1}{\delta\phi_{cl} \delta\phi_{cl}}\right) + \int d^4x \delta \mathcal {L}(\phi_{cl}). \ee
To first order in $\phi^4$ theory, \cite{Zee} computes the leading order expression for $\Gamma(\phi_{cl})$, (where $S_1 = \int d^4x \mathcal {L}_1$) and counter terms are not written:
\be \label{one loop correction} \Gamma(\phi_{cl}) = S_1(\phi_{cl}) + \frac{\hbar}{2}tr \ln [-\partial^2 + U''(\phi_{cl})]  + {\cal O}(\hbar^2). \ee
In a constant field configuration the second term on the right hand side of (\ref{one loop correction}) reduces to (in phase space):
\be \label{one_loop} \frac{\hbar}{2}tr \ln [-\partial^2 + U''(\phi_{cl})] = \frac{\hbar}{2} \int d^4x \int \frac{d^4k}{(2\pi)^4} \ln \left[\frac{k^2 + U''(\phi_{cl})}{k^2}\right], \ee
where we have divided the factor $k^2 + U''(\phi_{cl})$ in the logarithm by a factor $k^2$ to avoid taking the logarithm of a dimensionful  quantity.  Combining this with (\ref{eff Potential expansion}), one obtains to first order and where $B\phi_{cl}^2$ and $C\phi_{cl}^4$ are the counter terms quadratic and quartic in the fields, contained within $\delta \mathcal {L}(\phi_{cl})$-containing term in (\ref{Legendre_perturbative}).
\be \label{one_loop_effective} U_{eff}(\phi_{cl}) = U(\phi_{cl}) + \frac{\hbar}{2}\int \frac{d^4k}{(2\pi)^4} \ln \left[\frac{k^2 + U''(\phi_{cl})}{k^2}\right] + B\phi_{cl}^2 + C\phi_{cl}^4. \ee
The integral on the right-hand-side of (\ref{one_loop_effective}) contains divergences quadratic in the cutoff which can be absorbed by the counter terms (again only first order in $\hbar$ terms are considered here).  Here $B$ and $C$ are co-efficients containing divergent corrections to the powers of $\phi_{cl}$ to absorb the cutoff dependence in the integral on the right-hand-side.  Evaluating (\ref{one_loop_effective}), the following is obtained, with the integral taken up to $k=\Lambda$ with $\hbar$ set to 1 for brevity (a convention largely followed in the remainder of the thesis).  
\be \label{one_loop_effective2} U_{eff}(\phi_{cl}) = U(\phi_{cl}) + \frac{\Lambda}{32\pi^2}U''(\phi_{cl}) - \frac{[U''(\phi_{cl})]^2}{64\pi^2} \ln\frac{\sqrt{e}\Lambda^2}{U''(\phi_{cl})} + B\phi_{cl}^2 + C\phi_{cl}^4.  \ee
Later in this thesis we will study $\phi^4$ theory in the context of spontaneous symmetry breaking and other phenomena, taking $\mathcal {L}$ to be: 
\be \mathcal {L} = \frac{1}{2}(\partial \phi_{cl})^2 + \frac{1}{2}\mu^2(\phi_{cl})^2 - \frac{1}{4!}\lambda(\phi_{cl})^4 + A (\partial \phi_{cl})^2 + B\phi_{cl}^2 + C\phi_{cl}^4. \ee
Here $A (\partial \phi_{cl})^2$, $B\phi_{cl}^2$ and $C\phi_{cl}^4$ are the counter terms and the bare potential can be considered as $U(\phi_{cl}) = - \frac{1}{2}\mu^2(\phi_{cl})^2 + \frac{1}{4!}\lambda(\phi_{cl})^4 + B\phi_{cl}^2 + C\phi_{cl}^4$.  We consider the case of $\mu = 0$ where we have the condition  $\frac{d^2U''(\phi_{cl})}{d\phi_{cl}^2}|_{\phi_{cl} = 0} = 0$.  This will be of interest later in the thesis as it represents the transition point between a purely convex potential to a double well shape.  We can evaluate (\ref{one_loop_effective2}) in this context (to quadratic order in $\lambda$):
\be U_{eff}|_{\mu = 0} = \left(\frac{\Lambda}{64\pi^2}\lambda + B \right)\phi_{cl}^2 + \left(\frac{1}{4!}\lambda + \frac{\lambda^2}{(16\pi^2)}\ln \frac{\phi_{cl}^2}{\Lambda^2} + C \right)\phi_{cl}^4. \ee
We want to absorb the dependence of the cut-off $\Lambda$ in the counter terms $B$ and $C$ and impose renormalisation conditions.
\bea \label{renormal_conditions} \frac{d^2U_{eff}(\phi_{cl})}{d\phi_{cl}^2}|_{\phi_{cl} = 0} &=& 0 \nn
 \frac{d^4U_{eff}(\phi_{cl})}{d\phi_{cl}^4}|_{\phi_{cl} = m} &=& \lambda(m).\eea
The first condition in (\ref{renormal_conditions}) implies that $\mu=0$ (or the renormalised mass-squared term vanishes) at $\phi_{cl} = 0$ in the $\phi^4$ theory and that $B = -(\Lambda^2/64\pi^2)\lambda)$.  In the second condition, since $\frac{d^4U_{eff}(\phi_{cl})}{d\phi_{cl}^4}$ is not defined at $\phi_{cl} = 0$ due to the log term, we choose an arbitrary energy scale $m$.  Here we are left with:
\be U_{eff}|_{\mu = 0} =  \left(\frac{1}{4!}\lambda + \frac{\lambda^2}{(16\pi^2)}\ln \frac{\phi_{cl}^2}{\Lambda^2} + C \right)\phi_{cl}^4, \ee 
which if plugged into the second of (\ref{renormal_conditions}) gives:
\be \label{easy derivative} \lambda(m) = \lambda + K_1 \lambda^2 \ln \frac{m}{\Lambda} + K_2, \ee
where $K_1 =\frac{3}{16\pi^2} $ and $K_2$ is a constant not computed here.  These two conditions can eliminate the $\Lambda$ dependence. (We do not derive the value of $C$ required to remove the $\Lambda$ dependence above).
\be \label{phi4_eff_pot} U_{eff}(\phi) = \frac{1}{4!}\lambda(m)\phi_{cl}^4 + \frac{\lambda(m)^2}{(16\pi)^2}\phi_{cl}^4\left(\ln\frac{\phi_{cl}^2}{m^2} - \frac{25}{6}\right) + {\cal O}(\lambda(m)^3), \ee
where we now have the effective potential in terms of the classical field and the coupling $\lambda(m)$ for which we have a beta function relationship, which follows from (\ref{easy derivative}) to order ${\cal O}(\lambda(m)^3)$.
\be \label{beta_lambda} m \frac{\partial \lambda(m)}{\partial m} = \frac{3}{16\pi^2}\lambda(m)^2 + {\cal O}(\lambda(m)^3). \ee
\section {Spontaneous Symmetry Breaking} \label{section SSB}
Considering a particle at position $q$ in one dimension defined by the classical Lagrangian, with velocity $\dot{q}$ and with $k'$ and $\lambda$ two real constants with $\lambda >0$, we have the Lagrangian (here in Minkowski spacetime).
\bea \label{SSBQM} \mathcal {L} &=& \frac{1}{2}\left[\dot{q}^2 - k' q^2 \right] - \frac{\lambda}{4}(q^2)^2 \nn  
 &=& \frac{1}{2}\dot{q}^2  - U(q) \eea
This exhibits a discrete reflective symmetry, invariant under the operation $q\rightarrow - q$. In both classical physics and quantum mechanics the equilibrium and ground states respectively are found by minimizing the potential. 
\be \frac{\partial U}{\partial q} = 2 k' q + \lambda q^3 \ee
If $k' > 0$ there is one real minimum at $\phi = 0$ and the ground state respects the reflective symmetry and the system represents a damped harmonic oscillator. However when $k' < 0$ there is a local maximum at $q = 0$, (corresponding to a potential of $U_{max}$) and two minima at $q = \pm \sqrt{\frac{2k'}{\lambda}}$, which represents the double well or Mexican hat potential in one dimension.  
\begin{figure}[h] \centering
\includegraphics[scale=0.5]{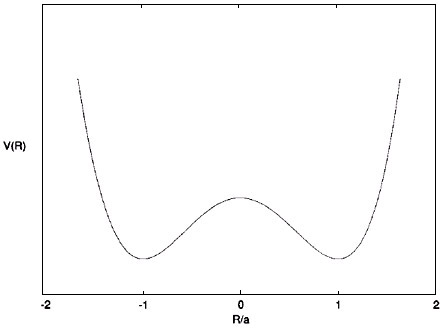}
\caption[A one dimensional double well potential]{A one dimensional double well potential}\label{strings1}
\end{figure}
At energies less than $U_{max}$, classically the particle must be in one of the two minima, thus breaking the reflective symmetry of the system.  Since we have not added any terms by hand to do this, it is termed spontaneous symmetry breaking.  (It should be noted that given a large number of particles $q_a$ initially with potential energy $>U_{max}$ this reflective symmetry will be restored in a sense as there will be equal probability that a particle will end up in either minima in this one dimensional model).\\\\ We now promote the particle position variable $q$ to $N$ dynamical fields $\phi_n(x)$ with $x$ the space time coordinate in $D$ dimensions, $n= 1, 2...N$ and $k'$ replaced by $-\mu^2$   

\bea \label{SSB} \mathcal {L} &=& \frac{1}{2}\left[(\partial\phi_n)^2 + \mu^2\phi_n^2 \right] - \frac{\lambda}{4!}(\phi_n^2)^2 \nn  
 &=& \frac{1}{2}(\partial\phi_n)^2  - U(\phi_n). \eea
This now exhibits a continuous $O(N)$ symmetry with the fields transforming as an $N$-dimensional vectors under the $O(N)$ transformations.  The potential is minimized at values of $\phi_{n}$ such that
\be \label{SSB2}(\phi_{n-min})^2 = \frac{\mu^2}{\lambda}.\ee
The length of the vector $\phi_{n-min}$ is defined but its phase or direction is arbitrary.  If one chooses the direction to be in the $N$ direction, we have the set of shifted fields such that:
\be \phi_{Nmin} = (0,0,0....\phi_0),\ee
where here $\phi_0 = \frac{\mu}{\sqrt{\lambda}}$.  The system can be perturbed around $\phi_0$ such that $\phi_N(x) = \phi_0 + \sigma (x)$, and define a new set of fields:
\be \phi_N'(x) = (\phi_1(x), \phi_2(x),....,(\phi_0 + \sigma(x))). \ee
We can now rewrite (\ref{SSB}) in terms of $\phi_N'(x)$ and we obtain:
\begin {multline}\label{classicalSSB} \mathcal {L} = \frac{1}{2}(\partial\phi_n)^2 + \frac{1}{2}(\partial\sigma)^2 - \frac{1}{2}(2\mu^2)\sigma^2 - \sqrt{\lambda}\mu\sigma^3  - \sqrt{\lambda}(\phi_n)^2\sigma - \frac{\lambda}{4}\sigma^4 - \frac{\lambda}{2}(\phi_n)^2\sigma^2 - \frac{\lambda}{4}(\phi_n)^4. \end{multline}
This can be interpreted as the perturbation field $\sigma(x)$ with a mass factor $\sqrt{2}\mu$, with the fields $\phi_n(x)$ for $n= 1,2....,(N-1)$ being massless degrees of freedom.  The $O(N)$ symmetry has been spontaneously broken to an $O(N-1)$ symmetry.  Goldstone's theorem states that for every spontaneously broken continuous symmetry a massless bosonic degree of freedom results.  In this case the $O(N)$ symmetry (which has $N(N-1)/2$ generators of the symmetries between the $N$ fields) is broken by setting the field in the $N$ direction, resulting in $(N-1)(N-2)/2$ symmetries between the fields.  The difference in the number of symmetries before and after $(N-1)$ is also the number of resulting Nambu-Goldstone bosons, \cite{Peskin}. 
\\\\
The above description does not incorporate quantization.  This will affect the terms present in (\ref{SSB}) and (\ref{classicalSSB}) and thus will impact the symmetry breaking process.  Considering the case of $N=2$ and breaking of $O(2)$ to $O(1)$, in the ground state we have $\phi_1 = 0$ and $\phi_2 = \phi_0$, with the $\phi_1$ field representing the massless Goldstone boson.  Considering quantum fluctuations about the ground state of $\phi_1$ we can consider the mean square of these fluctuations, in momentum space, $k$. (\cite{Zee}, IV.1).
\bea \label{SSB QFT} < (\phi_1(0))^2 > &=& \frac{1}{Z}\int {\cal D} \phi (\phi_1(0))^2 \exp( i S(\phi)) \nn
&=& \lim_{x\rightarrow0} \frac{1}{Z}\int {\cal D} \phi  (\phi_1(0))(\phi_1(x))\exp (i S(\phi)) \nn
&=&  \lim_{x\rightarrow0} \int\frac{ d^Dk}{(2\pi)^D}\frac{\exp i\vec{k}\vec{x}}{k^2}, \eea
where if $\phi_1$ represented a massive field $\frac{1}{k^2}$ would be replaced by $\frac{1}{k^2 + \mu^2}$. $D$ is the spacetime dimensionality. There is an infrared divergence for $D\leq 2$ leading to the Coleman-Mermin-Wagner theorem which states that spontaneous symmetry breaking cannot occur for $D\leq 2$ dimensions, \cite{no goldstone}.  This illustrates the effects quantization can have on classical spontaneous symmetry breaking.  We now consider a variation of (\ref{SSB}) - a complex scalar field theory invariant under a $U(1)$ transformation (denoted by $\phi \rightarrow \exp i\theta \phi$):
\be \label{comlex scalar theory} \mathcal {L} = \partial_{\mu}\phi^* \partial^{\mu}\phi + \mu^2\phi^*\phi - \lambda (\phi^*\phi)^2. \ee
We now promote the global symmetry $\exp i\theta$ to  $\exp i\theta(x)$ (that is, to a locally variant gauge symmetry).  The invariant Lagrangian is as follows, where $A_{\mu}$ is the gauge field and $D_{\mu}\phi = (\partial_{\mu} - i e A_{\mu})\phi$ is the operator required to preserve symmetry.
\be 
\label{comlex scalar gauge theory} \mathcal {L} =  -\frac{1}{4}F_{\mu\nu}F^{\mu\nu}+(D_{\mu} \phi)^* D^{\mu}\phi + \mu^2\phi^*\phi - \lambda (\phi^*\phi)^2. \ee
Spontaneous symmetry breaking of (\ref{comlex scalar theory}) results in one massless boson.  However, a similar analysis, \cite{Zee}, shows that spontaneous symmetry breaking of (\ref{comlex scalar gauge theory}) results in the gauge field $A_{\mu}$ acquiring mass and one degree of freedom which it acquires from the massless scalar which disappears.  This is the basis of the Higgs mechanism which is postulated to be the mass-acquiring mechanism for the particles in the standard model. Spontaneous symmetry breaking in a Lagrangian such as (\ref{SSB}) at the classical level may be accompanied by further explicit breaking of the symmetry by quantum effects.  Additional interaction terms involving the Goldstone boson may arise in the Lagrangian and it can acquire mass. This contrasts with the Higgs mechanism for gaining mass which is purely a classical effect.

\section{Wilsonian renormalisation Group Theory}\label{renormalisation Group Theory}
The descriptions below are based on \cite{Peskin}, \cite{Zee}, \cite{Ryder} and where noted.

\subsection {The Renormalisation Group Equations}\label {RGE}
The functional approach we use to derive exact evolution equations for the QCD axion and the bosonic string (described in Section (\ref{An Alternative, Exact Approach}) has its roots in the exact renormalisation group equations which in turn rely on the renormalisation group techniques pioneered by Wilson in the early 1970s.  In this thesis we thus briefly introduce the latter and then the former prior to describing our technique.\\\\
Initial misgivings about divergences in quantum field theories were dispelled by the application of regularization and renormalisation techniques in the second half of the 20th century to extract meaningful information from both infrared and ultraviolet divergences.  In regularization, an infinitesimal, usefully definable, spatial distance $\epsilon$ (or $\Lambda \sim\frac{1}{\epsilon}$ in momentum space) is set and calculations performed in the limit $\epsilon \rightarrow 0$.   In physical terms, regularization equates to not allowing the internal lines in Feynman diagrams to fluctuate at momentum $k > \Lambda$. If the resulting expression is definable it may contain terms proportional to $\Lambda$. Once regularization has been achieved, the renormalisation process refers to techniques to determine whether these $\Lambda$-proportional terms are significant (a non-renormalisable theory) or may be canceled (a renormalisable theory).  A non-renormalisable theory such as gravity in four dimensions, for example, may not be tackled using perturbative quantum field theory techniques.  For a renormalisable theory, a limited number of measurable field parameters can be calculated which are independent of $\Lambda$, which is an arbitrary energy scale above which even the bare theory may or may not be applicable.    
%
%
Until the late 1960s, it was not well understood why high energy self interactions in renormalisable theories had so little impact at lower, infrared energy scales where measurements could be taken, and thus why the techniques worked so well.  renormalisation group theory emerged to address this and it describes how the physical couplings (e.g. $\lambda_P$ in $\phi^4$ theory, where $P$ denotes physical) vary with momentum scale $k$ where $k < \Lambda$.  It is useful in investigating the high and low energy behaviour of quantum field theories.  To look at this concept further, we consider the generating functional of a scalar field theory in Euclidean space time with field parameter $\phi$ (source term here is zero for brevity):
\be \label{barepartition} Z[J]_{\Lambda} = \int [{\cal D} \phi]_{\Lambda} \exp (-S[\phi]), \ee
with $[{\cal D} \phi]_{\Lambda} = \prod_{|k| < \Lambda} d\phi(k)$.  Wilsonian renormalisation takes shells of $\phi(k)$ at $|k|< \Lambda$, integrates out these field degrees of freedom.  The "group" transformations are energy scale transformations, $k \rightarrow k + \delta k$, and the impact of these on the effective parameters of the theory, such as $\lambda_P$, are studied.  In a scale invariant theory the physics remains the same under such transformations.  However, in all useful, fully quantized field theories of the standard model, scale invariance appears to be broken, \cite{Peskin}.\\\\ In implementing renormalisation group techniques, the starting point is a cutoff at scale $k = \Lambda$ which represents the bare or classical theory, as in (\ref{barepartition}).  To arrive at an expression for an effective action at some energy scale $k < \Lambda$ the high energy field degrees of freedom $\phi(\Lambda)\rightarrow \phi(k)$ are integrated out.  To do this we divide the field degrees of freedom into two sharply defined sectors:
\bea\label{phidef2}
 \phi(k) &= \Phi(k)&\text {for}\qquad {|k|\leq \Lambda-\delta\Lambda}\text {     (0 otherwise)}\\
\phi(k) &= \varphi(k)&\text {for}\qquad{\Lambda-\delta\Lambda<|k|\leq \Lambda}\text { (0
otherwise),}
\eea
where $\phi(k) = \Phi(k) + \varphi(k)$.  The generating functional up to
momentum scale $\Lambda$ is:
\be
\label{E002ii} Z_{\Lambda}= \int_{\Lambda} {\cal D}(\phi)\exp (-S_{\Lambda}[\phi]) = \int _{0<k\leq \Lambda-\delta\Lambda}
{\cal D}(\Phi)\int_{\Lambda-\delta\Lambda<k\leq \Lambda} {\cal D}(\varphi)\exp (-S[\Phi+\varphi]). \ee
This can be written as:
\be \label{E002iii}Z_{\Lambda}=\int
_{0<k\leq \Lambda-\delta\Lambda}
{\cal D}(\Phi)\exp (-S_{\Lambda-\delta\Lambda}(\Phi)), \ee
where the Wilsonian effective action $S_{\Lambda-\delta\Lambda}(\Phi)$ is defined as involving only the components of $\phi(k)$ with $|k| < \Lambda-\delta\Lambda$:
\be \label{Wilson_effaction}  \exp (-S_{\Lambda-\delta\Lambda}(\Phi))=  \int_{\Lambda-\delta\Lambda<|k|\leq
\Lambda}{\cal D}(\varphi)\exp (-S[\Phi+\varphi]). \ee
In $\phi^4$ scalar theory, the effective action $S_{\Lambda-\delta\Lambda}(\Phi)$ includes corrections proportional to powers of the bare coupling $\lambda$ which contain information on the quantum interactions of the large $k$ components in the sector $\varphi(k)$. Wilsonian group theory leads to expressions for the variation of the effective parameters of a theory with energy scale $k$, with, as before $k < \Lambda$.  The $\beta$ functions are differential flow equations, where $\lambda_i$ are the effective couplings:
\be \label{beta functions} k\partial_k\lambda_i(k) = \beta (\lambda_i).\ee
Fixed points of a theory are defined where $k\partial_kS_{k}=0$, that is, the effective action is unchanged by a slight transformation, $\delta k$ in energy scale.  The simplest case of a fixed point is the free-field scalar theory where the action is given by $S_{k}=\int d^Dx\frac{1}{2}(\partial_\mu\phi)^2$ (also known as a trivial or Gaussian fixed point).  Here the couplings vanish and in the vicinity, the form of the $\beta$ functions can be studied.  Couplings that grow in strength approaching this fixed point are termed relevant while those that die out are termed irrelevant.  Quantum field theories with irrelevant couplings are non-renormalisable.

\subsection{The Exact Renormalisation Group Equation}\label{The Exact renormalisation Group Equation}
The equation (\ref {Wilson_effaction}), depending on the nature of $S$ the effective action, is generally solved by perturbative methods, where one or two loop approximations may suffice.  In this thesis, we are interested in very high energy theories where perturbation theory may not be well established and where exact expressions are useful.  The exact renormalisation group (ERG) techniques, \cite{non-perturbative renormalisation Flow} build on Wilsonian methods. The resulting equations are non-perturbative in that they capture all quantum fluctuations in an effective expression and are based on applying a cutoff as in (\ref{phidef2}) and integrating out higher energy degrees of freedom.  Expanding $S_k(\Phi + \varphi)$ around $\Phi$ gives, (where the phase space volume within $p\rightarrow \Lambda$ is $V = \int \frac{d^Dp}{(2\pi)^D})$:
\be \label{taylor_1} S_k(\Phi + \varphi)= S_k(\Phi) + \frac{1}{V}\int _k
\frac{\delta S_k(\Phi)}{\delta \Phi (p)}\varphi(p)
\,+\,\frac{1}{2V^2}\int _k\int _k \varphi
(p)\frac{\delta^2S_k(\Phi)}{\delta \Phi (p)\delta \Phi
(q)}\varphi (q)  +  {\cal O} (\varphi)^3\ee
From (\ref{taylor_1}) one can obtain an expression for the evolution of the Wilsonian effective action in a constant field $\Phi_0$ configuration:
\be \partial_k \, S_k (\Phi_0)=\lim _{\delta
k \rightarrow 0}\, \frac{(S_{k}(\Phi_0)-S_{k-\delta
k}(\Phi_0))}{\delta k}.\ee
Evaluation of the Gaussian integrals in (\ref{taylor_1}) leads to the Wegner-Houghton equation, \cite{Wegner-Houghton}, where here $|p| = k$ and $\Omega_D$ is the solid angle in $D$ dimensions.
\be \label {010_1}\partial_k S_k (\Phi_0) =
-\frac{1}{2}\Omega_D k^{D-1} \frac{V_D}{(2\pi)^{D}}\ln
\left[\frac{\delta^2S_k(\Phi_0)}{\delta \Phi (p)\delta \Phi (-p)} \right].
\ee
A full derivation of this is provided in Appendix \ref{appendix Exact renormalisation} where we note that the term containing $\frac{\delta S_k(\Phi)}{\delta \Phi (p)}$ vanishes as we consider a saddle point where $\frac{\delta S_k(\Phi)}{\delta \Phi (p)} = 0$ and the field $\Phi$ is constant with spacetime.  A key point to note concerning the derivation of eq (\ref{010_1}) is that we consider an infinitesimal momentum shell $\delta k << k$ and the equation becomes an exact expression as we take the limit $\frac{\delta k}{k}\rightarrow 0$.  We make two points about the introduction of the infrared cutoff, $\Lambda$.  Firstly, if the cutoff is sharp, the Wilsonian exact renormalisation group methods may only be used to study the behaviour of the effective potential, as the derivative, kinetic terms will prove problematic at the sharp boundary.  In this instance, a smooth cutoff will be required, \cite{Wetterich:1992yh}.  Secondly, the introduction of an energy scale cutoff is not in general gauge invariant, as gauge invariance implies energy scale invariance.  In our treatment, we note that we use a running variable other than energy scale or momentum, which may overcome gauge invariance issues associated with evolution with energy scale as in the original Wilsonian treatments, \cite{Wegner-Houghton}.

\subsection{An Alternative Exact Approach}\label{An Alternative, Exact Approach}
Following this brief description of the exact renormalisation group, we describe an alternative non-perturbative technique for deriving an exact form for an evolution equation for the effective action.  Motivated by the exact renormalisation group equations they were developed in \cite{non-perturbative renormalisation Flow} and utilised in \cite{Alexandre1}, \cite{Alexandre2}, \cite{Alexandre3}, \cite{Alexandre4} and \cite{Alexandre5}.  These results, along with the results derived in Section \ref{The Effective Action and Potential} on the effective potential are key building blocks in the non-perturbative techniques used in this thesis.  We are interested in the evolution of quantized parameters like $\Gamma(\phi_{cl})$.  \cite{Alexandre0} describes a functional method to arrive at an evolution equation of the effective action with a parameter of the bare theory and the equivalence of such an approach with the evolution using the exact renormalisation group.  As an example, we consider a bare scalar theory, in Euclidean space $\phi \equiv \phi(x)$:
\be \label{scalar_theory}  S(\phi) = \int d^Dx \frac{1}{2}\partial_{\mu}\phi \partial^{\mu}\phi + \frac{1}{2}bm^2\phi^2  - \frac{\lambda}{4!}(\phi^2)^2, \ee
to which we have added a unit-less parameter $b$ to illustrate our non-perturbative evolution technique.  When $b=\infty$ the theory is classical because the particle becomes infinitely massive.  The decrease of $b$ coincides with the increased importance of quantum fluctuations, the full extent of which are apparent when $b \rightarrow 1$.  We note the analogy with the exact renormalisation group derivations where the cut off $k$ is taken from $\Lambda \rightarrow 0$ to capture quantum effects.   Starting with (\ref{legeffact}), we take the partial derivative with respect to $b$ and we consider $b$ and $\phi_{cl}$ to be independent variables of $\Gamma [\phi_{cl}]$ and consider a constant field configuration so that $\frac{\partial \phi_{cl}}{\partial b} = 0$.
\be \label{legeffact_b_derivative} \partial_b\Gamma [\phi_{cl}] =  \partial_b W[J] + \int d^4x \frac{\partial W[J]}{\partial J(x)}\frac{\partial J(x)}{\partial b} - \int d^4x
\phi_{cl} \, \frac{\partial J(x)}{\partial b} = \partial_b W [J].  
\ee
where we have used the definition of the classical field in (\ref{classicalfield}) and performed a chain rule manipulation similar to that used in (\ref{using_chain_rule}).  We have from our definitions (\ref{partitiondefi}) and (\ref{def_of_W}):
\be \label{evolution_W} \partial_b W [J(x)] =  \frac{1}{Z}\partial_b Z = \frac{1}{Z}\int {\cal D} \phi \exp \left[-S(\phi) - \int d^Dx J(x)\phi(x)\right] \partial_b S(\phi), \ee
and can evaluate $\partial_b W [J]$ directly, using (\ref{classicalfield}) with (\ref{evolution_W}).
\bea \label{scalar_theory_b_evolution} \partial_b W [J]  &=& -\frac{m^2}{2}\int d^4x <\phi(x)\phi(x)> \nn &=& -\frac{m^2}{2}\int d^4x\int d^4y \delta^4(x-y) <\phi(x)\phi(y)>.\eea
Using the results (\ref{classicalfield}), (\ref{secondderivative}), (\ref {inverse relationship1}) and (\ref{legeffact_b_derivative}) the following is obtained.
\bea \label{scalar_theory_b_evolution1} \partial_b\Gamma [\phi_{cl}] &=& -\frac{m^2}{2} \int d^4x d^4y \delta^4(x-y)\left[ \left(\frac{\delta^2 \Gamma [\phi_{cl}]}{\delta
\phi_{cl}(x)\delta \phi_{cl}(y)}\right)^{-1} + \phi_{cl}(x)\phi_{cl}(y)\right] \nn &=&
-\frac{m^2}{2}\left[ \int d^4x (\phi_{cl}(x))^2  + \mbox{Tr}\left(\frac{\delta^2 \Gamma [\phi_{cl}]}{\delta \phi_{cl}(x)\delta \phi_{cl}(y)}\right)^{-1} \right], \eea
where we define $\mbox{Tr}[...] = \int d^4x d^4y \delta^4(x-y)[...]$.  We note that as with the exact renormalisation result obtained in (\ref{010}) the equation (\ref{scalar_theory_b_evolution1}) is an exact, non-perturbative expression for the effective action. While (\ref{010_1}) uses the momentum cutoff $k$ as an evolution parameter, the result (\ref{scalar_theory_b_evolution1}) is not restricted to this form and uses any suitable parameter, the evolution of which captures all quantum fluctuations.  Typically equations such as (\ref{scalar_theory_b_evolution1}) are not directly solvable and an assumed form for the effective action is used in conjunction with this to investigate viable solutions.  For example, the gradient approximation for scalar theory was described in the context of the exact renormalisation group in the Appendix in (\ref{phi4i}), which may be applied here, (in this $\phi \equiv \phi_{cl}$).
\be \label{Alt_evolution_assumed_form} S_{eff}(\phi)=\int
d^Dx\,\left(\frac{1}{2}Z(\phi)\partial_\mu\phi\partial^\mu\phi + {\cal O}(\partial_\mu\phi)^4 + 
U_{eff}(\phi)\right)\equiv \Gamma,\ee
where $U_{eff}(\phi)$ and $Z(\phi)$ contain quantum fluctuations and $\Gamma$ is the Legendre effective action.  Here, $\partial_b S(\phi)_{eff}$ may then be compared with (\ref{scalar_theory_b_evolution1}) and viable solutions found.  This is the approach we take with effective theories for the QCD axion and bosonic string in this thesis.

\section {Convex Potential and Spinodal Instability} \label {Section Spinodal Instability} 
\begin{figure}[h] \centering
\includegraphics[scale=1.8]{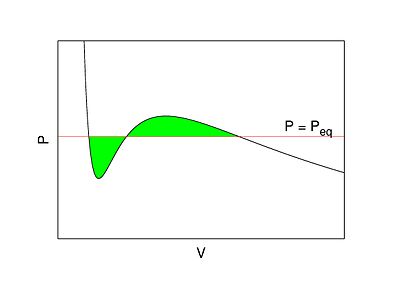}
\caption[Maxwell Construction effect]{The curve is an isotherm in a real-gas model (such as that associated with a van der Waal's equation of state). The unstable concave region and the convex minimum are replaced by the straight line at constant pressure P.  \cite{Maxwell Construction}.}\label{maxwell_construction}
\end{figure}

\begin{figure}[h] \centering
\includegraphics[scale=0.8]{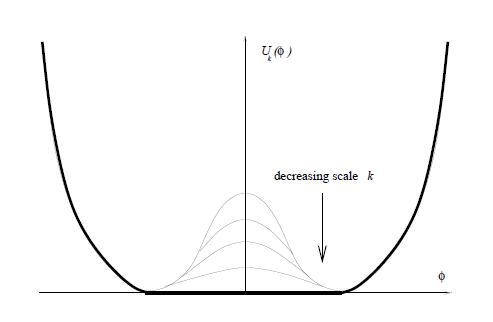}
\caption[Spinodal Instability.]{The flattening of a double well bare potential in one dimension by quantum effects, \cite{Inflaton and spinodal Instability}.  In the context of exact infinitesimal
Wilsonian renormalizatin group studies, the spinodal instablity is
compensated, at each step, by the presence of a non-trivial saddle
point in the path integral defining the blocking transformation, \cite{Instability Induced renormalisation}. The
result of this cancellation is the flattening of the Wilsonian effective
potential (the "average effective potential"), as the observational
scale $k$ goes to 0, as shown in the figure. It is noted that non-trivial saddle
points in the path integrals defining the renormalisation group
transformations imply a "tree-level renormalisation", since the
bare non-convex part of the initial potential is removed by
quantum fluctuations.}\label{spinodal_instability}
\end{figure}

The convex shape of the effective potential in a scalar field is a well known effect in classical physics and we begin this section with a brief discussion of the Maxwell construction, \cite{Maxwell Construction}.  Figure (\ref{maxwell_construction}) represents an isotherm predicted by the Van der Waals equation of state (the oscillating curve), where $P$ is pressure and $V$ is the volume occupied by the gas at that pressure.  The middle section of the curve where $\frac{\partial P}{\partial V} > 0$ is not physical in that the pressure is decreasing with increasing volume.  The two sections of the curve with$\frac{\partial P}{\partial V} < 0$ bounding the green shading are known as meta stable states.
In practice the two green sections are replaced by the straight line of constant pressure $P=P_{eq}$ where a phase transition from liquid to gas is occurring and both phases are present in the system.  This two-phase system remains at constant pressure with increasing volume until all the liquid has vapourised (where the isotherm in Figure (\ref{maxwell_construction}) intersects the $P=P_{eq}$ line on the right hand side).  In the Maxwell construction which is used to resolve the unphysical nature of the equation of state isotherms, it is shown that the internal energy of the system (assumed to have a fixed numbers of particles) depends only on the volume for a given isotherm.  It is shown that the change in energy between the two outer volume points where the $P=P_{eq}$ line intersects the isotherm may be calculated along either the $P=P_{eq}$ line of constant pressure or along the Van der Waals isotherm - both give identical results.  The $P=P_{eq}$ line represents the physical reality and thus the concave point of the isotherm is flattened with the straight line.   The method, which is inserted by hand rather than analytically, relies on the fact that the Gibbs free energy of the two phases must be equal when they coexist and the two areas shaded in green are of equal area.  It may be used in any thermodynamical system with $P$ and $V$ replaced with any pair of conjugate variables which can be used to express the internal energy of the system.  We now consider this in the context of quantum field theory.
\begin{figure}[h] \centering
\includegraphics[scale=.2]{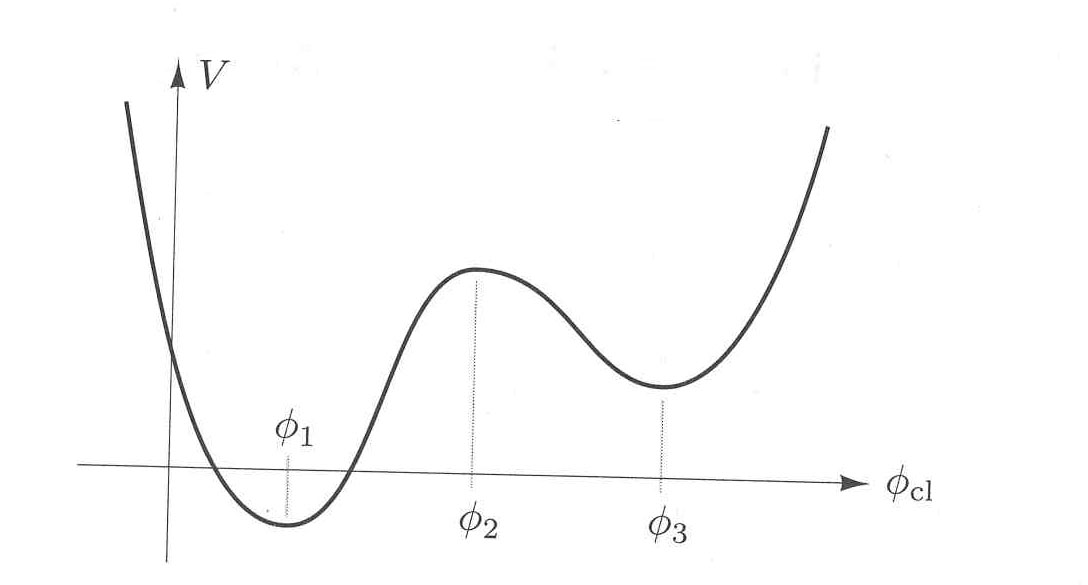}
\caption[Convexity of the Effective Potential - 1]{The curve indicates a possible form for the effective potential of a scalar field, with extrema at the three points $\phi_{1,2,3}$. In our notation $U_{eff} = V$  \cite{Peskin}.}\label{convex_1}
\end{figure}
\begin{figure}[h] \centering
\includegraphics[scale=.2]{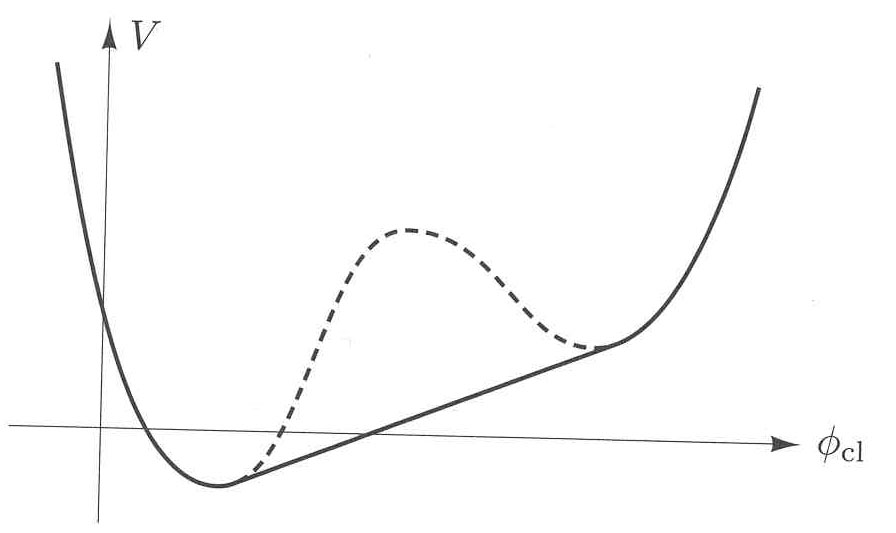}
\caption[Convexity of the Effective Potential - 2]{In practice the concave part of the effective potential is flattened. In our notation $U_{eff} = V$ \cite{Peskin}.}\label{convex_2}
\end{figure}
\\\\ \cite{Peskin} notes that there is a field theory analogy for the Maxwell construction in thermodynamics in which the conjugate variables are the effective potential $U_{eff}$ and classical field $\phi_{cl}$. Figure (\ref{convex_1}) represents a form of $U_{eff}(\phi_{cl})$ with an absolute minimum at $\phi_1$, a local minimum at $\phi_3$ and local maximum at $\phi_2$, and a constant background field configuration.  In eq. (\ref{effective_potential_theory}) it was noted that the extrema of $U_{eff}(\phi_{cl})$ represent the stable quantum states of the theory, in this case $\phi_1$, $\phi_3$ and $\phi_2$.  In the case of our potential form in figure (\ref{convex_1}), it is clear that the stable states are the minima: $\phi_1$ and $\phi_3$; with $\phi_3$ being a locally stable state which could decay to $\phi_1$ via tunneling.  $\phi_2$ does not represent a stable state even though it is an extrema of $U_{eff}(\phi_{cl})$.  In a manner analogous to the Maxwell construction, if we consider a value of $\phi_{cl}$ between $\phi_1$ and $\phi_3$, it can be described by a superposition of the two stable vacuum states (where $U_{eff}$ is a minimum) such that, where $a$ is between $0$ and $1$.
\be \phi_{cl} = a\phi_1 + (1-a)\phi_3. \ee
We may also describe the average value of $U_{eff}$ at $\phi_{cl}$ as:
\be U_{eff}(\phi_{cl}) = a U_{eff}(\phi_1) + (1-a)U_{eff} (\phi_3), \ee
whose line is represented by the bold line in Fig (\ref{convex_2}).  Thus our requirement to see that the true vacuum state is represented by the minimum rather than maximum of $U_{eff}$ leads to a flattening of the concave part of $U_{eff}$.  In effect this implies the condition:
\be \label{convex suppression} \frac{\delta^2 U_{eff}(\phi_{cl})}{\delta \phi_{cl}^2} > 0. \ee
It should be stressed that this scenario applies to a configuration with source $J$ set to zero and in a constant field configuration as described by eq. (\ref{effective_potential_theory}).  As with the theoretical form of the isotherm in Figure (\ref{maxwell_construction}) yielding unphysical regions where phase transitions mean that the Maxwell construction is used to remove these by hand, unphysical regions in the effective potential's form as in figure (\ref{convex_1}) are removed by hand.  The value of the local minima are unaffected by this.\\\\
The shape of figure (\ref{convex_1}) is similar to the double well scalar potential which gives rise to spontaneous symmetry breaking as described in section \ref{section SSB}.  From the very qualitative account above it should be that when quantized, a concave scalar potential (with source set to zero and constant field configuration) should be flattened when the system is quantized.  This is indeed a well known phenomenon and one that we utilize in our discussion of the flattening of the axion's effective potential prior to any interactions in section \ref{Flattening of the Axion Potential}.  In practice this flattening is due to spinodal instability effects which are briefly outlined, with this qualitative account taken from \cite{Instability Induced renormalisation}, \cite{Spinodal Instabilities and the Dark Energy Problem}, \cite{Tachyonic Instability}, \cite{Spinodal Instability and Confinement} and \cite{Inflaton and spinodal Instability}.\\\\  We consider the potential:
\be \label{tachyonic SSB} U(\phi) = - \frac{m^2}{2}\phi^2 + \frac{\lambda}{4!}\phi^4 + \frac{m^4}{4\lambda}.  \ee
In (\ref{tachyonic SSB}), $U(\phi)$ has a local maximum at $\phi=0$ (when $U'' = -m^2$) and two minima at $\phi = \pm v$ ($v^2 = \frac{2 m^2}{\lambda^2}$).  The equation of motion for the scalar field fluctuations about the $\phi = 0$ is quoted without derivation, where $\phi_k$ are the momentum ($k$) modes of the field and $U''$ is with respect to the fields and where the momentum squared in Euclidean coordinates can be expressed in terms of the time and spatial components, $k^2 = \omega^2 + \vec{k}^2$.
\be \label{tachyonic SSB2}(k^2 + U'')\phi_k = 0. \ee
While symmetry is still evident at $\phi = 0$ and time $t = 0$, the field is massless for the quantum fluctuations $\phi_k = \frac{1}{\sqrt{k}}\exp (-i\omega t + i\vec{k}\vec{x})$ (that is with $m$ and $\phi$ terms inactive).  As $\phi$ deviates slightly from 0, the $m$ term is "turned on", from when $|\vec{k}| < m$, the $k$ term in (\ref{tachyonic SSB2}) results in exponential growth of these modes implying a lack of restorative force driving the small quantum fluctuations.  Qualitatively, \cite{Tachyonic Instability} notes that these $|\vec{k}| < m$ wave modes grow in amplitude and reach energy $\sim \frac{m^4}{\lambda}$ which is of similar size to the initial maximum  of the bare potential $U(0)=  \frac{m^4}{4\lambda}$.  Thus a large part of $U(0)$ is transferred to the kinetic energy of the field as it rolls to the minimum at $U(\phi)$.  The flattening of the concave classical potential in the quantum model occurs to avoid exponential build up of negative mass squared-originating (hence tachyonic) undamped fluctuations (Figure \ref{spinodal_instability}).  It is again stressed that at this point that this analysis is limited to non-interacting scalar field models.
The graph in figure (\ref{spinodal_instability}) was inspired by results computed numerically in \cite{Instability Induced renormalisation}.  Here the authors combine the non-perturbative methods outlined in section (\ref{The Exact renormalisation Group Equation}) with the saddle-point (or steepest descent) approximation method for evaluation of divergent integrals.  They found that when the spinodal instabilities occur near $k^2=m^2$, non-trivial saddle points appear in the integrals of the type expressed in (\ref{taylor_1}).  At tree level this gives $k^2-m^2+\frac{\lambda}{2}\phi^2 = 0$, and a flat potential for any value of $k$ down to the IR limit $k=0$, within the two minima of $U_k(\phi)$.

\chapter{Axion Theory} \label{chapter axion}

\section{Axion Related Topics}
Axion physics involves relationships between a number of concepts and phenomena in quantum field theory and particle physics which are reviewed here briefly both to introduce these relationships and terminology used.  Note: in this and the next Chapters the term "axion" refers to the scalar field postulated by Peccei-Quinn to explain the CP problem, as opposed to the term "string-axion" used in Chapters 5 and 6.  Section (\ref{string-axions}) of this chapter briefly mentions string-originating axions, however.

\subsection {QCD and its Symmetries} \label{QCD and its Symmetries}
Quantum chromodynamics (QCD) is a highly symmetric quantum field theory and much of QCD research has centered around how these symmetries are broken or preserved and the resulting phenomenological implications.  Initially the strong interaction was believed to be explained by a new quantum number, isospin, described by $SU(2)_I$ symmetry in which the proton and neutron existed in the fundamental representation and the thee pions, postulated to be Goldstone bosons of the theory, in the adjoint representation of the $SU(2)_I$ group.   Because the pions were shown to have a light mass, it was postulated that this symmetry was slightly broken explicitly.  Modern interpretations refer to a more fundamental, quark-based model with associated symmetries of the Lagrangian and vacuum, the key ones of which are described. 

\subsubsection{SU(3) Colour Symmetry}
This is the defining symmetry of QCD and results in a non-Abelian gauge theory:
\be \label{QCD LAG} \mathcal {L}_{QCD} =
\overline{q_i}(i\gamma^\mu\partial_\mu-m)q_i -
g(\overline{q_i}\gamma^\mu T_{ij}^a q_j)G_\mu^a -
\frac{1}{4}G^{a\mu\nu}G^a_{\mu\nu}. \ee
Here $q_i (x)$ and $\overline{q_i}(x)$ are quark and anti-quark fields of $i$ flavours in the fundamental representation of $SU(3)$.  $G_\mu^a$ are the eight gluon
fields which lie in the adjoint representation of $SU(3)$ with $T_{ij}^a$ the generators for this representation and $\gamma^\mu$ are the Dirac matrices and $m$ is the quark mass matrix.
$G^a_{\mu\nu}$is the gauge covariant field strength tensor given by:
\be \label {QCDLagrangian} G^a_{\mu\nu}=\partial_\mu G_\nu^a-
\partial_\nu G_\mu^a-g f^{abc}G_{b\mu} G_{c\nu}, \ee
where $g$ is the coupling constant of the theory and $f^{abc}$ are
the structure constants.  Throughout we use a trivial summation notation for the $a,b,c$ and $i,j$ indexes, with the "`up-down"' convention respected for the space time indexes.  This gauge symmetry is an exact one which remains unbroken and along with electroweak $SU(2) \times U(1)$ is one of the fundamental symmetries of unified theory.

\subsubsection{Baryon Number}
Baryon number is defined as $B=\frac{1}{3}(n_q - n_{\bar{q}})$ where $n_q$ and $n_{\bar{q}}$ are the number of quarks and anti quarks in a baryon.  It is a conserved quantum number and hence exhibits a $U(1)_B$ symmetry, analogous to the $U(1)$ electromagnetic symmetry.  If baryon number is conserved precisely, it is an exact symmetry.  We consider a generic Lagrangian with a quark and anti quark doublet:
 \be Q = \begin {pmatrix} u  \\ d \end
{pmatrix} ; Q_L = \begin {pmatrix} u_L  \\ d_L \end
{pmatrix} ; Q_R = \begin {pmatrix} u_R  \\ d_R \end
{pmatrix}. \ee
Here we have the relationships:
\be Q_L = \left( \frac{1-\gamma^5}{2}\right) Q ; \,\, Q_R = \left( \frac{1+\gamma^5}{2}\right) Q, \ee
where $\gamma^5$ is the chirality matrix:
\be \gamma^5 = \begin {pmatrix} -1 & 0  \\ 0 & 1 \end
{pmatrix}. \ee
There is a conserved vector current $j^{\mu} = \bar{Q}\gamma^{\mu}Q$ from the $U(1)$ transformation $Q \rightarrow (\exp i \theta) Q$.  If $m_u = m_d = 0$ the left and right hand fields transform independently in the chiral symmetry that transforms $Q \rightarrow \exp (i \beta \gamma^5 )Q$ where $\beta$ is a phase degree of freedom introduced by the symmetry.  If we consider $m_u\simeq m_d\neq 0$ and $m_u,  m_d \ll \Lambda_{QCD}$, the quarks have a small mass meaning the chiral (also termed "axial vector") symmetry is explicitly broken as the quarks acquire mass.   The chiral current in this model is characterised by:
\bea j^{\mu 5} &=&  \bar{Q}\gamma^{\mu}\gamma^{5}Q \nn
\partial_{\mu}j^{\mu 5} &=& 2 i m \bar{Q}\gamma^{5}Q, \eea
and the chiral current is conserved in the massless case.  However, (see for example \cite{Peskin}, 19.2) when the above classical symmetries are quantized the Adler-Bell-Jackiw anomaly in four dimensions results in breaking of the chiral $U(1)_B$ symmetry even in the massless case of (\ref{QCD LAG}) such that:
\be \label{chiralanomaly}
\partial_\mu
j^{\mu5}= - \frac{Ng^2}{16\pi^2}\widetilde{G}^{\mu\nu}G_{\mu\nu}, \ee 
where $N$ is the number of quarks in the model, and $\widetilde{G}^{\mu\nu}$ is the dual of $G^{\mu\nu}$ and equal to $\frac{1}{2}\epsilon^{\alpha\beta\mu\nu}
G_{\alpha\beta}$.  Detailed analysis \cite{'t Hooft} by 't Hooft and others in the early 1970s revealed the chiral anomaly in a four space time dimensional QCD vacuum to be a one-loop effect, breaking what would have been a fundamental $U(1)_B$ chiral symmetry.  The experimental absence of a pseudo-scalar of mass smaller than $\sqrt{3}m_{\pi}$, \cite{weinberg 1}, indicative of a spontaneously broken $U(1)_B$ chiral symmetry is known as the "$U(1)$ problem".  

\subsubsection{Flavour Symmetry}
The idea of $SU(2)_I$ isospin symmetry gave way to flavour symmetry $SU(N_f)$ where $N_f$ is the number of quark flavours (denoted as $i$ in (\ref{QCD LAG})).  At present it is thought there are six flavours of quarks.  If all the flavours are massless there is also a chiral $SU(N_f)$ symmetry. For our analysis of axion physics we consider only the lightest two quarks, $u$ and $d$, such that $m_u\simeq
m_d\ll \Lambda_{QCD}$, where $\Lambda_{QCD}$ is the energy scale at which confinement occurs, $\sim 200 MeV$.  As with the $U(1)_B$ symmetry there is a $SU(2)$ vector symmetry and if $m_u = m_d = 0$  a chiral $SU(2)$ symmetry.  The vector symmetry is one of rotational invariance between $m_u = m_d$ and if the masses vary slightly it is an approximate symmetry. The vector and chiral currents are:
\bea j^{\mu a} &=&  \bar{Q}\gamma^{\mu}\tau^a Q \nn
j^{\mu 5a} &=&  \bar{Q}\gamma^{\mu}\gamma^{5}\tau^a Q , \eea
with $\tau^a$ the Pauli matrices.  One may calculate the quantum anomaly contribution to the divergence of the $SU(2)$ chiral current as, \cite{Peskin}:
\be \label{chiralanomaly2}
\partial_\mu
j^{\mu5a}=-\frac{g^2}{16\pi^2}\widetilde{G}^{\mu\nu}G_{\mu\nu} \cdot tr[\tau^a t^c t^d] = -\frac{g^2}{16\pi^2}\widetilde{G}^{\mu\nu}G_{\mu\nu}tr[\tau^a] tr[ t^c t^d] = 0, \ee 
as the trace over a single $\tau^a$ vanishes.  It is noted that $\tau^a$ is an isospin matrix while $t^c$ is a colour matrix.  Thus unlike in the case of the $U(1)_B$ chiral symmetry, in $SU(2)$, the chiral anomaly does not break the symmetry.  Due to the presence of a triplet of low mass mesons (the pions), it is postulated however that the chiral $SU(2)$ symmetry is spontaneously broken. This is due to the fact that even when the mass of the quarks is set to zero, the QCD vacuum is thought to consist of a condensate of quark-anti quark pairs arising from the vacuum such that:
\be <0|\overline{Q}Q |0> = <0|\overline{Q}_LQ_R + \overline{Q}_RQ_L |0> \neq 0. \ee
This non-zero expectation value of the quark-anti quark pairs in the condensate causes breaking of the $SU(2)$ chiral symmetry, if we consider a model with two quarks, up and down.  The quark-anti quark pairs act as scalar bound states and three massless quark-anti quark Goldstone bosons result - the pions.  Since in reality the up and down quarks have a small mass the chiral $SU(2)$ symmetry is also explicitly broken and these pions acquire mass.

\subsubsection{Scale Invariance}
Scale invariance is a sub-symmetry of the conformal symmetry group.  It occurs in a massless field theory, with field parameter $\varphi(x_{\mu})$ and with no dimensionful coupling constants that is invariant under the following transformation, \cite{Peskin}.
\be \label{scale invariance} \varphi(x_{\mu}) \rightarrow \exp (-D\sigma) \varphi(e^{-\sigma} x_{\mu}),  \ee
where $D$ is the canonical mass-dimension of the field $\varphi $ and the space time coordinate transforms as $x_{\mu} \rightarrow (\exp -\sigma) x_{\mu}$.  (We write the constant in the form $\exp -\sigma$ as $\sigma$ will be promoted to $\sigma (x_{\mu})$ in the context of later string-related discussions of conformal transformations).  These are equivalent to the renormalisation group transformations discussed in Section \ref{renormalisation Group Theory}. Classically this symmetry is evident in massless QCD, as in (\ref{QCD LAG}) with $m=0$ and where the coupling $g$ is dimensionless. Classically, the physics of such theories remain invariant if the length scales are multiplied by a common factor.  In quantum field versions of such a theory one can equate such invariance with length scales inverse to mass scales.  Following renormalisation, the effective Lagrangian is dependent not only on the classical coupling parameters but also an arbitrary mass scale $M$ introduced through the regularization of ultraviolet divergences. The scale invariance has potentially been broken by the chosen cut off.  This is evident in $\phi^4$ theory by the flow in the coupling $\lambda$ with the arbitrary mass parameter $M$, \cite{Peskin}.
\be \lambda \rightarrow \lambda + \frac{\lambda^2}{4(\pi)^2}(N+8)\cdot \frac{\delta M^2}{M^2}, \ee
where $O(N)$ is the classical symmetry of the theory.  In general this leads to a formulation for the beta function of the coupling which describes the variation with energy scale $\mu$.
\be \beta(\lambda) = \mu \frac{\partial \lambda}{\partial \mu} \ee
If the beta function vanishes, the quantum field theory is scale invariant - no physically useful examples of these are known in the standard model, \cite{Peskin}.  If $\beta(\lambda) > 0$, as in QED, the coupling approaches zero at low energies to allow for perturbative solutions in this region.  If $\beta(\lambda) < 0$, the theory is said to be asymptotically free, a phenomenon discovered in the early 1970s as applying to non-Abelian gauge theories like massless QCD where the form of the beta function was found to be, \cite{asymmtotic}:
\be \beta (g) \sim - \beta_0 \frac{g^3}{16 \pi^2}, \ee
with $\beta_0$ a constant and $g$ the strong coupling.  Thus the coupling $g$ is high at low energies making effective perturbation theory difficult in this region.  Conversely, high energy behaviour of such theories is predictable via perturbation methods.  An outcome predicted by asymptotic freedom is that the energy required to separate particles governed by such interactions could tend to infinity, potentially explaining why quarks are not seen alone in nature.  

\subsubsection{CPT Symmetry}
CPT symmetry is now considered to be a fundamental symmetry of the standard model, including QCD.  Parity is a discrete transformation which sends $(\textbf{x}, t)$ to $(-\textbf{x}, t)$, that is a reflection in space.  A discrete time transformation on the other hand sends $(\textbf{x}, t)$ to $(\textbf{x}, -t)$.  Charge conjugation is a discrete transformation which changes a particle into its antiparticle.  The CPT Theorem, \cite{CPT theorem}, states that any quantum theory in flat space time is symmetric under the combined CPT transformations provided the theory respects Lorentz invariance, locality and conservation of probability (unitarity).  Theories may break C, P, T or double combinations thereof individually. The couplings of the SU(2) gauge bosons in the QCD Lagrangian violates C and P symmetries individually (\cite{Peskin}, 20.3) but the combined CP transformation is a symmetry of the QCD Lagrangian as it stands in (\ref{QCD LAG}), as is T, the time reversal transformation.  There are, however, possible terms discussed in Section \ref{SectionInstantons} which can break CP symmetry in the QCD Lagrangian, and the fact that this has not been observed in practice is known as the CP problem.

\subsection {Instantons} \label {SectionInstantons}
In the context of a non-linear field theory, solitons are stable, well-behaved solutions to the classical theory, which are stable against decay (or topologically distinct from) to the trivial solution.  Most solitons are exact and non-perturbative. The stability of these solutions arises, \cite{Ryder}, as a result of constraints imposed by boundary conditions of the coordinate space being considered.  This boundary has a non-trivial homotopy group associated with it, which has a mapping with the coordinate space.  This can result in an infinite number of topologically distinct solitons which can be degenerate.  Solitons are often termed topological defects in this context and examples are kinks \cite{Rajantie1}, domain walls and 't Hooft-Polyakov monopoles, \cite{Rajantie2}.  Mathematical discovery of the latter trigged searches for magnetic monopoles.  They are non-linear solutions to gauge theories (such as $U(1)$ electromagnetic gauge theory incorporating the Higgs field) which represent objects with finite energy, localized around a particular point in space, and can be shown to possess magnetic charge.  Here, however, we focus on a form of soliton localized in space and time, hence instanton.  The following is sourced from \cite{Ryder} and where noted.\\\\In the double well potential illustrated by the Lagrangian (\ref{SSBQM}) where the constant $k'$ obeys the condition $k'<0$, one may apply the time independent Schrodinger equation simply as (with factors of $\hbar$ left out for brevity):
\be \frac{d^2\psi}{dq^2} = (2m(V(q) - E)) \psi, \ee
where $\psi$ is the particle's wave function, $q$ its position and $E$ its total energy and $V(q)$ the potential.   A solution is $\psi = \exp (-ik' q)$ where $k' = \sqrt{2m (E-V)}$.  If $k'$ is real one obtains a familiar plane wave solution, however if $k'$ is imaginary, that is $E<V(q)$, as is the case with the particle being in one of the two double wells, one obtains an exponential (i.e. $\exp (-k' q)$) solution which represents quantum mechanical tunneling between the two wells with amplitude proportional to $\exp -(\int_{-q}^q dx \sqrt{2m(V(q)-E)}$.  If we apply the path integral approach to this two dimensional example using a Wick rotation in which $it \rightarrow \tau$, one obtains as the amplitude, with $H$ the Hamiltonian:
\bea < -q |\exp (-H\tau) |q> &=& \int dq(\tau) \exp (-S_E) \nn
S_E &=& \int_{\tau (-q)}^{\tau (q)} d \tau \,\left(\frac{m\dot{q}}{2}\right) + V(q), \eea
where one notes $V(q)$ changes sign under the Wick rotation and the double well is inverted.  This transforms the process from $-q$ to $q$ as the particle traveling from the top of one maximum and rolling to the other.  It is noted that this quantum field theory approach differs from that shown in (\ref{SSB QFT}) in that it does not involve perturbation theory.  Rather this so-called instanton solution is based on applying the path integral approach in Euclidean form to a classical solution in a fixed time span and localised area.  Instantons add alternative solutions to the vacuum structure of a theory, in this case suggesting that the particle may reside with equal probability at both $-q$ and $q$, two areas in the theory disconnected classically and by conventional perturbative quantum field theory, but connected by quantum mechanical tunneling.  \cite {instantons} notes that the result in Euclidean space-time can be equated to Minkowski space time in the path integral approach to a good approximation.  Instanton solutions to four dimensional non Abelian gauge (Yang Mills) theory were discovered \cite{'t Hooft} in the early 1970s shedding new light on the QCD vacuum.  While a rigorous derivation is beyond the scope of this thesis, the major steps are outlined, drawing from \cite{Ryder}, \cite{'t Hooft} and \cite {instantons}.  
\begin{itemize}

\item Considering the Euclidean action in 4D of the kinetic term of the pure Yang-Mills theory, \cite{Zee}:
\be \label{Euclidean QCD} S_E =  \frac{1}{g^2}\int d^4x \, \mbox{Tr} G_{\mu\nu}G_{\mu\nu}, \ee
with $G_{\mu\nu} = G^a_{\mu\nu}T_a$, and $T_a$ are the generators of the symmetry group, $G^a_{\mu}$ are the gauge fields, $G_{\mu}$ is the gauge potential equal to $G_{\mu}^aT_a$ and  $\tilde{G}^{\mu\nu} = \frac{1}{2}\epsilon^{\mu\nu\rho\sigma} G_{\rho\sigma}$.  We take the symmetry group as the $SU(N_f)$ flavour symmetry of a two quark model, where $N_f = 2$.  This is a local, continuous and therefore a gauge symmetry.  Unlike Abelian $U(1)$ theory, this contains cubic and quartic terms representing self interactions of the gauge bosons, $G^a_{\mu}$.  For $S_E$ to be finite in the integral as $|x| \rightarrow \infty $, the potential must be pure gauge in such a configuration:
\be G_{\mu} = U^{-1}\partial_{\mu} U, \ee
where $U$ is an element of the symmetry group, (e.g. in $SU(2)$, $U = \exp i \theta^a T^a$ with $T^a$ the generators in the adjoint representation) and it is noted $G_{\mu\nu} = 0$ with this gauge at the $|x| \rightarrow \infty $ boundary.

\item Euclidean space time in 4D has as its boundary the 3-sphere, or $S^3$.  Meanwhile the group $SU(2)$ may be represented by:
\be U = U_0 + i \sum_{j=1}^3 U_j\sigma_j, \ee
where $\sigma_j$ are the Pauli matrices.  $U$ is unitary with $U_0^2 + U_1^2 + U_2^2 + U_3^2 = 1$, which is also the equation for the 3-sphere, $S^3$.  Thus one can say that the gauge potential at $\infty$ describes a map from group space to physical space $S^3 \rightarrow S^3$, with the mapping defined by an integer $q$ (known as the Pontryagin index).  A solution with one value of $q$ is termed stable if it cannot be continuously transformed into a different $q$ solution, and the mapping is non-trivial.    

\item We can define a total divergence, $\partial_{\mu} K_{\mu}$:
\be \frac{1}{4}\mbox{Tr} \tilde{G}_{\mu\nu}G_{\mu\nu} = \partial_{\mu} \left[ \epsilon^{\mu\nu\rho\sigma}   \mbox{Tr} ( \frac{1}{2} G_{\nu}\partial_{\rho}G_{\sigma} - \frac{i g}{3} G_{\nu}G_{\rho}G_{\sigma})\right] = \partial_{\mu} K_{\mu}. \ee
Applying the classical equations of motion for (\ref{Euclidean QCD}), that is, $D_{\mu}G_{\mu\nu} = 0$ and applying Gauss' theorem gives (where $K_{\bot}$ is the component of $K_{\mu}$ normal to the surface of the volume under consideration):
\be \int d^4x \, \mbox{Tr}\, \tilde{G}_{\mu\nu}G_{\mu\nu} = 4 \int d^4x (\partial_{\mu} K_{\mu}) = 4 \oint_{S^3} d^3x K_{\bot},\ee
which implies that the integral $\int d^4x \, \mbox{Tr}\, \tilde{G}_{\mu\nu}G_{\mu\nu}$ depends only on the homotopy of the mapping $S_3 \rightarrow S_3$.  With our pure gauge condition on the $|x|^2 \rightarrow \infty $ boundary (i.e. $G_{\mu} = U^{-1}\partial_{\mu} U$), it can be shown that:
\be \label{instanton anomaly} \frac{1}{g^2} \int d^4x \, \mbox{Tr}\, \tilde{G}_{\mu\nu}G_{\mu\nu} =  \frac{16 \pi^2}{g^2}, \ee
and we define $q$ here as equal to $\frac{g^2}{16 \pi^2} \int d^4x \, Tr\, \tilde{G}_{\mu\nu}G_{\mu\nu}$ as a measure of the degree of mapping of the $SU(2)$ group space to physical space $S_3 \rightarrow S_3$, \cite{Ryder}, which is the Pontryagin index, $q$ noted above (sometimes referred to as the winding number), \cite{instantons}. The solution of the equations of motion of (\ref{Euclidean QCD}) with the pure gauge condition at the $|x|^2 \rightarrow \infty$ boundary is an instanton.  It represents the transition as Euclidean time evolves from negative to positive infinity, from one vacuum (represented by a homotopy class $n-1$ to another in homotopy class $n$, with Pontryagin index, $q$ in this case equal to $n-(n-1) = 1$.  The non-trivial mapping $S_3 \rightarrow S_3$ is represented by the set of integers which physically means there are an infinite number of identical but topologically separate vacuum configurations.  The $q=1$ case is known as the BPST instanton, \cite{BPST}.  The barrier penetration is given by $\exp (-S_E)$, or $\exp (- \frac{16 \pi^2}{g^2})$ in this case.

\item Thus the QCD vacuum is infinitely degenerate with non-zero transition amplitudes between the vaccua belonging to different homotopy classes.  If $|n>$ is a vacuum described by homotopy class $n$
the real vacuum should be invariant under a transformation (termed a "`large gauge"' transformation) which maps the $n$ vaccua onto one another. A gauge invariant vacuum state, parameterised by $\theta$ is thus constructed as a superposition of the $n$ homotopy class vaccua:
\be \label{theta vacuum} |\theta >  = \sum_{n = -\infty}^{\infty} \exp i \pi \theta |n>, \ee
with $\theta$ a phase with period $2\pi$. $\theta$ parametrizes the degree of tunneling between the $n$ vacuum configurations which occurs in the true vacuum state.  If $\theta\neq 0, 2n\pi$, instanton effects are present and the vacuum state is complex and is not invariant under CP transformations (discussed in more detail in Section \ref{section CP Problem}).  This instanton parameter $\theta$ can be accounted for in the QCD Lagrangian by adding a $\theta$ containing term:
\be \label{CP Breaking term} \mathcal {L}_{\theta} =  -\theta \frac{g^2}{16 \pi^2} \mbox{Tr}\, \tilde{G}_{\mu\nu}G_{\mu\nu}. \ee
$\theta$ parametrizes the degree of tunneling between the $n$ vacuum configurations which occurs in the true vacuum state.

\item Considering the Lagrangian (\ref{QCD LAG}) with $m=0$ and $i=2$, that is a massless two quark model, \cite{'t Hooft} computes the chiral anomaly associated with the $U(1)_A$ symmetry breaking as  (\ref{chiralanomaly})and notes this is a one-loop effect.  Comparing this with the result in (\ref{instanton anomaly}) gives:
\be \int d^4x \, \partial_\mu j^{\mu5} = N \, q, \ee
where $N$ is the number of quarks in the model and $q$ is the Pontryagin index.  This implies that in an instanton background there is non-conservation of the charge associated with axial current.  This implies the possibility of decays which violate both baryon and lepton number (the $U(1)$ axial charge relevant here), such as:
\be p + n \rightarrow e^+ + \bar{\nu}_{\mu}. \ee
The probability of such decays is small, however on the order of $\exp -(\frac{16\pi^2}{g}) \sim 10^{-262}$,  \cite{Ryder}.

\item While the computations leading to (\ref{instanton anomaly}) involved the gauge fields only, they were specific to the gauge group $SU(2)$ relating to a two quark model.  In a more general $SU(N_f)$ model, the corresponding $SU(2)$ subgroup of this will produce the same result, \cite{instantons}.  If both quarks are massless there is also a chiral $U(1)$ symmetry relating to conservation of baryon number.  The $U(1)$ problem mentioned in Section \ref{QCD and its Symmetries} may be resolved by considering that this $U(1)$ chiral symmetry is dynamically broken by instanton effects resulting in the chiral anomaly.  This is an inherent feature of the quantized theory and  chiral $U(1)$ is not a true symmetry of the theory and hence the lack of pseudo-Goldstone bosons whose mass vanishes in the limit $m_u = m_d \rightarrow 0$.

\end{itemize}

\subsection{The CP Problem} \label{section CP Problem}
As noted in \cite{CPT theorem}, under the CPT Theorem, CPT symmetry may only be violated in the case where there is violation of either Lorentz symmetry, locality or unitarity, none of which we wish to consider here.  We thus start by the assumption that CPT symmetry is preserved in strong interactions. In the 1960s, CP violation was observed in weak interactions which implied T violation in order to preserve CPT symmetry.  In strong interactions, instanton effects result in an additional phase degree of freedom to the QCD Lagrangian manifest in (\ref{CP Breaking term}).  This term potentially violates parity (as immediately seen by the four space time indices in the full form $\frac{1}{2}\epsilon^{\mu\nu\rho\sigma} G_{\rho\sigma}G^{\mu\nu}$) while preserving charge symmetry, \cite{Peccei2} and hence can violate CP symmetry.  Why CP and separately T symmetry violation has not been observed in strong interactions, despite being theoretically possible under the standard model, is the CP problem.  In a simplistic view, one may rotate away the CP-violating $\theta$ term via a chiral transformation in the massless $m=0$ QCD theory (\ref{QCD LAG}) and thus preserve the CP and CPT symmetry, \cite{axion dark matter}.
\be \label{large gauge trans}  \theta \rightarrow \theta - \sum_{i=1}^N \alpha_i, \ee
where $i$ is the flavour index.  We consider a two quark model with $m_u = m_d \neq 0$ where  \cite{Peccei1996} notes that the quarks acquire mass from the Higgs mechanism and the quark mass matrix $M_{ij}$ need not necessarily be real nor diagonal.  Chiral transformations similar to (\ref{large gauge trans}) can be performed to diagonalize the mass matrix to derive meaningful mass parameters, \cite{Peccei1996}.
\bea q_{iR} &\rightarrow & \exp (i\alpha_i \frac{\gamma_5}{2}q_{iR}) \\ q_{iL} &\rightarrow & \exp (-i\alpha_i \frac{\gamma_5}{2}q_{iL}). \eea
However, the price paid for this is that CP breaking terms remain in the Lagrangian as the quark mass parameters transform as $m_i \rightarrow \exp (-i\alpha_i) m_i$.  The complex mass term can be factored into one of the quarks, \cite{Witten}. 
\be M_{ij} \label{complex_mass} = \begin {pmatrix} m_u & 0 \\ 0 & m_d \end
{pmatrix} \rightarrow
\begin {pmatrix}  e^{i\overline{\theta}} m_u & 0 \\ 0 & m_d \end
{pmatrix} \ee
This results in the quark-mass phase $\alpha_i$ being added to the $\theta$ parameter, \cite{Peccei1996}:
\be \overline{\theta} = \theta + N_i \alpha_i = \theta + \arg \det M_{ij}. \ee
Attempts to rotate away $\overline{\theta}$ by further transformations as in (\ref{large gauge trans}) will result in complex mass, and CP symmetry violating terms in the Lagrangian.  \\\\The neutron's electric dipole moment (hence nEDM) is a measure of the separation of the centers of negative and positive charge within the neutron and should be a consequence of the CP-violating $\overline{\theta}$ term in the effective Lagrangian when evident as a complex quark mass term: \cite{Peccei2}.  
\be \mathcal {L}_{CP-viol}  = i \overline{\theta} m_q \left[q_i\frac{\gamma_5}{2}\bar{q}_i\right].\ee
The nEDM can be computed by considering this $\overline{\theta}$ term as proportional to a one-loop correction in $n\overline{n}$-meson coupling, \cite{Axion_Kim_2010_Summary} and a relationship of the following arrived at, \cite{Peccei2}:
\be d_n \sim 2.7-5.2 \times 10^{-16} \overline{\theta}, \ee
with the range depending on precise couplings considered.  \cite{NEDM_survey} in 2006 concludes that the phenomenological accuracy puts the nEDM, $d_n < 2.9 \times 10^{-26} e$ cm.  The measurement method compares the Larmor frequency of the neutron spin polarisation in applied electric and magnetic field when $\vec{E}$ and $\vec{B}$ are parallel and anti parallel.  Thus the term $\overline{\theta}$ is limited by the lack of observational evidence of the nEDM to be of the order $\overline{\theta} < 10^{-10}$.  $\overline{\theta}$ is a phase originating as the sum of two unrelated terms ($\theta_{QCD}$ and the electroweak-QCD interaction related term $\arg \det M_{ij}$).  Having period $2\pi$ it could feasibly take any value from $0 \sim \pi$. Why it should be so close to $0$ is the CP problem.  The solution pertinent to this thesis is a theorized additional $U(1)$ symmetry and scalar field termed the axion.  For completeness, several alternatives are outlined, \cite{Axion_Kim_2010_Summary}, \cite{Peccei2}, \cite{Witten}.

\begin{itemize}

\item \textbf{Massless quarks}: As pointed out above, in a massless quark model, one may rotate away with a chiral transformation the CP breaking $\theta$ term to eliminate (\ref{CP Breaking term}), implying that in this case $\theta$ is not a physical parameter within the theory.  \cite{Axion_Kim_2010_Summary} notes that it is sufficient that the mass of the up quark vanish (as evident in (\ref{complex_mass})), but also notes that Weinberg's up/down mass quark ration $Z=\frac{m_u}{m_d} = \frac{5}{9}$ has historically ruled this out, and more recently \cite{Mass_Up_Quark} showed in 2003 through lattice calculations that $Z = 0.410 \pm 0.036$.

\item \textbf{Spontaneous CP breaking }: \cite{Peccei2} postulates that the CP symmetry which is theorized to be broken by the $\overline{\theta}$ term is actually spontaneously broken.  At the bare Lagrangian level one may set $\overline{\theta} = 0$.  However the same source notes that the $\overline{\theta}$ CP symmetry breaking term reappears at the one-loop level and complex Higgs vacuum expectation values are needed to set the quantized CP breaking terms to zero.

\end{itemize}

\section{The Axion} \label{The Axion Section}
  
\subsection {Axion Models} \label{axion_models}
The leading candidate for solving the CP problem is the axion.  The idea was first put forward in two papers by Peccei and Quinn in 1977, \cite{Original Peccei Quinn}.  The theory proposes a new global $U(1)$ "PQ" symmetry for the standard model (later termed $U(1)_{PQ}$ with phase $\ap$).  \cite{Original Peccei Quinn} showed that the condition for CP conservation in eq. (\ref{CP conservation}) below, i.e.  $\overline{\theta} + \alpha = 0$, could be naturally achieved in the quantized Lagrangian, as the effective potential is minimized.  The mass of the up quark is rotated back to the real plane (achieving CP symmetry) and there is no need to set $m_u=0$.  $U(1)_{PQ}$ is spontaneously broken and a Goldstone boson produced from one of the Higgs degrees of freedom.  Here the CP-breaking phase is $\overline{\theta}$ and the phase associated with $U(1)_{PQ}$ is $\alpha$.
\be \label{CP conservation} M = \begin {pmatrix} m_u & 0 \\ 0 & m_d \end
{pmatrix} \rightarrow
\begin {pmatrix}  e^{i\overline{\theta}} m_u & 0 \\ 0 & m_d \end
{pmatrix} \rightarrow \begin {pmatrix}  e^{i\overline{\theta} + \alpha} m_u &
0 \\ 0 & m_d \end {pmatrix}\ee
Peccei termed this spinless scalar field the axion $a(x)$ such that when $U(1)_{PQ}$ is broken at energy scale $f_a$ (known as the scale factor, or decay constant) it is transformed as follows:
\be a(x) \rightarrow a(x) + \alpha f_a.  \ee
In effect, when added to it, the axion promotes the phase $\overline{\theta}$ (which is arbitrary) to a dynamical parameter, or equally, the axion as a dynamical phase can be redefined to absorb $\overline{\theta}$. \cite{Peccei3} notes that expressed in terms of the chiral anomaly (\ref{chiralanomaly}), the invariant effective Lagrangian has the following $\overline{\theta}$ and $a$-containing terms (with axion interactions not included here).
\be \label{axion_lagrangian}  \mathcal {L}_{\overline{\theta},a} = - \frac{1}{2}\partial_{\mu}a\partial^{\mu}a + \overline{\theta} \frac{g^2}{16 \pi^2} \mbox{Tr}\, \tilde{G}_{\mu\nu}G_{\mu\nu} + \xi\frac{a}{f_a} \frac{g^2}{16 \pi^2} \mbox{Tr}\, \tilde{G}_{\mu\nu}G_{\mu\nu}, \ee
where a kinetic term for the axion field has been added by hand.  We note the negative sign in front of this kinetic term as a convention deployed in \cite{Peccei3}.  $\xi$ is the $U(1)_{PQ}$ chiral anomaly co-efficient defined by (\ref{chiralanomaly}):
\be \label{chiralanomalyPQ}
\partial_\mu
j^{\mu5}_{PQ}= - \xi \frac{g^2}{16\pi^2} \mbox{Tr}\, \widetilde{G}_{\mu\nu}G_{\mu\nu}, \ee 
which appears when the $U(1)_{PQ}$ symmetry is explicitly broken by QCD instanton effects, \cite{Witten}.  Mass acquisition by the axion provides a natural mechanism for the minimization of the now-dynamical $\overline{\theta}$ (as explained below).  This initial model triggered a search for the axion and many variations of it, most of which involve the axion acquiring mass via coupling to other fields within the standard model. They key axion models are outlined below.

\subsubsection{Peccei-Quinn-Wilcek-Weinberg Theory}
This model arrived soon after Peccei-Quinn's initial papers following input from Weinberg and Wilcek, \cite{PQWW}. In addition to the standard model Higgs doublet, the PQWW model proposes an additional Higgs doublet, in which one field, $\phi_u$ couples to up quarks and the other,  $\phi_d$ to down quarks with no cross coupling.  The model requires that the quarks acquire their mass from the neutral components of the new Higgs fields, $\phi_u^0$ and $\phi_d^0$.  
\be \label{quarkmass} \mathcal {L}_{PQWW-m}= y_u \overline{u}_{Li} \phi_u^0 u_{R} + y_d \overline{d}_{L}
\phi_d^0 d_{R} + h.c., \ee
where $m_u = y_u v_u$.  The potential of the model is, \cite{axion dark matter}:
\be \label{PCCW_potential} U(\phi_u, \phi_d) = - \mu^2_u \phi_u^{*}\phi_u - - \mu^2_d \phi_d^{*}\phi_d + h.c.\ee
With the $U(1)_{PQ}$ symmetry, the Higgs, $N$ quark fields ($u$ and $d$) and  $\overline{\theta}$ parameter transform as:
\bea \label {axion_trans} \phi_u, \phi_d &\rightarrow & \exp (i2\alpha_u)\phi_u, \, \exp (i2\alpha_d)\phi_d \\
u, d &\rightarrow & \exp (-i\alpha_u \gamma_5)u, \, \exp (-i\alpha_d \gamma_5)d \\
\overline{\theta} &\rightarrow & \overline{\theta} - N(\alpha_u + \alpha_d). \eea
When electroweak symmetry is spontaneously broken the
neutral Higgs components $\phi^0_{u,d}$ acquire vacuum expectation values
$\langle \phi^0_{u,d} \rangle$ and hence Nambu Goldstone fields.  
\bea \langle \phi_u^0 \rangle &=& v_u\exp (i\frac{P_u}{v_u}) \\
\langle \phi_d^0 \rangle &=& v_d\exp (i\frac{P_d}{v_d}). \eea
In PQWW model, a linear combination of the two fields resulting from
the neutral components of the Higgs fields results in the Z boson,
while its orthogonal combination results in the axion field, $a$.
\be a  = \sin (\beta_v) P_u + \cos (\beta_v)  P_d, \ee
where $\beta_v$ is the angle between $P_u$ and $P_d$, the Goldstone fields. The axion couples to the quark fields resulting in complex quark mass which via the transformations in (\ref{axion_trans}) can be transferred to  $\overline{\theta}$ to give:
\be \overline{\theta} \rightarrow \overline{\theta} - \frac{N(\frac{v_u}{v_d} + \frac{v_d}{v_u})}{\sqrt{v_u^2 + v_d^2}}a, \ee
a change which can be absorbed by a redefinition of $a$.  \cite{axion dark matter} notes that non-perturbative QCD effects explicitly break $U(1)_{PQ}$ and result in the axion anomaly term (third term on the right-hand-side of (\ref{axion_lagrangian})).  \cite{Peccei3} notes that it may be regarded as an  effective potential for the axion.
\be \label{axion bare potential} U(a)_{eff} = \xi\frac{a}{f_a} \frac{g^2}{16 \pi^2} \mbox{Tr}\, \tilde{G}_{\mu\nu}G_{\mu\nu}. \ee
The addition of the axion to the Lagrangian (\ref{axion_lagrangian}) allows the promotion of $\overline{\theta}$ to a dynamic variable $(\overline{\theta} + \xi\frac{a}{f_a})$.  This will have a minimum when the expectation value of the axion field $< a > = - \frac{f_a}{\xi} \overline{\theta}$.  The fact that $a$ is a dynamic variable provides a natural means for this potential and the phase $(\overline{\theta} + \xi\frac{a}{f_a})$ to be minimized, addressing the CP problem, where the $<...>$ is the vacuum expectation value operator.
\be < \frac{\partial U(a)_{eff}}{\partial a} > =  - \frac{\xi}{f_a} \frac{g^2}{16 \pi^2} < \mbox{Tr}\, \tilde{G}_{\mu\nu}G_{\mu\nu} > |_{< a > = - \frac{f_a}{\xi} \overline{\theta}} = 0. \ee
Differentiating again with respect to the axion field, \cite{Peccei3} provides an expression for the mass squared matrix for the axion:
\be \label{axion_mass_squared} M_a^2 = < \frac{\partial^2 U(a)_{eff}}{\partial a^2} > = - \frac{\xi}{f_a} \frac{g^2}{16 \pi^2} \frac{\partial}{\partial a}< Tr\, \tilde{G}_{\mu\nu}G_{\mu\nu} > |_{< a > = - \frac{f_a}{\xi} \overline{\theta}}. \ee
\cite{Peccei1996} and \cite{axion mass2} show that due to inherent difficulties in computing low energy effective QCD quantities, axion mass computations are more practical if the axion degrees of freedom in (\ref{axion bare potential}) are transferred into effective interactions of the axion with QCD (the $\pi$ and $\eta$ mesons), and the term quadratic in $a$ equated to the mass.  The PQWW axion mass has the following form (where $m_{\pi}$ and $f_{\pi}$ are the mass and scale factor of the $\pi$ and $v$ is the electroweak energy scale, $\sim 250GeV$),  \cite{Peccei3}:
\be \label{PQ axion mass} m_{a-PQWW} = \frac{m_{\pi} f_{\pi}}{v}\frac{\sqrt{m_um_d}}{m_u + m_d} \cong 25KeV. \ee
The PQWW model and its variations are firmly linked to the electroweak scale as the field $a$ is coupled with the Z boson. Experimental evidence soon ruled this set of models out, but the axion dynamics remain valid in general for subsequent models, which are outlined qualitatively below.

\subsubsection{Invisible axion models}\label{Invisible axion models}
The PQWW model assumes that the $U(1)_{PQ}$ symmetry breaks at the electroweak scale $v$ resulting in a relatively heavy, coupled axion which was not found.  If $f_a >> v$ the axions are light ($m_a \sim \frac{1}{f}$), weakly coupled and invisible.  Making $f_a$ a free parameter allows the axion to be a candidate for cosmological phenomena: cold dark matter and to a lesser extent dark energy.  In the Kim-Shifman-Vainshtein-Zakharov (KSVZ) model \cite{KSVZ}, the axion is the phase of a new electroweak singlet scalar field and couples only to a heavier quark, $Q_h$ with interactions of the type $ - h\overline{Q}_{hL}\phi Q_{hR} - h^*\overline{Q}_{hR}\phi^* Q_{hL}$, where $\phi$ is an electroweak scalar field singlet and $h$ a coupling.  As with all axion models, a chiral anomaly term originating from (\ref{chiralanomalyPQ}) arises.  In KSVZ, rather than coupling directly to the ($u$ and $d$) quarks (as in PQWW), it couples to a heavier quark and the axion couplings are then induced by the interactions of this heavier quark with other fields.\\\\The Dine-Fischler-Srednicki-Zhitnitsky model, \cite{DFSZ}, like the PQWW model requires a doublet of two non-standard model complex Higgs scalars.  Like the KSVZ model it also has an electroweak scalar singlet which transforms under the $U(1)_{PQ}$ symmetry and whose phase results in the dynamical axion field.  This axion field then couples with the Higgs doublet and the complex degrees of freedom are transformed to the chiral anomaly term as in the PQWW model.  The two invisible axion models share similarities.  Firstly, they contain an electroweak ($SU(2)\times U(1)$) scalar singlet which spontaneously breaks the $U(1)_{PQ}$ symmetry at some arbitrary energy scale $f_a >> v$ with the axion degree of freedom resulting from the phase $\alpha$.  QCD instanton effects explicitly break $U(1)_{PQ}$ at some energy scale $\mu$ less than $f_a$ resulting in a chiral anomaly term of the form $\alpha \frac{g^2}{16 \pi^2} \mbox{Tr}\, \tilde{G}_{\mu\nu}G_{\mu\nu}$ which can be regarded as a potential for the axion field, \cite{Peccei1996} which minimises to eliminate the CP breaking $\overline{\theta}$ term.  Crucially, the axion may acquire mass via direct coupling to heavy particles other than the light quarks at an energy scale less than $\mu$.  While not derived in this paper a result is quoted from \cite{Peccei3}:
\be \label {KVSF_mass} m_{a-KVSF} = \frac{m_{\pi} f_{\pi}}{f_a}\frac{\sqrt{m_um_d}}{m_u + m_d} \cong 6.3\times \left[\frac{10^6GeV}{f_a}\right]eV, \ee
where the lack of dependence on $v$, the electroweak energy scale is noted.  The DFSZ axion mass has a similar form (not quoted here).  $f_a$ is a free parameter in both and (\ref{KVSF_mass}) may be expressed in the general form for an invisible axion model:
\be m_a = \frac{\mu^2}{f_a}, \ee
where $\mu$ is an energy scale related to QCD confinement $\Lambda_{QCD} \sim 0.2 GeV$.  This can set bounds for the value of $f_a$ via $m_a$.  These can be tested by considering the interactions a QCD axion is likely to have and the resulting cosmological implications of these. 

\subsubsection{String Axions}\label{string-axions}
String axion models arise from string compactifications generating
PQ symmetries which can be spontaneously broken. These involve
natural origins for the PQ symmetry unlike in non-string axion
models. Model-independent string axions, \cite{Model Independent}, arise from the antisymmetric tensor field of the bosonic and 
heterotic string theories (we consider this axion later in the thesis, eq (\ref{string_axion})). The properties of the string axion do not heavily depend
on the details of the compactification. \cite{Model Independent} computes
 the theoretical value of $f$ as given by:
\be f = \sqrt{2}\frac{\alpha_U}{4\pi}M_{P} \sim 10^{16} GeV, \ee
where $M_{P}$ is the reduced Planck mass, and $\alpha_U$ is
proportional to the square of the unified gauge coupling of ten-dimensional superstring theory compactified to four dimensions.  The value of $f$ arises from the
theory. The mass acquisition scale $\mu$ is a free parameter.
\cite{Model Independent} notes that in
order for these models to address the CP problem, QCD instantons
must be the dominant form of mass acquisition, thus $\mu \sim 0.2
GeV$.  The authors note that higher energy scale instantons could also play
a role. Model-dependent string axions arise from the zero modes of the
antisymmetric tensor field, \cite{Model Independent}.  The $f$ values are more variable in these models with a
typical value of $f \sim 10^{17} GeV$ noted, but with $f \sim
10^{15} GeV$ possible with fine tuning of the string action
parameters.  As with the model independent string axions, $\mu$, the energy scale where the $U(1)_{PQ}$ symmetry is explicitly broken so that the axion acquires mass, can be a
free parameter of the model, depending on the energy scale of
instantons responsible for explicit symmetry breaking.  
\subsubsection{Dark energy and axions}\label{Dark energy motivated axions}
This brief review of dark energy is sourced from \cite{Dark Energy in the Universe}, \cite{Frontiers of Dark Energy}, \cite{Copeland_Dark_Energy2} and where noted.  Following the discovery of the acceleration of the expansion of the universe in 1998, dark energy in the form of a homogeneous energy density, contributing almost three quarters of the universe's mass-energy, permeating all space and exerting a negative pressure was postulated.  A key revelation for theoretical physics of the newly observed phenomenon was that it appeared as if particle physics developments of the early universe were effecting current-era cosmology.   \cite {Spinodal Instabilities and the Dark Energy Problem} and \cite{Copeland_Dark_Energy2} note a required density for dark energy in the current era of $U_{DE} \sim (10^{-3} eV)^4$ to fit with observations of the known mass of the universe and acceleration of the most distant objects.  A positive cosmological constant, interpreted as a universal vacuum energy, of $\Omega_{\Lambda}=0.7$ was immediately proposed as the simplest explanation of the observed accelerating expansion.  The source of this vacuum energy in the light of the much larger fundamental energy levels associated with quantum theory remains unclear.  A vacuum energy originating from quantum theory, if one considers very early universe energy levels close to the Planck scale would be 120 orders of magnitude higher than the this required level.  One may, as in renormalisation, introduce counter terms to cancel the high vacuum energies but this requires ad-hoc fine tuning.  These and other difficulties have led to a second class of theories being proposed grouped under the term "dynamical scalar field models", the most well known of which are quintessence models, although others include tachyon fields and dilatonic dark energy. \cite{Copeland_Dark_Energy} notes that current observations are unable to rule in favour of either a cosmological constant or dynamical scalar fields (or another form of theory) as the cause.  Quintessence models currently are the most favoured in theoretical dark energy research.  They are represented by an scalar field coupled to gravity with a potential $U(\phi)$ which may explain the dynamical aspects of dark energy and perhaps other dynamical aspects of the $\Lambda$-CDM model of cosmology, such as inflation, \cite{Copeland_Dark_Energy2}.
\be \label{quint_action} S_Q = \int d^4x \sqrt{-g}\left[-\frac{1}{2}(g^{\mu\nu}\partial_{\mu}\phi\partial_{\nu}\phi)^2 - U(\phi)\right].\ee
Initial quintessence models used a potential such as:
\be \label{qunit_potential} U(\phi) = \frac{M^{4+\alpha}}{\phi^{\alpha}}, \ee
with $\alpha$ a positive parameter, and $M$ an energy scale.  The use of energy scales observed in particle physics such as $M = 1 GeV$ can result in the required energy density of the $\phi$ field without the need for further fine tuning, and provide dynamical variation.  We do not attempt to describe the details of the full range of quintessence models here, but focus on axion-related dark energy theories. \\\\

Axion based quintessence models have emerged in the last decade.  The quintessence axion model \cite{Quintessence1}, (1999, 2000) uses four new
pseudo scalar Goldstone bosons created by additional $U(1)$ symmetries. Two
of these relate to axions, the other two make contact with hidden
sector quarks to provide mass to the axions.  The two axions, $f_q
\sim \Lambda_{plank}$ and $f_a \sim 10^{12} GeV$, describe quintessence and the conventional CDM-QCD axion respectively.  Mass acquisition occurs at $\mu \sim \Lambda_{QCD}$
for the $f_a$ axion and at $\mu \sim 10^{-12} GeV$ for the $f_q$
quintessence axion. The latter results in an ultra light mass of $m
\sim 10^{-33} GeV$ which is equated with quintessence.  The
mechanism of the explicit symmetry breaking is via the two
additional bosons which have hidden sector interaction at the
intermediate SUSY scale ($10^{13}GeV$) and electroweak scale
($10^{2}GeV$) respectively.  The quintaxion \cite{Quintaxion} (2002, 2009) builds from the
quintessence model and seeks the qualities of a very large $f$ value
and a slow roll of the potential to current times.  It also relies
on a number of pseudo scalar Goldstone bosons, three in this case.
Two of these represent invisible axions and one is a model-independent
string axion, $a_{MI}\equiv a_q$ and the other a composite axion $a_{comp}$
which is the QCD axion with $f \sim 10^{12} GeV$ and $\mu \sim
\Lambda_{QCD}$.  $a_{MI}$ is the quintaxion with $f \sim
\Lambda_{Planck}$  and $m\leq 10^{-32}GeV$ and slow roll potential
$U \sim \Lambda (\frac{a_q}{f_q})$, where $\Lambda \sim f^2_{\pi}
m^2_{\pi} \sim 10^{-1} GeV$.\\\\Several variations of the so called false vaccua theory postulate that the axion field does not correspond to its true value, and this false vacuum
can act as dark energy provided its lifetime is longer than the age
of the universe.  \cite{Axion Cosmology Revisited}
suggests an "unstable axion quintessence" model in which the minimum
of the axion potential is negative.  
\subsubsection{Heavy axion models}
The heavy axion model in \cite{HeavyAxion} is motivated by the lack of observational evidence for the axion mass in the ranges predicted by the invisible axion and suggests axion physics could fall under a superstring
force, dubbed "QC'D" which operates in parallel and with similar
properties to QCD.  This model puts a lower bound on $m_a$ of 
$\sim 5 \times 10^{-2} GeV$, with the explicit symmetry breaking scale $\Lambda_{QC'D} \equiv \mu
\sim 3 GeV$ and $f \sim 2 \times 10^2 GeV$.  \cite{HeavyAxion1} (1997) builds on this idea.  A toy GUT model with $S(U)5
\times S(U)5$ gauge symmetry is considered with the second $SU(5)$ a
mirror of the first, but breaking at lower energies and resulting
in a QCD scale $\Lambda_{mirror} > \Lambda_{QCD}$.  The axion
acquires mass from mirror interactions and is equivalent to a PQWW
model in this mirror sector.  The mass of the axion in this mirror
PQWW-like model is given by
\be m_a \sim \left(\frac{\Lambda_{mirror}}{\Lambda_{QCD}}\right
)^{\frac{3}{2}} \cdot\left( \frac{v}{v_m}\right )^{\frac{1}{2}}
\cdot m_{aPQWW}, \ee
where $v$ and $v_m$ are the Higgs VEV in the standard model and
mirror sectors and $m_{aPQWW}$ is the axion mass as calculated by
the PQWW model.  A $m_a \leq 10^3 GeV$, with $\Lambda_{mirror} \leq
10^5 GeV$ is proposed.  The value of $f$ in this instance we have $f \sim
10^7GeV$.  The model in \cite{HeavyEtaAxion}, (1993) uses as its basis a "Walking Technicolor" model, which results in sextet quark-axion state,
the $\eta_6$, which the author notes has the properties of a
conventional Pecci-Quinn axion, but with higher color instantons
providing additional mass contributions.  It suggests a $m_a \sim 60
GeV$ with $\mu \sim 10^2 GeV$ suggesting a value of $f$ also of this
order, although a value for $f$ is not specifically referred to.  In \cite{Heavy Axions From Strong Broken Horizontal Gauge Symmetry},
(1992), as with other heavy axion models \cite{HeavyEtaAxion}, \cite{HeavyEtaAxion1},
breaking of the PQ symmetry occurs at just above the electroweak
scale such that $f \sim 2 \times 10^2 GeV$.  Also as with other
heavy axion models this uses non-Higgs EW symmetry breaking, in this
case via a heavy top quark and includes four quark flavours in
total. A value of $\mu \sim f \sim 2 \times 10^2 GeV$ is used and a
$m_a \sim 10^3GeV$ is computed.  

\subsection{Axion Physics}
\subsubsection{Axion interactions}\label{section_axion_interactions}
\begin{figure}[h] \centering
\includegraphics[scale=0.8]{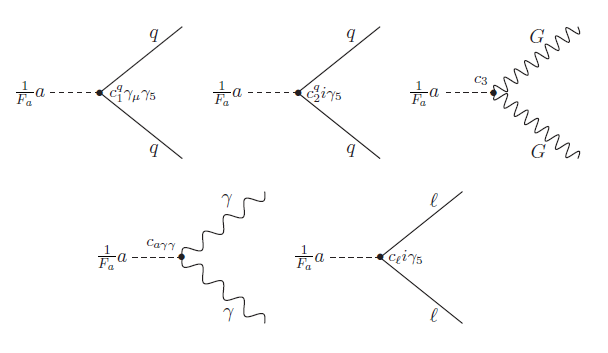}
\caption[Axion Couplings.]{Axion couplings represented by Feynman diagrams, with G the gluon, $\gamma$ the photon, $q$ the quark and $l$ leptons, \cite{Axion_Kim_2010_Summary}.}\label{axion_coupling}
\end{figure}

We focus here on invisible axions which involve an electroweak singlet and thus experience the electromagnetic and weak as well as the strong nuclear forces and exhibit, depending on model, relevant interactions in the $\theta$ Lagrangian.   In a comprehensive review of axion physics
\cite{Axion_Kim_2010_Summary} a generalisation of the
$\theta$-containing terms in the Lagrangian is presented, (here the overline notation previously used is dropped so that the $\overline{\theta}$ of section (\ref{axion_models}) is now taken as simply $\theta$ and where it is understood it is now dynamic and incorporates the pseudo-scalar axion field $a$ such that we may set $\theta \equiv \frac{a}{f_a}$ in the following descriptions).
\be \begin {split} \label {general theta terms} \mathcal{L}_{\theta} &=
\frac{1}{2}f_a^2
\partial^{\mu}\theta\partial_{\mu}\theta -
\frac{1}{2g_c^2}G_{\mu\nu}G^{\mu\nu} + (\overline{q}_L\imath
D_{qL} + \overline{q}_R\imath
D_{qR})\\
&c_1(\partial_{\mu}\theta)\overline{q}\gamma^{\mu}\gamma_5 q -
(\overline{q}_L m q_R \exp (\imath c_2 \theta) + h.c.) \\
&+ c_3 \frac{\theta}{16 \pi^2}\mbox{Tr} G_{\mu\nu}\widetilde{G}_{\mu\nu} + c_{0\gamma\gamma} \frac{\theta}{32 \pi^2}\mbox{Tr} F_{\mu\nu}\widetilde{F}_{\mu\nu} + 
\mathcal{L}_{leptons, \theta}, \end {split} \ee
where the term $c_1$ is the coupling of the interaction term derivative in
$\theta$, $c_2$ is the phase in the quark mass matrix, $c_3$ is the
coupling in the CP symmetry restoring term, and $c_{0\gamma\gamma}$ is the coupling of an electromagnetic anomaly term analogous to (\ref{chiralanomalyPQ}) with $F_{\mu\nu}$ the electromagnetic field strength tensor.   $\mathcal{L}_{leptons, \theta}$ contains axion interactions with leptons.  $c_1$, $c_2$ and $c_3$ are
couplings below the scale $f_a$ and the quark mass matrix $m$ is real.  (\ref{general theta terms}) is constructed as a general
expression and by assigning a non-zero values to combinations of
$c_1$, $c_2$ and $c_3$, well known axion models result (e.g. the
PQWW axion is given by $c_1=0$, $c_2\neq 0$ and $c_3=0$, and the
KSVZ axion \cite {KSVZ} by $c_1=0$, $c_2= 0$ and $c_3\neq 0$).  \cite{Axion_Kim_2010_Summary} notes there are also axion couplings to the electroweak bosons of the form $\theta W\widetilde{W}$ and $\theta Z\widetilde{Z}$ which are not shown here.  It is noted that through $\theta = \frac{a}{f_a}$ all axion couplings are relatively weak, suppressed by a large $f_a$.  \cite{Axion_Kim_2010_Summary} notes that the CP symmetry restoring term can be represented by a three quark instanton diagram
of the kind first suggested by 't Hooft \cite{'t Hooft}.  Couplings are represented graphically, (Figure \ref{axion_coupling}).

\begin{itemize}

\item \textbf{Hadron couplings}: These are the basis of low-energy laboratory based axion detection efforts and axion physics in supernovae. They are governed by $c_1$ and $c_2$ containing terms in (\ref{general theta terms}).  

\item \textbf{Photon couplings}: \cite{Peskin} (19.2) describes the one loop QED chiral anomaly $\partial_{\mu} j^{\mu 5} = - \frac{e^2}{16\pi^2} \mbox{Tr} F_{\mu\nu}\widetilde{F}_{\mu\nu}$ where $e$ is the electric charge and $F_{\mu\nu}$ the electromagnetic field tensor.  In a manner analogous to the derivation of the axion-gluon coupling term, a axion-photon coupling term arises, as noted in (\ref{general theta terms}): 
\be \mathcal{L}_{\theta\gamma\gamma} = c_{0\gamma\gamma} \frac{\theta}{32 \pi^2}\mbox{Tr} F_{\mu\nu}\widetilde{F}_{\mu\nu}. \ee
This may further be expressed in terms of the axion field $a$, and the electric and magnetic fields $\textbf{E}$ and $\textbf{B}$, \cite{axion dark matter}:
\be \mathcal{L}_{a\gamma\gamma} = c_{a\gamma\gamma} a \textbf{E}\cdot
\textbf{B}. \ee
with:
\be c_{a\gamma\gamma}  = \frac{c_{\gamma} \alpha}{\pi f_a}, \ee
with $c_{\gamma}$ containing the model-dependence and $\alpha$ the electromagnetic fine structure constant.  For the DFSZ invisible axion model, $c_{\gamma} = 0.36$ and for KSVZ, $c_{\gamma} = -0.97$, \cite{axion dark matter}.  This coupling results in the $a\longrightarrow \gamma\gamma$ decay, which \cite{Axion_Kim_2010_Summary} estimates has a lifetime of:
\be \label{axion_photon_decay} \tau(a\longrightarrow \gamma\gamma) = \frac{64 \pi^3 f_a^2}{c_{a\gamma\gamma} \alpha_{em}^2 m_a^3}, \ee
and notes that for $c_{a\gamma\gamma} \sim 1$, an axion with $m_a = 24eV$ has a lifetime of the order of the age of the universe $\tau_U \sim 4.35 \times 10^{17} s$.

\end{itemize}

\subsubsection {Effective Potential of the Axion Field} \label{axion effective potential}
In this research we utilize without formal derivation the commonly
quoted form (e.g. eq 29, \cite{axion dark matter}) for
the effective potential of the axion field where $m$ is a mass scale:
\be \label{axion effective potential form}U(a)_{eff} = m^4( 1- \cos
\frac{a}{f_a}). \ee
In (\ref{general theta terms}),\cite{Axion_Kim_2010_Summary} provides an expression for the $\theta$ dependence in the axion's Lagrangian, and notes that the cosine effective potential of the form (\ref{axion effective potential form}) is determined by two fundamental properties of the axion - periodicity with period $2\pi f_a$ and minima at $a=(0, 2n\pi f_a..)$, with $n$ an integer.  However, we note that (\ref{general theta terms}) does not represent an effective Lagrangian for all the $\theta$ terms, for example, the $\theta$ photon and lepton coupling and kinetic terms have been added by hand. \cite{Axion_Kim_2010_Summary} notes that while the cosine form for the effective potential may be a simplification the vast majority of cosmological axion models use the initial cosine term as in (\ref{axion effective potential form}). \\\\ The history of computations of the effective axion potential is reviewed briefly. Prior to the initial suggestion of a new dynamical axion field in 1977, \cite {Original Peccei Quinn}, 't Hooft, \cite{'t Hooft} and Weinberg,\cite {weinberg 2} described the non-perturbative nature of the QCD vacuum. Based on this, Peccei-Quinn began with the
following Lagrangian, where $g'$ and $h$ are coupling constants and
$\psi$ and $\phi$ are fermion (quark) and Higgs scalar fields
respectively.
\begin{multline} \mathcal{L} = -\frac{1}{4}G^a_{\mu\nu}G^{a\mu\nu} + \imath
\overline{\psi}D_\mu \gamma^\mu\psi +
\overline{\psi}\left[g'\phi\frac{1+\gamma_5}{2} +
g'^*\phi^*\frac{1-\gamma_5}{2}\right]\psi -\\
\mid\partial_{\mu}\phi\mid^2 - \mu^2\mid\phi\mid^2 -
h\mid\phi\mid^4 \end{multline}
Peccei-Quinn find the effective potential of this axion model can be
stated to a good approximation to leading order in $g'$ and $h$ and $g'\lambda$, where $\phi$'s vacuum expectation value is defined by
$\langle\phi\rangle= \lambda \exp \imath \beta$, with $\beta$ and
$\lambda$ being real constants.
\be \label{PecceiQuinnPotential} V_{\theta}(\phi) = U(\phi) - K\mid
g'\phi \mid \cos \theta, \ee
where $K$ is a real and positive constant, and $\theta$ is the dynamical axion phase.  \cite {Original Peccei
Quinn} notes the limitations on this result in that $g'$ and $h$ and
$g'\lambda$ are to leading order and the region of validity being
that these three constants be small.  However the authors note that
for constant $\phi$ and using the dilute gas approximation (where
a dilute gas of instantons is approximated by considering a
superposition of one-instanton solutions at great distances from one
another) (\ref{PecceiQuinnPotential}) is valid to all orders of
$g'\lambda$.  In recent research \cite{Axion Cosmology Revisited} the authors note
that exact analytical solutions to forms for the the QCD axion
effective potential are limited by the presence in the action of
strongly coupled terms, while numerical methods are limited by the
imaginary nature of the action.  They strive to demonstrate using
the so-called "interacting-instanton-liquid-model" a cutoff
independent formulation for the axion potential and mass. It is
noted that difficulties with these calculations arise at low energy
levels. \cite{QCD Vacuum and Axions},
(2002) presents an analysis of the form of the axion effective
potential in light of new understandings of the QCD vacuum using
supersymmetric gauge and brane theories.  The authors note that the potential as in 
(\ref{axion effective potential form}) may have
higher cosine powers and in general is a smooth periodic function of
$\theta$ with period $2 \pi$.  They conclude that with
axion models (including invisible axion models) using light quarks such that $m_q \ll \Lambda_{QCD-cutoff}$ (where $\Lambda_{QCD-cutoff}$ is the cut off used to integrate out the QCD degrees of freedom), the form (\ref {axion effective
potential form}) is valid.  We make comments on our use of this form in our research in section (\ref{Full_Quantization of the Axion}).

\subsection{Axion Phenomenology}\label{Axion Phenomenology}

\subsubsection{Axion Production}
Section \ref{axion_models} describes how the axion field arises as a result of spontaneous breaking of the $U(1)_{PQ}$ symmetry at some energy scale $f_a$ and then acquires mass close to $\Lambda_{QCD}$.  The production of massive axions from this method is known as vacuum realignment, \cite{axion dark
matter}, with the axions characterised as non-relativistic and largely non-interacting. If $f_a >> \Lambda_{QCD}$, as in invisible axion models, the axion may be considered in the context of the $\Lambda CDM$ model of the universe as a candidate for cold dark matter.  Considering the cooling of the universe following the big bang at $T_0$ the key temperature milestones are inflation reheating at $T_R$, the temperature at which the $U(1)_{PQ}$ is broken at $T_{PQ}$ and the QCD scale temperature at $T_{QCD}$ when the axion acquires significant mass.  \cite{axion dark matter} notes two additional plausible methods of axion production from the vacuum which may contribute to an axion density with the condition $T_R > T_{PQ}$.  These are axion strings and domain wall decay.  We comment only on the former here for brevity. Axion strings \cite{Axion from Strings}, \cite{Cosmic String Axions}, are theorised to arise from the $U(1)_{PQ}$ breaking as topological defects in a similar fashion to cosmic strings arising as a result of a global $U(1)$ symmetry breaking.  Axions are produced in the string oscillations until the strings decay.  This decay continues until the axions acquire mass at the QCD temperature scale \cite{cold axion populations}, \cite{axions from wall decay}.  If $T_R > T_{PQ}$, the axion field is not homogenized by inflation and axion strings are produced from $T_{PQ}$. \\\\We here consider only vacuum realignment as a production mechanism.  At the $T_{QCD}$ the axion acquires a temperature dependent mass $m_a$ and an effective potential given by \cite{axion dark matter}, \cite{cold axion populations}, \cite{Pseudo-Nambu-Goldstone Bosons}:
\be \label{axion eff potential} U_{eff} = m_a^2(T)f_a^2 (1-\cos
\theta).\ee  
If $f_a$ is very large and with suitable choice of $a(x)$ the oscillations can be damped and the potential varies only with $m_a^2(T)$.  \cite{axion dark matter} considers the solution to
$\theta(x) \equiv a(x)/f_a$ equation of motion in the FRW metric.
\be \ddot{\theta} + 3H(t)\dot{\theta} -
\frac{1}{S^2(t)}\nabla^2\theta + m_a^2 T(t) \sin (\theta) = 0, \ee
where $T(t)$ is the time dependent temperature, $S(t)$ is the scale
factor, $H(t)$ the Hubble parameter equal to $\frac{\dot{S}}{S}$ and dot is derivative with
respect to time $t$. For $T_R > T_{PQ}$, \cite{axion dark matter}
derives a value of $f_a \simeq 10^{12}
GeV$ which from (\ref{KVSF_mass}) results in a lower bound for $m_a \simeq 6 \mu eV$.  
Other models and scenarios for early universe axion production
\cite{axion dark matter}, \cite{cold axion populations}, \cite
{axion cosmology}, \cite{dark matter axions} present a range of
bounds.  In axion models based on
$U(1)_{PQ}$ symmetry breaking occurring after inflation (or in
inflation-less cosmological models) estimates place $f_a$ in the
range $10^{8} \sim 10^{12} GeV$, with the upper limit of $f_a$
constrained by existing levels of cold dark matter observed.  If $T_R < T_{PQ}$
estimates of the value of $m_a$ range from $meV$ to $neV$ levels
(resulting in $f_a$ in the range $10^{9} \sim 10^{15} GeV$), thus
raising the energy scale of $f_a$ to pre-inflationary levels
($\Lambda_{INFL} \sim 10^{14} GeV$, \cite {Probing a QCD String Axion}). Given an existing density of axions remaining from vacuum realignment, \cite{QCD Vacuum and Axions} notes that as well as the axion photon conversion, the following processes (based on couplings in (\ref{general theta terms})) can result in axion emission in astrophysical objects (stars):
\begin{itemize}
\item hadron-hadron bremsstrahlung-type interactions: $H + H \rightarrow H + H + a$;
\item photon-electron interactions: $\gamma + e^- \rightarrow e^- + a$;
\item electron-nucleus bremsstrahlung: $e^- + N \rightarrow e^- + N + a$;
\item photon fusion: $\gamma + \gamma \rightarrow a$.
\end{itemize}
Observations of stellar energy loss result in a lower bound on the invisible axion of $f_a \sim 10^9GeV$.

\subsubsection {Axion Detection} 
\begin{figure}[h] \centering
\includegraphics[scale=0.9]{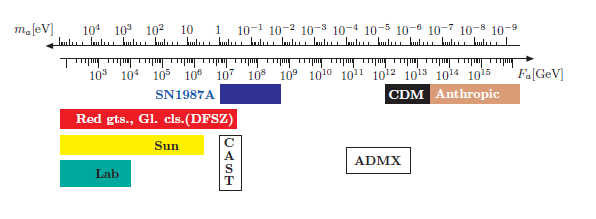}
\caption[Axion Production.]{The bounds of the scale factor $f_a$ and mass with axion production and detection methods, \cite{{Axion_Kim_2010_Summary}}.}\label{axion_production}
\end{figure}

In the original PQWW scheme for
axion production, the mass would have been of order $100keV$,
\cite{Peccei1996}, and thus within the bounds of laboratory testing.
Experimental studies of the PQWW axions have consisted of examining
radioactive decay and results to date \cite{axion dark
matter} have not detected an axion-like particle at this mass.  Subsequent axion detection has focused on the search for cold dark matter-motivated invisible axions.  Detection efforts have centered around the axion-producing version of the Primakoff effect.  It is predicted that axions
with energies of a few keV may be produced in the interaction
$a\longleftrightarrow \gamma + \gamma$ in the presence of electric and
magnetic fields within the solar plasma.  Helioscope experiments consist of dipole
magnets orientated towards the sun to catch the solar axions which
would be converted to photons in the form of x-rays. The largest
helioscope experiment is the CERN Axion Solar Telescope (CAST)
project \cite{Probing eV-scale axions with CAST} operational since
2003. Initial results provided an upper limit for the coupling $g_{a\gamma}$ of
$8.8\times 10^{-11} GeV^{-1}$ applying to an axion mass of $m_a \leq
0.02 eV$.  Phase II results increased the sensitivity of results to
$m_a \leq 0.4 eV$, with sensitivity of $m_a \leq 1.2 eV$ predicted
by 2011, \cite{Probing eV-scale axions with CAST}, \cite{Solar axion
search with the CAST experiment}.  There are several searches underway specifically aimed at axions being responsible in part for cold dark matter.  The Cryogenic Dark Matter
Search (CDMS), \cite{Search for Axions with the CDMS Experiment}
looks for possible solar-axion conversions to photos or
galactic-axion conversions to electrons within germanium crystal
detectors.  The Axion Dark Matter Experiment (ADMX), \cite{A High
Resolution Search for Dark-Matter Axions}, uses a microwave cavity
detector to search for CDM axions in the Milky Way galactic halo.
The mass detection zone for ADMX is $1.9<m_a<3.4 \mu eV$.  In \cite{Axion_Fermi}, the authors refer to the theory that axion-photon interactions can reduce the attenuation of very high energy gamma rays ($> 100GeV$) traveling over cosmological distances, but such effects have not been observed within the Fermi gamma ray experiment. \cite{AxionsSearch} summarises the recent laboratory,
astrophysical and cosmological limits placed on the invisible axion's mass and
$f$ values and regards the most conclusive evidence of axion
mass to come from stellar data. Restrictive limits arise from
observed neutrino signals from the SN 1987A Supernovae.  If the
axions are above the $m_a \sim10eV$ level, it is argued that they
would have been observed in the Cherenkov detectors used to count
the neutrino output.   In a 2010 survey of data from the Fermi Gamma Ray Space Telescope, looking at the high energy spectra from two specific sources, the authors conclude that there is no evidence for an axion-like particle attenuation effect.  To sum up, while axions have not yet been conclusively observed, Figure \ref{axion_production} cites the range in which they cannot be ruled out: $0.1 meV \leq m_a \leq 10meV$ corresponding to a range for $f$ of $10^{12} \geq f \geq 10^9 GeV$.

\chapter{Full Quantization of the Axion}\label{Full_Quantization of the Axion}

\section{Introduction}
The axion is proposed as an additional degree of freedom necessary to explain the lack of CP symmetry breaking observed when the action based on the bare QCD Lagrangian (\ref{QCD LAG}) is quantized resulting in a one loop anomaly term (\ref{CP Breaking term}).  The full axion Lagrangian is illustrated in (\ref{general theta terms}).  We note that the kinetic and non-QCD interaction terms for the axion in this have been added by hand.  Full quantization of the axion should then realistically involve quantization of the axion degree of freedom.  There have been some recent attempts to do this (see for example \cite{Bose}), in axion models and we propose to further this in this research to shed new light on fundamental physics arising from the axion in the early universe.  Finally, it is acknowledged that a large portion of the derivations in this section were published in \cite{Alexandre_Tanner}, (2010) by Alexandre and Tanner.  

\section{Quantization of the Axion}\label{Quantization of the Axion}
We begin with an action in which the non-axion degrees of freedom have been integrated out.  We thus assume that for some general action containing axion and non-axion fields, the following operation has been performed.
\be\label {QCD_quant}
Z[J]=\int{\cal D}[\phi]\exp(-S_{\phi}[\phi] -S_{\theta}[\theta]),\ee
where $\phi$ represent the non-axion fields (gauge bosons, leptons, hadrons) and $S[\theta]$ is the $\theta$ (axion) degree of freedom of the action. Section \ref{axion effective potential} describes attempts made at such computations and general results obtained.  We do not attempt such detailed computations in this research and denote the resulting $\theta$ dependent potential as $U(\theta)$ with the conditions only that it is periodic in $\theta$ with period $2\pi$ and $U(\theta) = 0$ at $\theta = 0, 2\pi,....2n\pi$, \cite{Axion_Kim_2010_Summary}.  We consider this our bare action in a general form.
\be\label{model} 
S_{\theta}=\int d^4x\left\lbrace
\frac{f^2}{2}\partial_\mu\theta\partial^\mu\theta +\sum_{n=1}^\infty
a_n(1 - (\cos\theta)^n)\right\}. 
\ee
This is based on discussions outlined in Section \ref{axion effective potential} in which the most generalised form of the axion potential is represented by a cosine power series, \cite{Axion_Kim_2010_Summary}. Full quantization will determine the coefficients $a_n$ and powers $n$ of the cosine. (The axion potential expression used commonly (\ref{axion eff potential}) is returned if only $a_1$ is non-zero).  We place the $f$ dependence in (\ref{model}) in the the kinetic term and make several remarks about our quantization approach. (Note: in this chapter, $\theta$ is the phase of the dynamic axion field defined in section (\ref{section_axion_interactions}), i.e. $a(x) = f_a\theta(x)$,  and also we hence drop the $a$ subscript such that$f_a \equiv f$). 

\begin{itemize}
\item We define an ultraviolet cutoff $\Lambda$ to impose on (\ref{model}), which will, along with $f$ will be a parameter of the theory.  We note at this point that $\Lambda$ is distinct from the cutoff used to integrate out the non-axion degrees of freedom in the operation conducted in (\ref{QCD_quant}).  This we term $\Lambda_{QCD-cutoff}$ and we assume that $\Lambda_{QCD-cutoff} \leq \Lambda$.  Further we point to recent research outlined in section (\ref{axion effective potential}) suggesting the form of the periodic axion potential is independent of $\Lambda_{QCD-cutoff}$ and we assume here that it is not a parameter of our theory. We further assume that the cutoff $\Lambda$ defines the upper limit of the theory and as such $f\leq\Lambda$.
\item  In terms of the the normalized scalar axion degree of freedom $a = f\theta$ in (\ref{model}) we wish to deploy non-perturbative methods derived in section \ref{An Alternative, Exact Approach}. Rather than an evolution with an energy scale $k$ (as in the Wegner-Houghton equation, (\ref{010})) we utilise another parameter of mass dimension, in this case $f$, to describe the exact evolution equation.  Such an approach is used in \cite{Alexandre1} and \cite{Alexandre2} and in the string cosmology section of this thesis, in the context of the bosonic string with the evolution parameter being the string mass scale.  While $f$ is generally not considered a variable parameter within conventional axion theory, it is not fixed in a physical sense, and we use the evolution of $f$ as a mathematical technique to arrive at our evolution equation.  With similar logic to \cite{Alexandre0}, as $f \rightarrow \infty$ the theory describing the axion represents the bare theory as the kinetic term is large compared to the potential term.  As $f$ decreases, quantum fluctuations begin dressing the system.  Thus evolution with $f$ in an exact equation similar to (\ref{scalar_theory_b_evolution1}) should serve our purpose of capturing all quantum effects in an effective expression.

\item We utilise results on effective field theory outlined in Section \ref{The Effective Action and Potential} for our quantization.  In this, the starting point is the generating functional, $Z[J]$ which defines the full quantum theory in path integral form (\ref{partitiondefi}).  Then via the definition of the connected graph generating functional $Z[J]=\exp (- W[J])$ we arrive at an expression for the effective action (\ref{legeffact}) which is expressed in terms of the classical field $\phi_{cl}$.  We want to deploy the same notation and results for the scalar axion field $a(x) = f \theta (x)$. We state the partition function
(or generating functional), $Z[J]$ in Euclidean 4-D space time.
\be \label{theta_partition}  Z[J]=\int D\theta \exp -\left  (S[\theta]
+ \int d^Dx J(x) \theta(x)\right ) = \exp - W[J].
\ee
We note that $\theta$, while a phase periodic in $2\pi$, is also a function of space time and make the following assumption about computing (\ref{theta_partition}).
\be \int D\theta \equiv \prod_x \int_0^{2\pi} d\theta(x). \ee
\end{itemize}

\section{Detailed Calculation}
Taking the derivative with respect to $f$ (represented by an overdot) of (\ref{legeffact}) we obtain (with $\phi_{cl}(x)$ replaced by $\theta_{cl}$, the classical field for $\theta$ for which we now drop the $cl$ subscript and denote simply as $\theta$):
\be\label{legeffact_f_derivative} \dot{\Gamma} [\theta] =  \dot{W}[J] + \int d^4x
\frac{\partial W[J]}{\partial J(x)} \, \frac{\partial J(x)}{\partial f} -  \int d^4x \frac{\partial \theta}{\partial f} J(x) = \dot{W}[J], \ee
following the methods used to arrive at (\ref{legeffact_b_derivative}).  We have used $\partial W[J]{\partial J(x)} = \theta$ as the definition of the classical field $\theta$ and consider a constant field configuration so that $\frac{\partial \theta}{\partial f} = 0$.  As was done in (\ref{scalar_theory_b_evolution}), combining (\ref{model}), (\ref{theta_partition}) and (\ref{legeffact_f_derivative}) we obtain:
\be \label{gamma_evolution} \dot{\Gamma} [\theta] = \dot{W}[J] = \frac{1}{Z}\dot{Z} = f \int d^4x <\partial_\mu\theta\partial^\mu\theta>. \ee
Which as derived in (\ref{scalar_theory_b_evolution1}) can be computed as:
\be \label{gamma_evolution1} \dot{\Gamma} [\theta]  = f\left[ \left(\int d^4x \partial_\mu\theta\partial^\mu\theta \right) + \mbox{Tr}  \left[ {\frac{\partial}{\partial _x}\frac{\partial}{\partial_y}
\left(\frac{\delta^2 \Gamma [\theta]}{\delta \theta_x\delta
\theta_y} \right)^{-1}}\right ]\right], \ee
where $\mbox{Tr} [....] \equiv \int d^4x d^4y [....] \delta^4 (x-y)$ and accounts for all quantum corrections up to the distances which are the space time equivalents of $\Lambda$, our regularization cutoff energy scale for the quantized axion theory. We now assume a generic form for $\Gamma$ as follows, with $f$ dependence also in $U_{eff}(\theta)$:
\bea \label{assumed form} \Gamma [\theta]= \int d^4x\left\lbrace
\frac{f^2}{2}\partial_\mu\theta\partial^\mu\theta +U_{eff}(\theta)\right\}.\eea
We consider an axion field constant in Euclidean space time, $\theta=\theta_0$.  Using (\ref{gamma_potential}) and taking the partial derivative with respect to $f$ of $\Gamma[\theta]$ we have (with $V$ the Euclidean space time volume):
\be \label{gammadot} \dot{\Gamma}[\theta] = V \, \dot{ U}_{eff}. \ee
which we combine with (\ref{gamma_evolution1}) to get:
\be  V \, \dot{ U}_{eff} = f \mbox{Tr}  \left[ {\frac{\partial}{\partial _x}\frac{\partial}{\partial_y}
\left(\frac{\delta^2 \Gamma [\theta]}{\delta \theta_x\delta
\theta_y}|_{\theta_0} \right)^{-1}}\right ]. \ee
We evaluate $\frac{\delta^2 \Gamma [\theta]}{\delta \theta_x\delta
\theta_y}|_{\theta_0}$ from (\ref{assumed form}) using a derivation similar to that performed in arriving at (\ref{second derivative1}):
\be \frac{\delta^2 \Gamma [\theta]}{\delta \theta_x\delta
\theta_y}|_{\theta_0} = [U_{eff}''(\theta_0)
- f^2\partial_\mu\partial^\mu] \delta^4 (x-y), \ee 
where the $''$
refers to a second derivative with respect to the axion field
$\theta_x$.  We evaluate the expression under the trace in (\ref{gamma_evolution1}) and convert to momentum space.  We use: $\partial_\mu\partial^\mu \equiv -p^2$, $F(p) =
\int d^4x e^{-i \textbf{p.x}}\,F(x)$ and $\int d^4xe^{-i \textbf{p.x}}\,\delta^4(0) = V /(2\pi)^4$.
\begin{multline} \label{result_Tr} f \mbox{Tr}  \left[ \frac{\partial}{\partial _x}\frac{\partial}{\partial_y}
\left(\frac{\delta^2 \Gamma [\theta]}{\delta \theta_x\delta
\theta_y}|_{\theta_0} \right)^{-1}\right ] =
f \int d^4x d^4y \delta^4(x-y)\left[ \frac{\partial}{\partial _x}\frac{\partial}{\partial_y}
\frac{1}{(U_{eff}''(\theta_0) - f^2\partial_\mu\partial^\mu)}\right ] \\ =
f \int d^4x d^4y \delta^4(x-y)\left[ \int \frac{d^4p}{(2\pi)^4}\frac{d^4q}{(2\pi)^4}
\frac{-p^{\mu}q_{\mu}\delta^4(p+q)}{(U_{eff}''(\theta_0) + f^2p^2)}  e^{-i(px+qy)}\right] \\ =
V f \int \frac{d^4p}{(2\pi)^4} \frac{p^2}{(U_{eff}''(\theta_0)
+ f^2 p^2)}, \end{multline}
where we have equated $\delta^4(0)$ in momentum space to the spacetime volume $V$.  Equating this (\ref{result_Tr}) with $\dot{\Gamma}$ as shown in (\ref{gammadot}) gives:
\be \label{final_result} \dot{U}_{eff} = \int \frac{d^4p}{(2\pi)^4} \frac{f p^2}{U_{eff}''(\theta_0)
+ f^2 p^2}.\ee
We evaluate the integral by considering $\int\frac{d^Dp}{(2\pi)^D}
\,=\,\int\frac{d \Omega_D
p^{D-1}dp}{(2\pi)^D}=\frac{\Omega_D}{(2\pi)^D} \int p^{D-1}dp$,
where the solid angle in four dimensions $\Omega_4 = 2\pi^2$. We use the substitution $x=p^2$ and
$\int p^3 dp f(p^2) = \frac{1}{2}\int p^2 d(p^2) f(p^2) =
\frac{1}{2} \int x f(x) dx$ to
compute the integral (\ref{final_result}), where $\Lambda$ is our high energy cutoff, above which our theory is not defined.\\
\bea \label{final_result_1} \dot{U}_{eff} &=& \frac{\Omega_4}{2(2\pi)^4} \int_0^{\Lambda^2} dx
\frac{f x^2}{x f^2 + U_{eff}''(\theta_0)},\nn &=& \frac{1}{16\pi^2}
\left[\frac{\Lambda^4}{2f} - \frac{U_{eff}''(\theta_0)\Lambda^2}{f^3} +
\frac{[U_{eff}''(\theta_0)]^2}{f^5}\ln \left(1 + \frac{f^2\Lambda^2}{U_{eff}''(\theta_0)}\right)\right]. \eea

\section{Flattening of the Axion Potential}\label{Flattening of the Axion Potential}
We refer to our description of the flattening of any concave potential in Section \ref{Section Spinodal Instability}. In this it was argued that given the result expressed in (\ref{convex suppression}), the shape of the effective potential must be convex, with concave features quickly suppressed by tachyon-mode fluctuations as the system undergoes quantization.  We note that the result arrived at in (\ref{final_result_1}) is generic and is in line with the derivations used to arrive at (\ref{convex suppression}).  We also point to our demand that the constraints on the effective potential in (\ref{assumed form}) are that it be periodic in $\theta$ and $U_{eff}(\theta) = 0$ at $\theta = 0, 2\pi,....2n\pi$, with $n$ an integer.  The periodicity and requirement that it be convex thus requires $U_{eff}$ in our theory to be flat, or only $a_0 \neq 0$ in (\ref{assumed form}).   We thus require $U_{eff}''(\theta) = 0$ and (\ref{final_result_1}) reduces to:
\be \label{final_result_2} \dot{U}_{eff} = \frac{\Lambda^4}{32\pi^2 f}. \ee
We make several comments on this result.  We can interpret $U_{eff}$, the effective potential of the axion in (\ref{final_result_2}), as a vacuum energy density associated with the axion at some point in its evolution.  It has general solution:
\be \label{general_solution} \int_f^{\Lambda} d U_{eff}(f) = \frac{\Lambda^4}{32\pi^2} \int_f^{\Lambda} \frac{df}{f}, \ee
and if we take the boundary condition that $U_{eff} = U_{\Lambda}$ when $f = \Lambda$:
\be \label{general_solution1} U_{eff}(f) = U_{\Lambda} + \frac{\Lambda^4}{32\pi^2} \ln \left(\frac{f}{\Lambda}\right). \ee
We note that the condition (\ref{convex suppression}) we have imposed on the solution to (\ref{final_result_1}) applies to a non interacting scalar theory, such as spontaneous symmetry breaking in the double well potential as outlined in section \ref{Section Spinodal Instability}.  We note that interactions are certainly present in axion theory as detailed in (\ref{general theta terms}) and Section \ref{section_axion_interactions}. (It is also acknowledged that the axion coupling to the metric $g_{\mu\nu}$ is not considered in this thesis).  In a dynamical axion theory, these terms arise following spontaneous symmetry breaking at energy scale $f$ and subsequent mass acquisition by the axion following contact with the QCD energy scale.   We use the following qualitative logic in the interpretation and relationships between $U_{\Lambda}$, the spinodal instability effect and the perturbative expansion of the quantized bare action (\ref{model}).

\begin{itemize}
\item The flattening of the potential induced by the spinodal instability is a quantum effect which by nature includes all quantum corrections in the resulting constant value of the effective potential, i.e. it is a tree level, exact, non-perturbative value.
\item In our evaluation of $U_{\Lambda}$ we are considering the potential at energy scale near the cutoff where $f = \Lambda \rightarrow \infty$.  In this regime, in the bare action represented by (\ref{model}), (and also by our scalar field approximation of this, represented by the Lagrangian density introduced in the next paragraph, (\ref{SSB_axion})) the bare action may be considered to be quadratic in the field $\theta$ with the interactions negligible, and hence free.  As such the second derivative of the bare potential is zero and the one loop correction is an exact expression of the full quantum theory, as with the spinodal instability effect.  
\item We thus interpret the flattening of $U_{eff}$ as an effect which provides a contribution to the evolution of the effective potential in the early phase of axion development, prior to mass acquisition and significant coupling effects.    
\end{itemize}
We consider the axion scalar field expressed by $\theta = a/f$, and are interested in the evolution of this field near the cutoff when $f\rightarrow \Lambda$.  Thus with small $\theta$, we may express the Euclidean potential $U(\theta)$ in (\ref{model}):
\bea \label{approximation_cosine} \sum_{n=1}^\infty
a_n(1 - (\cos\theta)^n) &=& \frac{k_1}{2}\theta^2 - \frac{k_2}{4!}\theta^4 + {\cal O}[\theta^6] \nn
&=& \frac{k_1}{2}(\frac{a}{f})^2 - \frac{k_2}{4!}(\frac{a}{f})^4 + {\cal O}[(\frac{a}{f})^6], \eea
where $k_1$ and $k_2$ are constants which are combinations of the factors $a_n$ in (\ref{model}) (note the units of $a_n$ and hence $k_1$ and $k_2$ are quartic in mass units).  To terms quartic in $\frac{a}{f}$, (\ref{approximation_cosine}) resembles the double well potential in (\ref{SSB}).  We may consider the axion initially as arising from a $U(1)_{PQ}$ symmetry which is spontaneously broken, a system whose classical Lagrangian for a scalar field $\phi \equiv \phi(x)\equiv f\theta$ is:
\be \label{SSB_axion} \mathcal {L} = \frac{1}{2}\left[(\partial\phi)^2 + \mu^2\phi^2 \right] - \frac{\lambda}{4!}(\phi^2)^2, \ee
where we now equate $\frac{k_1}{f^2}\equiv\mu^2$ and $\frac{k_2}{f^4}\equiv\lambda$.  The one loop effective potential of such a Lagrangian can be evaluated by reference to the Euclidean version of (\ref{one_loop}):

\be \label{one_loop_1} U_{eff}(\phi) = U(\phi) +  \frac{1}{2}\mbox{Tr} \ln [-\partial^2 + U''(\phi)], \ee
where $\phi$ is the classical scalar field and $U(\phi)$ is the classical potential and we thus have when $\Lambda = f$:
\be U_{\Lambda} = U(\phi) + \frac{1}{2}\mbox{Tr} \ln [-\partial^2 + U''(\phi)]. \ee
For completeness, we consider two cases, where $\mu=0$ and $\lambda\neq0$; and vice versa.

\begin{itemize}

\item At $\mu=0$, $\lambda\neq0$ (i.e. a vanishing renormalised mass squared condition) the system can be considered on the verge of spontaneous symmetry breaking as $\mu$ becomes $> 0$.  As such, it may be considered that the system is in the initial stages of the development of the axion field, discussed further in section (\ref{axion_Phenomenology and Discussion}).  In the derivation leading to (\ref{phi4_eff_pot}),\cite{Zee} provides a one loop effective potential of (\ref{SSB_axion}).  We state rather than derive the result here:
\be \label{quantized SSB} U_{eff}(\phi) = \frac{1}{4!}\lambda_m \phi^4 + \frac{\lambda_m^2}{(16\pi)^2}\phi^4\left( \ln \frac{\phi^2}{m^2} - \frac{25}{6}\right) + {\cal O}[\lambda_m^3], \ee
where $m$ is an arbitrary energy scale being considered and $\lambda_m$ the energy-scale dependent effective coupling.  Here $\phi$ is the classical field denoted by $\phi_{cl} \equiv < \phi >$ elsewhere in this thesis.  The term quadratic in $\lambda_m$ is the first order correction.  The flattening of the potential exhibited in (\ref{general_solution1}) is a tree level effect and we identify the term $U_{\Lambda}$ with the first order correction in (\ref{quantized SSB}).  In our initial assumptions stated in the text following section (\ref{Quantization of the Axion}) we consider a constant axion field configuration $\theta= \theta_0$.  We further set a high energy limit to our theory of $\Lambda$ and found that spinodal instability effects result in a flat effective potential.  With these conditions in mind, we consider a constant $\phi$ and also that $\phi, m << \Lambda$, thus satisfying our approximation in (\ref{approximation_cosine}).  We further now make an assumption that $m \approx \phi$ based on the fact that both are arbitrary for the purposes of our reasoning and both are small compared to $\Lambda$.   We note we are considering the case where $\Lambda = f$ giving from eq. (\ref{general_solution1}) $U_{eff} = U_{\Lambda}$.  We do not take the energy scale $m$ (and therefore $\phi$) as equal to $\Lambda$ as we wish to explore the behaviour at energy scales lower than the cut-off of our theory, $\Lambda$.  We wish to keep $m$ in our theory as a variable representing the energy scale below $\Lambda$ at which we are investigating.  We further note that in this logic we should include the $\frac{f}{\Lambda}$-containing term in eq. (\ref{general_solution1}) but we assume that the variation of $f$ from $\Lambda$ is not significant leaving the assumption $U_{eff} = U_{\Lambda}$.\\\\We now have the following.

\be \label{final result 3} U_{eff} = \frac{\Lambda^4}{32\pi^2}  \ln \left(\frac{f}{\Lambda}\right) + 
m^4 \left( \frac{1}{4!} \lambda_{m}  - \frac{25}{96 \pi^2}\lambda_{m}^2 \right), \ee
and with $f=\Lambda$:
\be \label{final result 3_1} U_{eff, f=\Lambda} = m^4 \left( \frac{1}{4!} \lambda_{m}  - \frac{25}{96 \pi^2}\lambda_{m}^2 \right). \ee
In an analysis of the parameter $\lambda_m$, we note that the use of the classical potential associated with (\ref{SSB_axion}) with $\mu = 0$ in quantization of the axion is similar to the approach taken in \cite{Bose} where it is shown that invisible axions form a Bose-Einstein condensate and in \cite{axion dark matter} and \cite{axion cosmology} where the axion field evolution is considered.  In \cite{Bose} (here the conventional, invisible QCD axion is considered as in section (\ref{Invisible axion models})) the authors use a $\phi^4$ scalar model for the axion and compute the effective scalar coupling constant as follows:
\be \label{scalar_coupling} \lambda = \frac{m_a^2}{f^2}\frac{m_d^3 + m_u^3}{(m_d + m_u)^3}\cong 0.35\frac{m_a^2}{f^2}, \ee
where $m_a$ is the axion mass and $m_d$ and $m_u$ are the up and down quark masses.  The authors state that this formula is obtained by using current algebra methods to derive an expression for the axion effective potential and equating the fourth-order coefficient to $\lambda$.  While $\lambda_{m}$ in (\ref{final result 3}) represents the scalar coupling at the energy scale $m$ and $\lambda$ in (\ref{scalar_coupling}) represents the coupling at the QCD energy scale, as a cursory approximation we take $\lambda \equiv \lambda_m$, which (with $f = \Lambda$) gives:
\be \label{final result 4}  U_{eff, f=\Lambda} \sim    4 \times 10^{-2}\, \frac{m^4}{f^4}m_a^4,  \ee
which, as our energy scale $m$ approaches the cut-off $\Lambda = f$ reduces to $ \sim 10^{-2}\, m_a^4$.  Here we have assumed $\lambda^2 << \lambda$.  The mass of the invisible axion has been experimentally reduced to a bound of $10^{-4}eV < m_a < 10^{-1}eV$, \cite{axion dark matter} and Figure \ref{axion_production}.  \cite {Spinodal Instabilities and the Dark Energy Problem} notes a phenomenologically required energy density for dark energy in
of order $U_{DE} \sim (10^{-3} eV)^4$, which is representative of commonly quoted values.  While the result in (\ref{final result 4}) is limited in usefulness by the inconclusive nature of the ratio $\frac{m^4}{f^4} < 1$, it could be that it does not vary by orders of magnitude with accepted values for $U_{DE}$.\\\\The beta function for the coupling constant in $\phi^4$ scalar theory can be expressed (\ref{beta_lambda}), (\cite{Peskin}, 12.2), with $m$ the energy scale:
\be \label{scalar-beta} \beta(\lambda) = m \frac{\delta \lambda}{\delta m} = \frac{3 \lambda^2}{16\pi^2} + O(\lambda^3). \ee
The result (\ref{final result 3_1}) depends on $\lambda_{m}$ and through this the energy scale $m$.  Using (\ref{scalar-beta}), fine tuning of (\ref{final result 4}) is likely necessary to arrive at a more accurate residual axion energy density.

\item We now consider the case of \textbf{$\lambda = 0$}, $\mu \neq 0$: Here we consider the following bare potential, from (\ref{SSB_axion}):
\be \label{bare_potential_zerolambda} U(\phi) = \frac{1}{2}\mu^2 \phi^2,  \ee

where $\mu$ is a mass parameter.  In we similarly consider the flattening of the potential as a tree level effect, the $\lambda = 0$ case for (\ref{one_loop_1}) may be solved directly.  
\bea \label{mu_0}
U_\Lambda&=&U(\phi) + \frac{1}{2}\mbox{Tr}\left\lbrace \ln\left[(-\partial^2+U(\phi)'')\right]\right\rbrace \\
&=&\frac{1}{2}\mu^2 \phi^2+ \frac{\Lambda^4}{64\pi^2}\left[ \left(
1-\frac{\mu^4}{\Lambda^4}\right) \ln\left(
1+\frac{\Lambda^2}{\mu^2}\right)
-\frac{1}{2}+\frac{\mu^2}{\Lambda^2}\right].
\eea 
We note $\mu$ is a mass scale associated with the bare axion potential and we make the assumption that this is at least several orders of magnitude less than the overall cutoff of our theory such that $\mu << \Lambda$ and we heglect the $\frac{1}{2}\mu^2 \phi^2$ term in eq. (\ref{mu_0}).   We thus have:
\be U_\Lambda \simeq
\frac{\Lambda^4}{32\pi^2}\ln\left(
\frac{\Lambda}{\mu}\right), \ee
and:
\be\label{vacpot} 
U_{eff}\simeq
\frac{\Lambda^4}{32\pi^2}\ln\left( \frac{f}{\mu}\right). 
\ee
Here we have taken $\phi$ the classical field such that $\phi < \Lambda$ and assume $\phi^2 \Lambda^2 << \Lambda^4$ in the evaluation of $U_\Lambda$ above.  If, as in (\ref{final result 4}) we refer to \cite{Bose} where $\mu$ is equated to the axion mass, we obtain, with $\Lambda$ a free parameter of the theory:
\be\label{vacpot} 
U_{eff, f=\Lambda} = 
\frac{\Lambda^4}{32\pi^2}\ln\left( \frac{\Lambda}{m_a}\right). 
\ee
\end{itemize}

We provide an interpretation of these results in section (\ref{axion_Phenomenology and Discussion}).
\subsubsection{Axion Interactions}\label{Axion Interactions_section}
The description of the flattening of any concave potential in a scalar effective potential when quantized  and its resulting flattening of the scalar potential was discussed in section \ref{Section Spinodal Instability} and refers to an effective potential where the source $J(\phi)$ is zero, that is, a theory with only self interactions.   As the axion makes contact with the QCD and electroweak energy scales a range of interactions emerge as outlined in section \ref{section_axion_interactions} and in (\ref{general theta terms}). This means that $J(\phi)$ is not necessarily zero and the convexity condition (\ref{convex suppression}) (that is $\frac{\delta^2 U_{eff}(\theta)}{\delta \theta^2} < 0$) may not hold and it may be avoided. For example we consider a non-QCD interaction, that of the axion with leptons: $gf\theta \bar{\psi} \gamma^5 \psi$.  Integrating out the lepton degrees of freedom results in an expression for $U_{eff}$ which contains the one loop correction to the lepton-axion interaction term.  In this, for simplicity, we assume the form of (\ref{assumed form}) is $m^4(1 - (\cos\theta))$, where $a_1 = m^4$, the infrared cutoff is $k'$, the coupling is $g$ and we neglect the mass of the leptons and make use of (\ref{one_loop_effective}).
\be U_{eff} = m^4(1 - (\cos\theta)) - \frac{1}{2}\int_0^{k'} \frac{d^4k}{(2\pi)^4} \ln \left[\frac{k^2 + g^2f^2\theta^2}{k^2}\right] \ee
We evaluate the second derivative of $U_{eff}$ at $\theta = \pi$ where the $m^4(1 - (\cos\theta))$ form is concave.
\begin{multline} \label{result 5} \frac{\delta^2 U_{eff}
(\theta)}{\delta \theta^2}|_{\theta=\pi} = -m^4 - \frac{g^2f^2k'^2}{16\pi^2}\frac{k'^2 + 3g^2f^2\pi^2}{k'^2 + 3g^2f^2\pi^2} + \\ \frac{g^2f^2}{16\pi^2}(3g^2f^2\pi^2) \ln \left(1 + \frac{k'^2}{g^2f^2\pi^2}\right). \end{multline}
For a high value of the cutoff, when $k' >> gf$ the result (\ref{result 5}) is less than zero and the spinodal instabilities arise, flattening the potential and it can be said that the lepton interactions add to the concave potential. If $k' < gf\pi$, a realistic assumption given the accepted value for $f \sim 10^{12}GeV$ and if $k' \sim 10^3 GeV$, the electroweak scale, the spinodal instability effects can be avoided.

\section{Summary and Discussion}\label{axion_Phenomenology and Discussion}
In section (\ref{Flattening of the Axion Potential}) we described how the spinodal instability in the axion's early stage quantized potential flattened it.  We thus derived a non-perturbative expression for the evolution of the effective potential with $f$ as in (\ref{final_result_2}). We then assigned a boundary condition to this differential equation such that the potential when $f = \Lambda$ is some value $U_{\Lambda}$ leading to:
\be \label{general_solution1.1} U_{eff}(f) = \frac{\Lambda^4}{32\pi^2} \ln \left(\frac{f}{\Lambda}\right) + U_{\Lambda}. \ee
Physically, at energy level $\Lambda$, which represents the upper limit cutoff of our theory, we consider that the evolution of the axion field in its earliest describable form.  In this sense, for very small $\theta$, the commonly used cosine form of the axion potential (\ref{axion effective potential form}) may be equated to a double well scalar potential as in (\ref{approximation_cosine}).  We consider that such a double well potential represents the origin of the spontaneously broken $U(1)_{PQ}$ symmetry responsible for the axion field.  An initial admission on this analysis is it does not incorporate any time parameter describing evolution of the field, but refers qualitatively to sequences of configurations.  We consider that our result (\ref{general_solution1.1}) describes the system on the verge of spontaneous symmetry breaking.  In the double well scalar potential (\ref{SSB_axion}) this occurs when $\mu$ changes from negative (representing a symmetrical "U" shaped potential) to positive, which is a form where spontaneous symmetry breaking can occur.  Thus we explore the $\mu = 0$ point as when the axion field is emerging.  At this point the axion has yet to evolve fully and the effects of the spinodal instability serve to flatten the potential, prior to any interactions.  We thus regard the potential as being represented by our result (\ref{general_solution1.1}), the form of which is impacted by the spinodal instability which is a tree level quantum effect.  We consider that the term $U_{\Lambda}$ in (\ref{general_solution1.1}) is a one loop correction to this.  We identify it with the one loop correction to the quantized form of (\ref{SSB_axion}) at  $\mu = 0$, which we compute using an established result quoted in \cite{Zee}.  The remaining parameters of our form for $U_{eff}(f)$, when we take $\Lambda = f$, is then $\lambda$, the scalar coupling constant of the quantized form of (\ref{SSB_axion}) and an energy scale $m$ used to compute the one-loop correction to the scalar field Lagrangian as in eq. (\ref{quantized SSB}).  We rely on work \cite{Bose} in which the authors compute (using current algebra methods) the effective scalar coupling constant in terms of $f$ and the axion mass $m_a$.  For $\Lambda = f$, the result which can then be expressed in terms of $m_a$ and the ratio of $\frac{m^4}{f^4}$ where $m$ is arbitrary and less than $\Lambda$ and $f$ is the scale factor equal to $\Lambda$ (thus the ratio is small).
\be \label{final result 4.1} U_{eff, f=\Lambda} \sim    4 \times 10^{-2}\, \frac{m^4}{f^4}m_a^4,  \ee
where $\frac{m^4}{f^4}$ is an undetermined ratio less than one (but may be close to one as we take $m \rightarrow \Lambda$).  We interpret this as an energy density associated with the axion at the earliest phase in its development, prior to significant interactions, and influenced by spinodal instability effects. (At later stages the axion acquires mass and resolves the CP problem as well as providing a candidate for cold dark matter).  With the caveat that the ratio $\frac{m^4}{f^4}$ is not determined by our theory, it could be that it is not orders of magnitude away from required values for dark energy ($U_{DE} \sim (10^{-3} eV)^4$, \cite {Spinodal Instabilities and the Dark Energy Problem}).
We make several further comments and refer to recent related research.

\begin{itemize}

\item The steps leading to (\ref{final result 4.1}) start with quantizing the dynamical field of the axion.  We next consider the quantum effect of the spinodal instability on the resultant field in its earliest phase of development.  Finally we use and existing result for the quantization of the double well potential approximating the early phase axion field as a boundary condition to add numerical estimates to our result.  Thus our theory is very much a full quantum theory of the axion.

\item Section (\ref{Dark energy motivated axions}) outlined quintessence theories involving a dynamical scalar field changing in space time, as evidenced by the accelerating expansion of the universe in the current era.  Key work in this area was conducted by Kim and Nilles, \cite{Quintessence1}, \cite{A Quintessential Axion}.  This focuses on linking quintessence with an ultra low mass axion whose potential has "slow rolled" down to a level associated with the required dark energy value.  The ultra low mass is obtained by considering a high $f$-valued string axion or an axion which acquires mass through contact with some hidden sector quark of ultra low mass.  In contrast, our result (\ref{final result 4.1}) does not rely on any non-standard model physics other than the proposed QCD axion.

\item The well cited works by the same authors in \cite {Pseudo-Nambu-Goldstone Bosons} and \cite {Pseudo-Nambu-Goldstone BosonsII} describe a family of particles termed "pseudo-Nambu-Goldstone-bosons" (PNGB), of which the axion is an example.  These particles exhibit spontaneously broken $U(1)$ symmetry at a scale $f$ and further explicit symmetry breaking at a lower scale $\mu$, and acquiring a mass $\sim \mu^2/f$.  \cite {Pseudo-Nambu-Goldstone Bosons} treats the neutrino as a PNGB, and attempts to link its dynamical field to an effective cosmological constant for several expansion times in the universe.  We note that this approach links the mass of the neutrino-PNGB at certain eras to achieve required energy densities. \cite {Spinodal Instabilities and the Dark Energy Problem} builds on the well-cited work in \cite {Pseudo-Nambu-Goldstone Bosons} but, as in our analysis, considers spinodal instability effects on the cosine form of the neutrino's effective potential resulting in a flat energy density of $M^4$, where $M$ is the mass of a light neutrino, corresponding to a dark energy like effect. In contrast our work, with the QCD invisible axion as the PNGB, results in $m_a^4$ being proportional to an energy density which we compare with dark energy.

\item We have characterized the early phase axion field as being flat due to spinodal instability effects.  Quintessence models require a dynamical scalar field, \cite{Copeland_Dark_Energy2}.  We have discussed only qualitatively the evolution of the axion field.  Further analysis of our result is necessary to determine how (\ref{final result 4.1}) could be shown to evolve into the current era.  In terms of the axion mass, $m_a$ which is a free parameter of our theory, \cite{Axion Cosmology Revisited}, for example, provides a discussion on how the axion mass may evolve with temperature scale $T$ in the early universe evolution (the result $m_a^2 = \alpha_a \Lambda^4 / f_a^2 (T/\Lambda)^n$) is quoted where $\alpha_a = 10^{-7}$ and $n = 6.68$).   Additionally the parameter $\lambda$ can be considered as a running coupling whose evolution with energy scale is governed by a beta function as in (\ref{scalar-beta}).

\end{itemize}

\chapter {Background to String Cosmology}\label{chapter String Cosmology}
Since its emergence in the 1970s and revival in the 1980s, string theories have occupied a central position in theoretical physics in the quest to unite gravity with the quantum field theories associated with the standard model.  Our motivation within the string context is to implement the non-perturbative approach outlined in section (\ref{An Alternative, Exact Approach}) to the simplest string-based quantum field theory - the bosonic string including the string-axion - and examine any cosmological predictions the solutions suggest.  As such, this thesis makes no attempt to review the breadth of topics within string theory such as advances in superstrings, D-brane and M-theories.  Rather we focus on the bosonic string as a conformal field theory and its applications in cosmology.  In section (\ref{QCD and its Symmetries}), scale invariance was discussed in the context of QCD where it was noted that the breaking of this symmetry results in the particular form of asymptotic freedom demonstrated by the strong interactions. Scale invariance is a specific case of wider conformal invariance, and this symmetry provides important constraints in forming many viable string cosmology configurations.  The following account is derived from \cite {polchinski} and
\cite {zwiebach}, and other sources where noted.

\section{Bosonic String Theory}

\subsection{Introduction to the Bosonic String}
\begin{figure}[h] \centering
\includegraphics[scale=0.7]{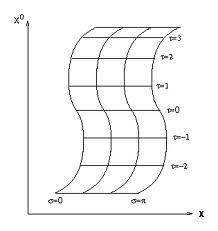}
\caption[The Worldsheet]{Worldsheet Parameters}\label{world_sheet_2}
\end{figure}

String theory provides what many consider to be
the leading candidate for incorporating the highly successful quantum field theories of the standard model with Einstein's equally successful theory of gravity.  \cite{polchinski} notes areas where the picture of our universe based on these two theories is incomplete.  Firstly, the standard model implies a particular pattern of fields and particles with certain initial conditions required.  These conditions are not naturally derived from the standard model's equations but rather need to be set by hand or from observation. Secondly, a quantum field theory of gravity is non-renormalisable in four spacetime dimensions.  Thirdly, classical gravity breaks down at its singularities.  With regards to the second point, the non-renormalisable nature of quantum gravity can be qualitatively seen by dimensional analysis.  We consider quantum gravity as the exchange of gravitons between two massive particles with associated higher order quantum corrections.
\begin{figure}[h] \centering
\includegraphics[scale=0.5]{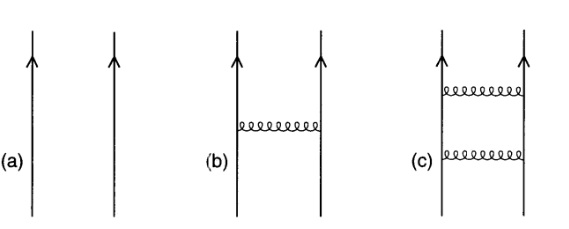}
\caption[Graviton Interactions]{Simple illustration of quantum gravity: (a) two free particles (zero graviton), (b) one-graviton correction and (c) two-graviton correction, \cite{polchinski}}\label{graviton}
\end{figure}
The one-graviton exchange is equivalent to the classical theory of gravity and its amplitude is proportional to $G_N$, Newton's constant. The ratio of the amplitudes of (b) to (a) must be dimensionless and thus proportional to $G_NE^2\hbar^{-1}c^{-5}$ where $E$ is the energy scale, as this is the only dimensionless combination of the parameters available.  With $\hbar=c=1$ one has by definition  $G_N \propto M_P^{-2}$ where $M_P$ is the Planck mass.  Thus the one-loop to tree level amplitude ratio is proportional to $(\frac{E}{M_P})^2$, meaning it is weak at the low energies of particle physics effective field theories but perturbative quantum gravity field theory breaks down at high energies $>M_P$.  The two-graviton to tree level ratio similarly diverges, proportional to $(\frac{E}{M_P})^4$. Such divergences have not been shown to be controllable via renormalisation as was successfully done with QED and other field
theories in the 1970s and a non-zero fixed point for the quantum field theory of gravity is not known. Alternatively it can be
considered that new physics emerges beyond the Planck energy
scale, and string theory is a leading candidate. To avoid
the high energy divergences, one dimensional strings (as opposed to point sources in conventional particle physics) are considered
to be the basic building blocks of nature, moving in a two
dimensional worldsheet.  This one dimensional string lives in $D$ dimensional spacetime, called target space.  
\begin{figure}[h] \centering
\includegraphics[scale=0.9]{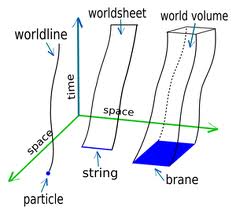}
\caption[Particle, String and Brane]{Particle, String and Brane in Two Space Dimensions}\label{world_sheet}
\end{figure}
Very qualitatively, high energy divergences
may be "smeared" out over the string to reduce them. The length of a string corresponds to
the Plank length $l_P$ which is considered fundamental.  In classical, non-string physics a zero-dimensional point source particle moves through spacetime and its motion can be described by $D-1$ functions of time $t= X_0$, (or $X^i(X^0)$ in Minkowski notation).  In a covariant form, the motion can be described by $D$ functions of $X^{\mu}(\tau)$ where $\tau$ is a newly introduced worldline parameter.  A key point is that the choice of $\tau$, as an additional, non-physical degree of freedom, should not affect the physics.  Thus:
\be X'^{\mu}(\tau'(\tau)) =  X^{\mu}(\tau), \ee
with the simplest Poincare-invariant action given by (where a dot indicates a derivative with respect to $\tau$):
\be \label{one_D_action} S_P = -m \int d \tau\sqrt{-\dot{X}^{\mu}\dot{X}_{\mu}}. \ee
This has the two local symmetries of worldline reparameterisation-invariance (i.e. $X'^{\mu}(\tau'(\tau)) =  X^{\mu}(\tau)$) and Poincare invariance.  The action (\ref{one_D_action}) may be expressed in an equivalent form with the introduction of an additional parameter $\eta(\tau)$ where $\eta = \sqrt{-\gamma_{\tau\tau}(\tau)}$ where $\gamma_{\tau\tau}$ is the worldline metric.  Now the action can be expressed:
\be S'_P \label{one_D_action_1} = \frac{1}{2}\int d\tau(\eta^{-1}(\dot{X}^{\mu}\dot{X}_{\mu} - \eta m^2), \ee
where $m$ can be considered as the particle's classical mass. While $S'_P$ contains an additional parameter $\eta$, it is quadratic in the derivatives of $X^{\mu}$ and its quantization in terms of a path integral can be more easily achieved.  The action in (\ref{one_D_action_1}), however turns out to be non-renormalisable in four dimensions as discussed qualitatively in terms of dimensional analysis above.  In \cite{polchinski_strings}, Polchinski notes that an elegant means of removing these divergences in that instead of a point we consider a one dimensional object as fundamental, a string, with an additional worldsheet parameter $\sigma$ as well as the $\tau$ in (\ref{one_D_action_1}), as illustrated in Figure (\ref{world_sheet_2}).
With a string as basic, a two dimensional worldsheet is traced out and one may define the an induced metric such that:
\be \label{induced_metric} h_{ab}=
\partial_a X^{\mu} \partial_b X_{\mu}, \ee
where $a$ and $b$ run over the two world sheet coordinates $\sigma$ and $\tau$.  The parameters $\tau,\sigma$ form a two dimensional vector we will denote as $\xi^a = (\tau,\sigma)$.  It is termed "induced" as it is superimposed on spacetime target space via the metric $g_{\mu\nu}$ and may be written:
\be \label{induced_metric_1} h_{ab}=
\partial_a X^{\mu} \partial_b X^{\nu} g_{\mu\nu} (X^{\alpha}). \ee
Thus the induced metric is defined by the spacetime manifold in which it is embedded and whose metric may be dependent on the spacetime coordinate $X^{\alpha}$.  It is noted that in eq (\ref{induced_metric_1}) the index $\alpha$ is not summed over and thus $X^{\alpha}$ acts as a scalar in this equation.  The simplest Lorentz invariant action (analogous to (\ref{one_D_action}) for a worldline based theory), the Nambu-Goto action is given by:
\be \label{nambu_goto} S_{NG} (h)= -\frac{1}{2\pi \alpha'}\int d\tau
d\sigma \sqrt{-det(h_{ab})}, \ee 
where $\alpha'$ is a parameter of the string theory with units of inverse energy squared (or length squared in a $\hbar = c = 1$ convention) which is inversely proportional to the tension of the string. ($\alpha'$ is termed the Regge slope).  One may express the action in a manner analogous to (\ref{one_D_action_1}) in the Polyakov action:
\be \label{polyakov_action} S_{Po} = -\frac{1}{4\pi \alpha'}\int d\tau
d\sigma (-\gamma)^{\frac{1}{2}} \gamma^{ab} \partial_a X^{\mu} \partial_b X_{\mu}, \ee 
where $\gamma^{ab}$ is a newly introduced independent worldsheet metric with $\gamma = \det \gamma_{ab}$ and $\gamma^{ab} = -\gamma_{ab}$. It is introduced in order to make the resulting Polyakov action more suited to the path integral formalism, while it introduces an additional degree of freedom compared to the Nambu-Goto action in that $\gamma_{ab}$ is proportional to $h_{ab}$ with the proportionality constant to be fixed by hand.  The Nambu-Goto action may be recovered from the Polyakov action by the relationship:
\be \label{induced_metric} h_{ab}(-h)^{-\frac{1}{2}} = \gamma_{ab}(-\gamma)^{-\frac{1}{2}}, \ee
and the definition of $h_{ab}$ in (\ref{induced_metric}).   As evident in the definition for $h_{ab}$ in eq (\ref{induced_metric_1}), $\gamma_{ab}$ also has an indirect dependence on the spacetime coordinate $X^{\alpha}$.   The form of (\ref{polyakov_action}), like (\ref{one_D_action_1}), does not contain a quadratic under a square root and is easier to quantize.  There are three important symmetries which are manifest in the form of $S_{Po}$ in terms of the invariance of the underlying spacetime physics.

\begin{itemize}
\item \textbf{Poincare invariance} of the spacetime coordinates (called target
space) in which the string is
embedded (where $\mu =1....D$, in D dimensional space time).  The worldsheet metric $\gamma_{ab}$ remains the same in any Lorentz frames.  This is a global symmetry of the action $S_{Po}$ and affects the spacetime parameter and worldsheet metric as follows.
\be X'^{\mu}(\xi) = \Lambda^{\mu}_{\nu}X^{\nu}(\xi) + k^{\mu}. \ee
\be \gamma'_{ab}(\xi) = \gamma_{ab}(\xi), \ee
\item \textbf{Diffeomorphism }(or diff) invariance of the two worldsheet dimensions implies that
physics is independent of the choice of worldsheet parameters $\tau, \sigma$
and can be expressed by the following transformations which leave the action invariant:
\be  X'^\mu(\xi') = X^\mu(\xi),  \ee
while the worldsheet metric transforms as a covariant tensor under a shift in coordinates $\xi^a(\tau, \sigma) \rightarrow \xi'^a(\tau', \sigma')$:
\be \label{diff-metric-transform}
\gamma_{ab}(\xi) \,\rightarrow\, \gamma'_{ab}(\xi')\,=\,\frac{\partial \xi^c}{\partial \xi'^a}\frac{\partial \xi^d}{\partial \xi'^b}\,\gamma_{cd}(\xi). \ee
Diff invariance implies that the physics on the target space $X^{\mu}$ is independent of the choice of worldsheet parameters.
\item \textbf{Weyl invariance} of the two worldsheet dimensions is a local rescaling of the worldsheet metric $\gamma_{ab}(\xi)$ which leaves the action and physics on the target space invariant:
\be \label{conformal_transformation} \gamma_{ab}(\xi) \,\rightarrow\,\gamma'_{ab}(\xi) = \exp
(2\omega (\xi))\gamma_{ab}(\xi), \ee
and  
\be X'^{\mu}(\xi) = X^{\mu}(\xi). \ee
This symmetry is not present in the Nambu-Goto action, and arises in $S_{Po}$ as the worldsheet metric $\gamma_{ab}$ is determined up to a scale factor represented by the proportionality relationship in (\ref{induced_metric}) which can be rephrased, \cite{zwiebach}:
\be \label{induced_metric_1} h_{ab} = f^2 (\xi)\gamma_{ab}, \ee
with $f (\xi)$ a non-vanishing worldsheet function.  The fact that this proportionality is governed by $f^2$ means that the signatures of the metrics  $h_{ab}$ and $\gamma_{ab}$ agree.  Choice of $f(\xi)$ is analogous to choice of gauge in quantum field theories. In the Nambu-Goto action, $f(\xi)$ is a constant and the symmetry is not evident.  (We note here that Weyl invariance is a generalised form of the scale invariance of QCD expressed in (\ref{scale invariance}), which is the global version of (\ref{conformal_transformation}) where $\omega(\tau,\sigma) =$ constant).  \\\\
Thus through the diff and Weyl invariance in the worldsheet metric $\gamma_{ab}$, the action (\ref{polyakov_action}) contains three redundant degrees of freedom.  On the other hand, the quadratic form of the $X^{\mu}$ "fields" in this action as compared to the Nambu-Goto action makes path-integral quantization possible.  These redundancies need to be fixed however, in the course of the quantization process.

\end{itemize}
The metric $\gamma_{ab}(\xi)$ is a symmetric two by two matrix meaning it has three degrees of freedom.  One may, via suitable choice of the diff transformation, fix two of these, as noted by the contractions in the transformation (\ref{diff-metric-transform}).  The remaining degree of freedom, associated with the Weyl invariance, may be set by choice of $\omega(\xi)$.  So one may define the worldsheet metric:
\be \label{fiducial_metric}
\gamma_{ab}(\xi)\,=\, \exp (2\omega(\xi)) \widehat{\gamma}_{ab}\,,
\ee
where here both $\widehat{\gamma}_{ab}$ and $\omega(\xi)$ are fixed. A common choice of $\widehat{\gamma}_{ab}$ is $\delta_{ab}$, the flat metric, known as the fiducial gauge.  The string parameterised by $\tau$ and $\sigma$ in Figure (\ref{world_sheet_2}) is an open string.  If the string has a defined length, $l$, one has for an open string:
\be \label{parameter_open_string} - \infty < \tau < \infty, \,\,\, 0 \leq \sigma \leq l, \ee
and the worldsheet is an infinite strip of width $l$.  One may also consider a closed string which forms an infinite cylinder such that it has the constraints:
\be \label{parameter_closed_string} X^{\mu}(\tau,0) = X^{\mu}(\tau,l); \,\,\, \gamma_{ab}(\tau,0) = \gamma_{ab}(\tau,l), \ee
that is, the parameter $\sigma$ is periodic with period $l$. So far in our discussion, the worldsheet metric $\gamma_{ab}$ has a Euclidean signature $(+, +)$.  However, one may have a worldsheet metric with a $(-, +)$ Minkowski signature and the $\tau$ parameter may be considered analogous to the time coordinate, $X^0$ in spacetime.  In this case, the indices $a,b$ may be raised or lowered by $\gamma_{ab}$ in conventional Minkowski fashion.   As a final general point \cite{polchinski} notes that all string theories contain closed strings but not all contain open strings as the former can be created from the latter.\\\\ As a note on convention, in this thesis, worldsheet indices will always be denoted by $a,b$ with spacetime indices left in the conventional Greek nomenclature $(\mu, \nu...)$.  
\subsection{Classical String Actions and Constraints} \label{Classical String Actions and Constraints}
In the Polyakov action  for the string, the spacetime coordinate $X^{\mu}$ and the worldsheet metric $\gamma_{ab}$ are treated as independent variables of $\xi$. In effect, $X^{\mu}$ can be regarded as a scalar field, (with $\mu$ an inactive index in (\ref{polyakov_action})), coupled to the metric $\gamma_{ab}$.  One may obtain two sets of constraints on the classical string via minimising $X^{\mu}$ and $\gamma_{ab}$ with respect to the action. 
\be \label{string_eom} \delta S_{Po} = \frac{\partial S_{Po}}{\partial \gamma_{ab}} \delta \gamma_{ab} = \frac{\partial S_{Po}}{\partial X^{\mu}} \delta X^{\mu}=0. \ee
We consider first the variation of $S_{Po}$ with $X^{\mu}$, \cite{zwiebach}.
\be \label{variation_X_Polyakov} \delta S_{Po}= -\frac{1}{2\pi \alpha'}\int d\tau
d\sigma \delta X^{\mu} \partial_a\left((-\gamma)^{\frac{1}{2}} \gamma^{ab} \partial_b X_{\mu}\right), \ee
with total derivatives set to zero and left out here.  This results in the constraint:
\be \label{total derivative constraint} \partial_a\left((-\gamma)^{\frac{1}{2}} \gamma^{ab} \partial_b X^{\mu}\right) = (-\gamma)^{\frac{1}{2}} \nabla^2 X_{\mu} =0, \ee
where $\nabla^2 \equiv \partial^a\partial_a$.  An additional surface term will result if boundary conditions such as (\ref{parameter_open_string}) are placed on the string. This surface term vanishes if:
\be \partial^{\sigma}X^{\mu}(\tau,0) =  \partial^{\sigma}X^{\mu}(\tau,l) = 0, \ee
i.e. the derivatives of $X^{\mu}$ with $\sigma$ vanish at the boundary (known as the Neumann conditions, \cite{polchinski}).  In the closed string the surface term vanishes naturally.  Variation of $\gamma_{ab}$ with respect to the action defines the stress energy tensor for the worldsheet, \cite{polchinski}:
\bea
\label{classical stress} T^{ab}(\tau,\sigma) &=&
-\frac{4\pi}{\sqrt{-\gamma}}\frac{\partial S_{Po}}{\partial
\gamma_{ab}} \nn &=& -\frac{1}{\alpha'}\left[\partial^a X^{\mu} \partial^b X_{\mu} - \frac{1}{2}\gamma^{ab}(\partial_c X^{\mu} \partial^c X_{\mu})\right].\eea
Diff invariance means that:
\be \label{stress_energy_conservation} \nabla_a(\partial^a X^{\mu} \partial^b X_{\mu}) = \nabla_a T^{ab} = 0. \ee
Weyl invariance as governed by the transformation in (\ref{conformal_transformation}) means that an infinitesimal variation of the parameter $\gamma_{ab}$ is (to first order in the expansion of $\exp 2\omega$):
\be \delta \gamma_{ab} = 2\omega \gamma_{ab}, \ee
which when substituted into the equation of motion for $\gamma_{ab}$ in (\ref{string_eom}) gives:
\be 2\omega \gamma_{ab} \frac{\partial S_{Po}}{\partial \gamma_{ab}}   = 0, \ee
which, since $\omega$ is arbitrary, leads to:
\be \gamma_{ab}\frac{\partial S_{Po}}{\partial \gamma_{ab}}  = 0, \ee finally leading to the trace of the stress energy tensor vanishing, a direct consequence and condition of Weyl invariance in the classical string.
\be \label{stress_energy-vanish}\gamma_{ab}\frac{\partial S}{\partial
\gamma_{ab}}=0 \Rightarrow T^{a}_{~a}=0. \ee 
Thus we have, in solving the equations of motion of the worldsheet metric, $\gamma_{ab}$ deployed two key symmetries (diffeomorphism and Weyl invariance) of the Polyakov action to derive results. Further, while our discussion relates to bosons only, fermions may be considered by the
addition of super-symmetry, and this forms the
basis of the current generation of superstring theories (which will not be discussed in this thesis). The Weyl and diff invariance of (\ref{polyakov_action}) are fundamental to the quantization of the classical string and in setting boundary conditions for use in string cosmology.
\subsection{Conformal Invariance}
Continuous conformal transformations on a manifold rescale the metric but preserve locally angles (and hence shapes). Considering $D$ dimensional geometry with metric $g_{\mu\nu}$ and line element $ds^2 = g_{\mu\nu} dx^{\mu} dx^{\nu}$, under a change of coordinates $x \rightarrow x'$ one has:
\be \label{conformal} g_{\mu\nu} \rightarrow g'_{\mu\nu}(x') = \frac{\partial x^{\alpha}}{\partial x'^{\mu}} \frac{\partial x^{\beta}}{\partial x'^{\nu}} g_{\alpha\beta}(x). \ee
The conformal group is the group of transformations which leave the metric invariant up to a position dependent scale factor such that $g_{\mu\nu} \rightarrow g'_{\mu\nu}(x') = \omega(x)g_{\mu\nu}$, \cite{CFT_Ginsparg}.  These are the transformations which preserve the angle between two vectors on the manifold.   The infinitesimal generators of  the conformal group can be found by considering the transformation $x^{\mu} \rightarrow x^{\mu} + \epsilon^{\mu}$, where $\epsilon^{\mu}$ is small.  To satisfy (\ref{conformal}) one must have (quoted here without derivation), \cite{CFT_Ginsparg}:
\be \label{conformal_condition} \partial_{\mu}\epsilon_{\nu} + \partial_{\nu}\epsilon_{\mu} = \frac{2}{D}(\partial \cdot \epsilon) \eta_{\mu\nu}. \ee
The following types of transformations  leave the metric invariant as in (\ref{conformal}) (where $a^{\mu}$ and $b^{\mu}$ are constant vectors):

\begin{itemize}
\item translations independent of $x$: $\epsilon^{\mu} = a^{\mu}$;
\item rotations: $\epsilon^{\mu} = \sigma^{\mu}_{\nu} x^{\nu}$;
\item scale transformations: $\epsilon^{\mu} = \lambda x^{\mu}$;
\item special conformal transformations: $\epsilon^{\mu} = b^{\mu}x^2 - 2x^{\mu}b\cdot x$.
\end{itemize}
The first two are a representation of the Poincare group ($\omega = 1$), the third are scale transformations ($\omega = \lambda^{-2}$) while the special conformal group has $\omega(x) = (1+2 b \cdot x + b^2x^2)^2$.  When dimensionality $D=2$, (\ref{conformal_condition}) results in :
\be \label{cauchy-Reiman} \partial_1\epsilon_1 = \partial_2\epsilon_2; \partial_1\epsilon_2 = -\partial_2\epsilon_1, \ee
which are the Cauchy-Riemann equations providing a sufficient condition for a differentiable function to be holomorphic. Thus we may write the above coordinates in the complex plane: $z, \bar{z} = x^1 \pm i x^2$ and the two dimensional conformal transformations are a representation of analytic coordinate transformations in $z, \bar{z}$:
\be z \rightarrow f(z), \bar{z} \rightarrow \bar{f}(\bar{z}), \ee
and $ds^2 = dz d\bar{z} \rightarrow \left|\frac{\partial f}{\partial z}\right|^2 dz d\bar{z}$ with $\omega = \left|\frac{\partial f}{\partial z}\right|^2$.  The local symmetry group is infinite dimensional in the special case of $D=2$ which provides for the complexity of a two dimensional conformal field theory such as the bosonic string theory under discussion here.  (We note that $D$ here refers to the dimensionality of the geometry in general and not necessarily physical spacetime dimensionality).  \cite{CFT_Ginsparg} notes that a combination of diff and Weyl invariance in a two dimensional field theory with a Euclidean metric as in Section (\ref{Classical String Actions and Constraints}) is equivalent to conformal invariance on a complex worldsheet such as the analysis leading to (\ref{cauchy-Reiman}).  We may restate the Polyakov action by expressing the worldsheet coordinates ($\tau, \sigma$) in the complex plane.  First we consider the Polyakov action (\ref{polyakov_action}) with a flat Euclidean worldsheet metric $\gamma_{ab} = \delta_{ab} = (1,1)$, \cite{polchinski}, and hence the change of sign compared with eq. (\ref{polyakov_action}):
\be \label{polyakov_action_1} S = \frac{1}{4\pi \alpha'}\int d^2\tau
( \partial_1 X^{\mu} \partial_1 X_{\mu} + \partial_2 X^{\mu} \partial_2 X_{\mu}). \ee 
The choice of $\gamma_{ab} = \delta_{ab}$ fixes the local gauge of the worldsheet metric to be flat, a commonly used configuration for quantization.  The complex worldsheet coordinates are:
\be \label{complex_coordinates} z = \sigma^1 + i \sigma^2, \,\, \bar{z} = \sigma^1 - i \sigma^2, \ee
and
\be \label{complex_derivatives} \partial_z = \frac{1}{2}(\partial_1 - i \partial_2), \,\, \partial_{\bar{z}} =\frac{1}{2}(\partial_1 + i \partial_2), \ee
with terminology $\partial \equiv \partial_z$ and $\bar{\partial} \equiv \partial_{\bar{z}}$ used.  Now (\ref{polyakov_action_1}) becomes:
\be \label{polyakov_action_2} S = \frac{1}{2\pi \alpha'}\int d^2z
( \partial X^{\mu} \bar{\partial} X_{\mu}). \ee 
The equation of motion obtained by minimising the $X^{\mu}$ field becomes:
\be \partial \bar{\partial} X^{\mu} (z, \bar{z}) = 0. \ee
In (\ref{stress_energy-vanish}) Weyl invariance was shown to result in a traceless stress-energy tensor.  In the the coordinates (\ref{complex_coordinates}) this becomes:
\be T_{z\bar{z}} = 0, \ee
while the equivalent of (\ref{total derivative constraint}) becomes:
\be \bar {\partial} T_{zz} = \partial T_{\bar{z}\bar{z}} = 0. \ee
We denote $T_{zz} \equiv T(z)$ and $T_{\bar{z}\bar{z}} \equiv \tilde{T} (\bar{z})$, and for a free massless scalar theory the analogy of (\ref{variation_X_Polyakov}) leads to the following forms for the stress-energy tensor.
\bea \label{non vanishing stress energy}
T_{zz}&=&T(z)\,=\,- \frac{1}{\ap} \,\partial X^\mu\,\partial X^\nu\,\eta_{\mu\nu} \nn
T_{\bar z\bar z}&=&\tilde T(z)\,=\,-  \frac{1}{\ap} \,\bar\partial X^\mu\,\bar\partial X^\nu\,\eta_{\mu\nu} \nn
T_{  z\bar z}&=&0\,,
\eea
which are equivalent to the requirements of diff invariance that the stress energy is conserved (\ref{stress_energy_conservation}) and from Weyl invariance that it is traceless (\ref{stress_energy-vanish}).  Invariance of the target space physics with respect to conformal transformations of the worldsheet metric is a highly desirable feature of a string theory attempting to describe the physical universe. We have stated that a test for conformal invariance in both the real and complex planes of the worldsheet space relate to the conservation of and trace-vanishing properties of the stress energy tensor.  Discussion thus far has remained at the classical level.  We now consider implications of quantization on the above results, and the constraints imposed on this process by the conformal conditions.

\subsection{Quantization of the Bosonic String}
Rather than offer a comprehensive account of the derivations, methods and tools involved in quantization of the bosonic string, common results are simply quoted.  Following the above account of how conformal invariance presents key constraints in the classical theory, our aim is to build a brief background leading to an explanation of Weyl anomalies in the quantized theory.  In a quantum field theory of the classical bosonic string, the fields are promoted to operators.  In (\ref {polyakov_action_2}) both $X^{\mu}$ and $\gamma$ become operators and are functions of $\xi$. The theory may be quantized using the path integral formalism, where the partition function of an action $S[X, \gamma]$ is: 
\be \label{string_path_integral} 
\mathcal{Z} =\,\int [dX d\gamma] \,\exp (-S[X, \gamma]), \ee
where this integral runs over all worldsheet metrics $\gamma$ and all values of $X^{\mu}(\xi)$ (where the index $\mu$ is not integrated over).  The integration over the world sheet $d\gamma$ is conducted over the three degrees of freedom in the metric $\gamma^{ab}$, and it is here that anomalies arise which set a limit on the spacetime dimensionality of the effective theory if conformal invariance is to be preserved. In the quantized theory, $T^{a}_{~a}$ does not necessarily vanish as it does in the classical theory, but is proportional to the central charge of the theory, $c$.  The value of $c$ is found through the summation of the central charges arising from various sectors of the theory.  For the bosonic string, the central charge has contributions from the spacetime (of dimension $D$) coordinate acting as a scalar field ($c_X = D$) and from the "ghost" sector ($c_g = -26$) so that (results stated rather than derived here):
\be c = c_X + c_g = D - 26.  \ee
The setting of $c=0$ provides for the well known $D=26$ constraint of the free bosonic string.  In quantizing the Polyakov action (\ref{polyakov_action}), we use a Euclidean metric $\gamma^{ab}$ of signature $(+, +)$ and with $d\sigma d\tau$ replaced with $d^2\xi$.
\be \label{polyakov_action_Euclidean} S_X = \frac{1}{4\pi \alpha'}\int d^2\xi  \sqrt{\gamma} \gamma^{ab} \partial_a X^{\mu} \partial_b X_{\mu}. \ee 
We generalise (\ref{polyakov_action_Euclidean}) by adding a term which is allowed by the Poincare and diff invariance. This results in a term ($S_R$) which may be added to $S_X$:
\be S_R \label{polyakov_action_additional} = \frac{1}{4\pi}\int d^2\xi \lambda \sqrt{\gamma}\,R^{(2)}. \ee
Here $R^{(2)}$ is the Ricci scalar constructed from the metric $\gamma^{ab}$ and $\lambda$ is a coupling (which we will later relate to the dilaton).  The term (\ref{polyakov_action_additional}) is invariant under Poincare, diff and Weyl transformations on a worldsheet without boundaries, \cite{polchinski_strings}, resulting in our most general classical action, (which we denote simply as $S$ here):
\be \label{polyakov_general} S = S_X + S_R = \frac{1}{4\pi \alpha'}\int d^2\xi  \sqrt{\gamma} \left( \gamma^{ab} \partial_a X^{\mu} \partial_b X_{\mu} + \lambda \alpha' R^{(2)} \right). \ee
The expression $\frac{1}{4\pi}\int d^2\xi \sqrt{\gamma}\,R^{(2)}$ is known as the Euler number, $\chi$, and depends only on the topology of the worldsheet.  Under a path integral operation as in (\ref{string_path_integral}) this term produces a factor of $\exp (-\lambda \chi)$ which impacts only the relative weightings the various topologies of the worldsheet hold in the path integral summation.  Interactions on the worldsheet, including string self interactions are determined by the topology as defined by $\gamma^{ab}$, which in turn has a unique Euler number.  This Euler term in the path integral quantization of the action (\ref{polyakov_general}) controls the coupling constants in string theory, with the couplings being related to $\exp(\lambda)$.  The path integral (\ref{string_path_integral}) contains redundancies resulting from configurations of $(X,\gamma)$ and $(X',\gamma')$ related by the diff and Weyl transformations describing identical spacetime physics and it needs to be gauge fixed to avoid unphysical divergences.  This can be done by the Faddeev-Popov method, \cite{Faddeev Popov}, used commonly in gauge theory, where slices of each identical gauge are taken only once.  A common gauge choice is to fix the metric such that $\gamma_{ab} = \delta_{ab} = (1,1)$.\\\\ In quantum field theories, Feynman rules for calculating cross sections denote a particle at external point $\textbf{x}$ with momentum $\textbf{p}$ by $\exp(-i \textbf{p}.\textbf{x})$.  Likewise, string interactions are usefully described with the aid of vertex operators.  In conformal field theory, (\cite{polchinski}, p 63) there is an isomorphism between the set of allowed states in the quantised theory and the set of local operators describing the quantum field theory.  While details of this concept will not be developed here, as an example, the lowest permissible energy state of the closed string spectrum can be described by a vertex operator in the string action:
\be \label{tachyon_action} V_0 = 2 g_c \int d^2\xi  \sqrt{\gamma} \exp(i k.X), \ee
where $X \equiv X^{\mu}(\xi)$ and $k$ is the momentum in worldsheet space, and $g_c$ is the string coupling.  In this particular example, the state is characterised by having negative mass squared and is known as the tachyon.  In this language, the lowest states of the closed string act as local operators and can be regarded as interacting background fields all of which depend on the $X^{\mu}(\xi)$ parameter (where the $\mu$ index remains inactive thus far).  The free bosonic string theory represented by the most general version of the Polyakov action (\ref{polyakov_general}) thus becomes an interacting theory with the vertex operators representing the lowest quantum states playing the role of interacting fields.  In the next section we will state the full action we will use for string cosmology, based on the expansion of the Polyakov action to include low level energy states of the string represented by local vertex operators. 

\subsection{The Bosonic String in Curved Spacetime}
We now replace $\partial_a X^{\mu} \partial_b X_{\mu}$ with $g_{\mu\nu}(X) \partial_a X^{\mu} \partial_b X^{\nu}$ and thus are considering the string in a curved target spacetime.  (In the notation $g_{\mu\nu}(X)$ we leave out the $\mu$ index on $X$ as it is still inactive). One may consider the curved spacetime metric $g_{\mu\nu}(X) = \eta_{\mu\nu} + \chi_{\mu\nu}(X)$, where $\chi_{\mu\nu}(X)$ is an incremental variation in the metric at point $X$.  The term $\chi_{\mu\nu}(X)$ can be equated to a vertex operator representing a coherent vibrational mode of strings known as the graviton, whose dynamics are denoted by $s_{\mu\nu}$ and where $g_c$ is a coupling:
\be \label{vertex_graviton} \chi_{\mu\nu}(X) = -4\pi g_c \exp(i k.X) s_{\mu\nu}. \ee
Expanding this concept to include the backgrounds of other massless open or closed string states we state the following bare action, denoted again here simply as $S$.

\bea \label{string action backgds}
\hspace{-0.5cm}S & =& \frac{1}{4\pi\ap} \int d^2\xi \sqrt{\gamma} \,\Bigl\{\left[\gamma^{ab} g_{\mu\nu} (X) +i\,\epsilon^{ab}B_{\mu\nu}(X)\right]\partial_a  X^\mu \partial_b  X^\nu \Bigr. \nn
&&\hspace{3cm} \Bigl.+ \ap R^{(2)} \Phi( X) + 4\pi\ap T( X) \Bigr\}\,,
\eea
Within this the following vertex operators represent string states acting as coherent fields (the tachyon vertex operator $T(X)$ was defined in (\ref{tachyon_action})). 
\bea \label{vertex_operators} g_{\mu\nu} (X) &=& \eta_{\mu\nu} - 4\pi g_c \exp(i k.X) s_{\mu\nu}, \\
B_{\mu\nu} (X) &=& 4\pi g_c \exp(i k.X) a_{\mu\nu}, \\
\Phi( X) &=& 4\pi g_c \exp(i k.X) \phi. \eea
We should make several comments at this stage.  

\begin{itemize}
\item  It was noted in (\ref{polyakov_general}) that an additional term in the free string action is allowed by the Poincare and diff invariance of the Polyakov action.  This term is described by the Euler number $\frac{1}{4\pi}\int d^2\xi \sqrt{\gamma}\,R^{(2)}$ and a coupling constant $\lambda$.  A basic premise of string theory is that there are no predetermined parameters which need to be set by hand.  The coupling $\lambda$ in curved spacetime is thus promoted to a dynamical parameter, known as the dilaton, a scalar field notated as $\Phi(X)$, which appears naturally in the spectrum of the massless closed string and plays a key role in the couplings associated with the background fields of the quantized string.
\item In (\ref{string action backgds}) there are two fields coupling to the derivative term $\partial_a  X^\mu \partial_b  X^\nu$.  The term $\gamma^{ab} g_{\mu\nu} (X)$ is present in the classical action when the spacetime metric is generalised to curved spacetime and this term represents the graviton string state.
\item The term containing $i\,\epsilon^{ab} B_{\mu\nu}(X)$ is known as the Kalb-Ramond field and is an antisymmetric tensor field (that is, antisymmetric in two spacetime indices $\mu$ and $\nu$).  The fact that this term is complex arises from the Euclidean continuation.  It is analogous (in a tensor form) to the vector gauge field $A_{\mu}$ of QED which couples to the fermion fields in the Yukawa fashion, that is: $i e \gamma^{\mu}A_{\mu} \psi \overline{\psi}$.

\item \cite{zwiebach} notes that the quantised state at fixed momentum of the closed string can be represented by an arbitrary matrix $M_{IJ}$ of size $D-2$ (where $D$ is the spacetime dimension) multiplied by basis vectors representing the massless string states.  Such a matrix $M_{IJ}$ may further be generalised as the sum of symmetric traceless part, a antisymmetric part and a scalar trace - giving rise to the three background fields noted above - the graviton, the antisymmetric tensor and the dilaton.
\end{itemize}
The action in (\ref{string action backgds}) is the starting point for our treatment of string cosmology, with the caveat that we do not consider the tachyon field in our study. 

\subsection{The Weyl Anomaly}
Equation (\ref{stress_energy-vanish}) states that the a condition of conformal invariance in the classical bosonic string is a traceless stress energy tensor, $T^{a}_{~a}=0$.  In our discussion of QCD it was noted that when renormalised, scale invariance (the global version of Weyl invariance, as shown by (\ref{scale invariance})) is broken resulting in an effective theory dependent on energy scale.  In standard model QFTs this is not a problem as one is concerned with the infrared scale, and one may compute beta functions such as $\beta(\lambda) = \mu \frac{\partial \lambda}{\partial \mu}$ (where $\mu$ is energy scale and $\lambda$ is the coupling) to monitor the variation with energy scale of the theory's parameters.  However in the quantization of the bosonic string, maintaining Weyl invariance is usually conducted so as to derive invariant target space results.  (A class of string theories (for the bosonic string) where the $D=26$ condition, is by-passed are known as non-critical string theories).  In the quantised bosonic string, $T^{a}_{~a}=0$ vanishes only under certain conditions and these form the basis of constraints for a critical, effective string theory actions.  Classically, the stress energy tensor is derived from the variation of the action with the metric, (\ref{classical stress}).  In the quantum field theory approach, the stress energy tensor is related to the infinitesimal variation of the partition function with the metric, \cite{polchinski}:
\be \label{quantized_stress_energy} T^{ab}_Q = \frac{4\pi}{\sqrt{\gamma}} \frac{\delta}{\delta \gamma_{ab}}\mathcal{Z} (\gamma), \ee
where $\mathcal{Z}$ is computed as in (\ref{string_path_integral}).  In the quantised theory, $T^{a}_{~a}$ must be computed and may not vanish.  \cite{polchinski_strings} shows that in a quantised theory, one has for a free string:
\be \label{Weyl_anomaly} T^{a}_{~a} = c R^{(2)}, \ee
where $c$ is a constant proportional to the central conserved charges of the fields in $\mathcal{Z}$ and $R^{(2)}$ is the Ricci scalar associated with the metric $\gamma_{ab}$.  In the free bosonic string theory represented by (\ref{polyakov_general}) the central charge from the ghost fields arising from the Faddeev-Popov methods to fix the gauge is $-26$.  The charge arising from the $X^{\mu}$ fields is simply $D$, the spacetime dimensionality.  This leads to the commonly quoted result that free critical bosonic strings require $26$ spacetime dimensions to display Weyl invariance.  The string action (\ref{string action backgds}) contains additional fields which may break Weyl invariance.  We quote without proof the result, \cite{low.ener.eff.}
\be
T^{a}_{~a}\,=\, \beta^i\,\frac{\delta}{\delta g^i} <S>\,
\ee
where $S$ is the action in (\ref{polyakov_general}) and $\beta^i$ are the beta functions for each coupling $i$ describing the flow of the couplings with the Weyl rescaling parameter $\omega(\sigma)$.  The notation $< (.....) >$ is defined by:
\be < (.....) > \equiv \frac{1}{\mathcal{Z}} \int [dX d\gamma] \,(.....)\,\exp (-S[X, \gamma]), \ee
where $\mathcal{Z}$ is defined in eq. (\ref{string_path_integral}).  Here $g^i$ is the set of couplings associated with the background fields ${g_{\mu\nu}, \Phi, B_{\mu\nu}}$ in the action (\ref{string action backgds}).  As in the case of the free string, Weyl invariance occurs when the $\beta^i$ expressions vanish.  The beta functions are perturbative expansions in the $\alpha'$, the Regge slope and we quote here from \cite{polchinski}:
\be \label{tquantum1}
T^{a}_{~a}=-\frac{1}{2\alpha'}\beta^g_{\mu\nu}\gamma^{ab}<\partial_a
X^\mu\partial_b X^\nu> -
\frac{i}{2\alpha'}\beta^B_{\mu\nu}\epsilon^{ab}<\partial_a
X^\mu\partial_b X^\nu > - \frac{1}{2}\beta^\Phi R^{(2)}. \ee
To zeroth order in $\alpha'$, the beta functions are as
follows,\cite{low.ener.eff.}:
\bea\label{beta_functions}
\beta^{g}_{\mu\nu} &=& R_{\mu\nu} +2\nabla_\mu\nabla_\nu\phi -\frac{1}{4}H_{\mu\kappa\lambda}H_{\nu}^{~\kappa\lambda}  + {\cal O}(\ap) \nn
\beta^{B}_{\mu\nu} &=& -\frac{1}{2} \nabla^\kappa H_{\kappa\mu\nu} +  \partial^\kappa\phi\,H_{\kappa\mu\nu}+  {\cal O}(\ap)\nn
\beta^\phi &=& \frac{D-26}{6\ap}-\frac{1}{2} \nabla^2\phi  +  \partial^\mu\phi\partial_\mu\phi - \frac{1}{24}H_{\mu\nu\rho}H^{\mu\nu\rho} + {\cal O}(\ap) \,,
\eea
where $R_{\mu\nu}$ is the target space Ricci tensor.  $H_{\mu\kappa\lambda}$ is the field strength of the antisymmetric tensor such that $H_{\mu\kappa\lambda}=\partial_{\mu} B_{\kappa\lambda}+\partial_{\kappa} B_{\lambda\mu}+\partial_{\lambda} B_{\mu\kappa}$.  From the form of $\beta^\phi$ it can now be seen that with background fields the space time dimension $D$ must no longer
necessarily be 26 to preserve conformal invariance as terms of $\ap$ or higher order could potentially be made to cancel the $\frac{D-26}{6}$ term.   In addition, the diff symmetry on the
world sheet means that the beta functions are invariant under the following transformation, which is a field rescaling: \cite{Mavromates}.
\be \label{local transformations} \beta^i \rightarrow \hat{\beta}^i = \beta^i  + \delta \beta^i, \ee 
where we defined $\delta \beta^i$ in some detail in (\ref{redefbeta}).   These modified
$\hat{\beta}^i$ functions are known as the Weyl anomaly coefficients
and are often used in place of the $\beta^i$ functions. \\\\In
setting the $\beta$ or $\hat{\beta}$ functions to zero (conditions
of conformal invariance) one is effectively writing down equations
of motion which set constraints on the field variables, and from
this one can determine an effective action in space time expressed in terms of the dilaton field, $H_{\mu\kappa\lambda}$ and the spacetime Ricci tensor (all of which are functions of $X^{\mu}$),  \cite{Mavromates}.
\be
S_{eff} \label{effective_string_action}   = -\frac{1}{2\kappa_0^2}\int d^DX \sqrt{g}e^{-2\Phi}\left(R +
4(\partial_\mu\Phi\partial^\mu\Phi) - \frac{1}{12}H_{\lambda\mu\nu}H^{\lambda\mu\nu}\\
-\frac{2(D-26)}{3\ap} + {\cal O}(\ap) \right), \ee
where $\kappa_0$ is the
gravitational constant in $D$ space time dimensions, and here $R$ is the spacetime Ricci tensor (note the target space Ricci tensor has been denoted by $R^{(2)}$).  We note at this point the minus sign in front of the dilaton kinetic term in this frame (know as the $\sigma$-model frame).  The term "frame" in this context refers to the spacetime metric configuration the action exists in.  The positive sign is restored for the dilaton kinetic term in the physically relevant Einstein frame, by making an allowed transformation of the metric.  A common practice is that the spacetime metric $g_{\mu\nu}$ is transformed in a spacetime Weyl transformation  such that: 
\be \label{Einstein_action} g_{\mu\nu}
\rightarrow g_{\mu\nu}^E=e^{-\frac{4}{D-2}\Phi}g_{\mu\nu}. \ee 
Under this transformation, the curvature scalar $R$ in eq. (\ref{effective_string_action}) will have the standard co-efficient (under general relativity) - i.e. $\frac{1}{2\kappa_0^2}$.  In the Einstein frame the effective action is, \cite{Mavromates}:
\begin {multline} \label{effective_string_action_E}
S^E_{eff}   = -\frac{1}{2\kappa_0^2}\int d^DX \sqrt{g^E}[R^E -
\frac{4}{D-2}((\partial_\mu\Phi\partial^\mu\Phi) - \frac{1}{12}e^{-\frac{8}{D-2}\Phi}H_{\lambda\mu\nu}H^{\lambda\mu\nu} \\
- e^{\frac{4}{D-2}\Phi}\frac{2(D-26)}{3\ap} + {\cal O}(\ap) ]. \end {multline}
In the Einstein frame the transformed parameters superscripted with an $E$.  In this frame, the spacetime curvature scalar is denoted $R^E$ and is canonically normalised with the coefficient $\frac{1}{2\kappa_0^2}$ so as to be relevant to physical observations.  Further, one may see that with $D=4$ the dilaton kinetic term changes sign and is now positive. \\\\With this brief and largely qualitative introduction to string theory, we are now ready to describe the basics of string cosmology.
\section{String Cosmology}
\subsection{Introduction}
The following is based on \cite{Copeland_String_Cosmology}, \cite{String_Cosmology_Review}, \cite{String_Cosmology_Venesiano}, \cite{Mavromates} and sources where noted.  The study of the early universe is highly concerned with the study of the gravitational field and its evolution.  If one is to
take string theory as representing the fundamental fabric of our universe and uniting gravity with quantum field theory, it would appear logical to take a bare action such as in (\ref{string action backgds}) as a starting point in studying the evolution of fields and particles in the universe.  Likewise, cosmological observations are also one of the few feasible testing grounds for string theory, and currently show more promise than earth-bound collider experiments in testing its predictions. Conventional cosmology is based around the $\Lambda$-CDM model incorporating the big bang, an inflationary scenario and cold dark matter/dark energy both of unknown origin at present. Most string cosmology models seek to incorporate the dynamics of the $\Lambda$-CDM model and shed light on some of the unexplained phenomena such as dark energy.   Moreover, the last twenty years has seen advances in string theory and allow it to more easily make contact with real-world observations.  One such advance is moduli stabilisation, \cite{String_Cosmology_Review}.  Conformal invariance dictates that critical string models may need to live in more than four spacetime dimensions.  String models deploy gravitationally coupled scalar fields, known as moduli, to compactify the additional dimensions to a level where they cannot be detected by current observations.  However, the moduli fields should also present cosmological observables and moduli stabilisation is the term given to theories which counter the effects of moduli to non-detectable levels. A further challenge of string cosmology is in isolating its predictions compared to those feasibly arising from conventional QFTs.  Given the degenerate nature of string solutions one tactic, \cite{String_Cosmology_Review}, is to limit string cosmology theories to those which produce outcomes which could not arise from conventional quantum field theories.  One such area (which we will not discuss in this thesis) are string-based topological defects, known as cosmic strings.  Prior to a description of our research, we will outline a variety of string cosmology approaches.

\subsection{String Cosmology Approaches}
There are many string cosmology models which address a range of scenarios such as inflation, the graceful exit and pre-Big Bang (see for example \cite{Copeland_String_Cosmology}).  We do not attempt a comprehensive review in this thesis but rather introduce an early string cosmology model which can describe a static and also expanding universe in four dimensions.  In the previous section, we described the path taken from a bare worldsheet action such as the Polyakov action in (\ref{polyakov_action_2}) representing a free string to an action representing the bosonic string in curved spacetime though to the effective action in $D$ spacetime dimensions useful for physical observations, (\ref{effective_string_action_E}). The derivation and evaluation of such an effective action motivated by early universe phenomena forms the basis of string cosmology.  As the observed universe is isotropic and homogeneous, one is generally interested in configurations where the spatial co-ordinates are fixed with the time component varying. The conformal invariance conditions (i.e. the terms in (\ref{beta_functions}) vanishing) determine the equations of motion for the background fields in target space.  A common starting point is the
the bare action (\ref{string action backgds}) with the background fields displaying variation only in the time ($X^0$) direction with $\mu, \nu = 0,.....D-1$.  We exclude for the time being the tachyon field, (but make some comments on this in the next section).  
\be
\label{stringaction1} S_\sigma=\frac{1}{4\pi\alpha'}\int_M d^2\sigma
\sqrt{\gamma}[(\gamma^{ab}g_{\mu\nu}(X^0) +
i\epsilon^{ab}B_{\mu\nu}(X^0))\partial_a
X^\mu\partial_b X^\nu + \alpha'R^{(2)}\Phi(X^0)]. \ee  
Here we again note that $R^{(2)}$ is the worldsheet Ricci scalar, and it is noted that the dilaton field $\Phi(X^0)$ is one order of $\alpha'$ higher than the $g_{\mu\nu}(X^0)$ and $B_{\mu\nu}(X^0)$ fields.  We present without derivation and as an illustrative example one configuration used in string cosmology, the so-called linear dilaton background which satisfies the conformal invariance conditions and also demonstrates unitarity of the spectrum, \cite{Mavromates}, \cite{NonCritical_Cosmology}.
\be \label{linear_background} g_{\mu\nu}(X^0) = \eta_{\mu\nu}, B_{\mu\nu}(X^0) = 0, \Phi(X^0) = -2QX^0, \ee
where $Q$ is a constant.  $Q$ is often referred to as the central charge deficit, \cite{Liouiville_Cosmology} and it describes the evolution of the dilaton in $X^0$.  The stress energy tensor in the notation of (\ref{non vanishing stress energy}) where $z$ is the complex worldsheet coordinate, in the linear dilaton configuration of (\ref{linear_background}) is given by:
\be \label{linear_dilaton_set} T_{zz} = - \frac{1}{2}\partial_z
X^\mu\partial_z X^\nu g_{\mu\nu}(X^0) + Q\partial^2_zX^0. \ee
The Weyl anomaly (central charge) associated with this tensor is:
\be \label{linear_dilaton_charge} c = D-12Q^2 = 26. \ee
Thus the spacetime dimensionality to satisfy vanishing Weyl anomalies is now also dependent on $Q$ (unlike the free string case where $D=26$ was a condition to cancel the Weyl anomaly) which represents the dilaton field strength.  String cosmology models where the spacetime dimensionality need not necessarily be $D=26$ (for the bosonic string) due to the presence of a non-zero $Q$ value are often termed non-critical string theories, \cite{NonCritical_Cosmology}.  In eq (\ref{linear_dilaton_charge}), it is noted, \cite{NonCritical_Cosmology}, that the spacetime dimensionality $D$ may be expressed as $D=d + c_I$ where $d$ is the number of large physically apparent dimensions and $c_I$ is the internal charge related to the compactified dimensions; further,  $c_I$ and $12Q^2$ need not be integers.  In the linear dilaton background the line element is, \cite{Mavromates}:
\be \label{linear_dilaton_line_element} ds^2_E=e^{\frac{4QX^0}{D-2}}\eta_{\mu\nu}dX^\mu dX^\nu, \ee
which may be expressed in the form of the Robertson-Walker metric, where $t$ is cosmic time:
\be \label{einsteint frame} ds^2_E= -(dt)^2 + t^2dX^idX^j\delta_{ij}, \ee
where the scale factor, $a(t) = t$.
\be t = \frac{D-2}{2Q}\exp \left(\frac{2Q}{D-2} X^0\right ). \ee
Since $\ddot{a(t)} = \frac{d^2 a(t)}{dt^2} = 0$ this describes a non accelerating, flat universe in which the dilaton field in terms of $t$ is given by:
\be \Phi(t) = (2-D) \ln \frac{2Qt}{D-2}.  \ee
This model may be generalised to one incorporating a curved spacetime background and a non zero value of $B_{\mu\nu}(X^0)$.  One can set the spacetime dimensionality in terms of observable dimensions to four and assume that any remaining dimensions as required by the constraint in (\ref{linear_dilaton_charge}) are compactified.  The antisymmetric field strength may be expressed in terms of a scalar field $h(x)$ (sometimes termed the pseudo-scalar axion field in this context, differing from the QCD-axion), \cite{Mavromates}:
\be \label{string_axion} H_{\rho\mu\nu} = \exp(2 \Phi)\epsilon_{\rho\mu\nu\lambda}\partial^{\lambda}h. \ee
The equations of motion associated with the Weyl anomaly in equations in (\ref{beta_functions}) are as follows:
\bea \label{Weyl_eq_of_motion} R_{\mu\nu} &=& \frac{1}{2}\partial_{\mu}\Phi\partial_{\nu}\Phi + \frac{1}{2}(\partial_{\mu})^2 +\frac{1}{2}e^{2\Phi}[\partial_{\mu}h\partial_{\nu}h - g_{\mu\nu}(\partial h)^2] \nn
0 &=& \nabla^2h + 2 \partial_{\mu}h \partial^{\mu}\Phi \nn
\delta c &=& 12 Q^2 = -3 e^{-\Phi} \left[-R + \nabla^2\Phi + \frac{1}{2}(\partial\Phi)^2 - \frac{1}{2}e^{2\Phi}(\partial h)^2 \right]. \eea
The effective action (termed here simply as $S_{eff}$) obtained when the (\ref{Weyl_eq_of_motion}) are satisfied, in these conditions is:
\be \label{effective_action_nonzero_B} S_{eff} = \int d^4X \sqrt{g^E}\left[R^E -
\frac{1}{2}(\partial_\mu\Phi)^2 - \frac{1}{2}e^{2\Phi}(\partial h)^2 - \frac{1}{3}e^{\Phi} \delta c \right], 
\ee
where this is in four non-compactified spacetime dimensions and where $\delta c$ is defined in the third line of eq (\ref{Weyl_eq_of_motion}).  This admits as a solution the line element in polar coordinates in the Einstein frame:
\be \label {metricpolar}
ds^2_E=-dt^2 + a(t)^2 \left(\frac{dr^2}{1-\kappa r^2}+
r^2(d\theta^2+ sin^2\theta d\phi)\right),\ee 
where $a(t)$ is the scale factor and $\kappa$ is a parameter describing spatial curvature.  Eq. (\ref {metricpolar}) and the solutions to (\ref{Weyl_eq_of_motion}) can be expressed in terms of the Hubble parameter $H(t)=\frac{\dot{a}(t)}{a(t)}$, which describes the evolution of the universe (equation quoted here from \cite{Mavromates}):
\be \label{Hubble_equation} \left(\frac{\ddot{H}+6\dot{H}H -
\frac{4\kappa}{a(t)^2}H}{\dot{H}+3H^2+\frac{2\kappa}{a(t)^2}}\right)^2=-4\dot{H}+\frac{4\kappa}{a(t)^2}-\frac{(\delta
c)^2h_0^2}{36a(t)^6}\cdot\frac{1}{(\dot{H}+3H^2+(\frac{2\kappa}{a(t)^2})^2}.
\ee 
There are two types of solutions to this equation, both of which
have $\kappa$ as non-negative and therefore describe closed
de Sitter universes.  One is the static Einstein universe and the other is a
linearly expanding universe, with $a(t)=t$, as in the linear dilaton case.  The above description of the linear dilaton non-critical string cosmology model is illustrative of an early model of an expanding and homogeneous universe in four dimensions.  We note our model differs in that we derive a logarithmic relationship between the dilaton and $X^0$, which nonetheless can describe an isotropic and linearly expanding universe by choice of our free parameters as described in section (\ref{optical_anisotropy_isotropic_config}).

\subsection{Other String Cosmology Approaches}

\subsubsection{Liouville String Theory}
String theory is a two dimensional field theory where a parameter of the theory is a length scale $l$ as in (\ref{parameter_closed_string}).  Theories which are independent of the length scale are termed critical string theories and are Weyl or scale invariant, \cite{polchinski}.  (Note: the term "non-critical" string theory is generally used in the context where $D$ does not have to be limited to $26$, such as Liouville models described here where the conformal invariance condition is relaxed.  However we note that our model is not limited to $D=26$ but we demonstrate conformal invariance). When the theory is dependent on length scale, the Polyakov classical Euclidean action becomes, with comparison to (\ref{polyakov_action_Euclidean}), and where the length scale $l \equiv \mu$:
\be \label{polyakov_action_non_critical} S_{NC} = \frac{1}{4\pi \alpha'}\int d^2\xi (\gamma)^{\frac{1}{2}}\left( (\gamma^{ab} \partial_a X^{\mu} \partial_b X_{\mu}) + \mu \right). \ee 
Diff invariance remains in the theory but now the metric $\gamma_{ab}(\xi)$ may now only have two of its three degrees of freedom fixed, unlike in our gauge fixing exercise as in (\ref{fiducial_metric}) where the fiducial metric was fully defined.  Now we have:
\be \label{non-critical_metric} \gamma_{ab}(\xi) = \exp(2\phi(\xi))\widehat{\gamma}_{ab}(\xi), \ee
where $\phi(\xi)$ is the scale factor of the Weyl transformation of the worldsheet metric.  In quantizing the critical string, the gauge was fixed and the Faddeev-Popov procedure applied to generate the action to be quantized in the path integral approach as in (\ref{string_path_integral}). In applying this to (\ref{polyakov_action_non_critical}) one has \cite{polchinski}, section 9.9:
\be \label{polyakov_action_non_critical_FP} S = \frac{1}{4\pi \alpha'}\int d^2\xi (\gamma)^{\frac{1}{2}}\left( (\gamma^{ab} \partial_a X^{\mu} \partial_b X_{\mu}) + \mu \exp (2\phi) + \frac{13 \ap}{3}(\widehat{\gamma}^{ab}\partial_a \phi \partial_b \phi + \widehat{R} \phi)\right).\ee 
While details are beyond the scope of this thesis, the scale factor now has a kinetic term and is promoted to a dynamical field in the worldsheet parameter, $\phi(\xi)$, known as the \textit{Liouville} field (not to be confused with the dilaton field also denoted by $\phi$ in this thesis).  In non-critical strings, this is kept as an extra degree of freedom to be integrated over in path integral quantization.  This idea was initially proposed in the late 1980s, \cite{David}, \cite{Distler}.  \cite{polchinski} notes that the target space physics of (\ref{polyakov_action_non_critical_FP}) remains the same as compared to critical string theory where the metric is fully fixed which can be intuitively seen by the fact that the gauge fixing of (\ref{fiducial_metric}) is arbitrary.  Non-critical string models for free strings are not limited by the $D=26$ constraint and the quantized model allows for alternative means of calculating string interaction amplitudes. In \cite{Mavormatas_Liouville} it is proposed that the Liouville field is a dynamical parameter which acts as a renormalisation group scale in the quantized string theory flowing from an infrared fixed point to an ultraviolet one which represents a critical string vacuum state. The field is further identified with physical time.  Liouville string cosmology using the non-perturbative quantum field theory methods employed in this thesis has been researched and demonstrated in \cite{Alexandre3}. 

\subsubsection{Tachyon Cosmology}\label{Tachyon Cosmology}
Here we comment briefly on the tachyon field $T(X)$ referred to in eq. (\ref{string action backgds}) and the surrounding text. While the graviton, antisymmetric tensor and dilaton fields represent massless modes in the lowest state of the most general quantized bosonic string spectrum, the tachyon is also in the lowest state but is not massless.  While we do not derive the result here, \cite{polchinski} notes that its mass squared is proportional to $2-D$, and thus in $D>2$, is negative.  In section (\ref{SectionInstantons}) on instantons, it was described how negative mass squared states in a two dimensional scalar field theory (which is analogous to the tachyon in (\ref{string action backgds})) result in divergent fluctuations around the minimum of the field.  Many, if not most, string cosmology models rely on string actions where the tachyon fields have been removed by adding constraints on the closed string, \cite{zwiebach} to avoid such instabilities.  Tachyon-containing string models with their instabilities controlled (by considering that $T(X)=0$ is not the true vacuum of the theory, \cite{Sen review}) have been deployed recently in early universe evolution modeling.  We utilise a closed bosonic string action and assume that it has been derived so as the tachyon does not arise.   In the cosmological discussion of our results in section (\ref{The Dynamics of the Universe}) we note that our configuration of the bosonic string without the tachyon cannot give rise to an inflationary scenario as our solutions lead to a power law dependence of the scale factors with cosmic time $t$ rather than the exponential relationship $a(t) \sim \exp(Ht)$ required for inflation.  This arises from the dilaton factor in the relationship between the string and Einstein metric expressed in 
eq. (\ref{Einstein_action}) with which one cannot obtain an inflationary scenario unless the value of the dilaton field is zero throughout.  We note here that the non-perturbative methods employed in this thesis and using a tachyon-containing model is shown in \cite{Alexandre_tachyon}.  In this, the authors demonstrate that a bosonic string model incorporating graviton, dilaton, ($\phi$), and tachyon, ($T$), background fields can be shown to be conformally invariant in four spacetime dimensions.  Further this is shown, with the cancellation condition $2 \phi + T = 0$, to lead to an Einstein frame action in the canonical form with no dilaton nor tachyon pre-factors and hence leading to a scale factor of the form $a(t) = a_0 \exp (H t)$ where $H$ is the Hubble parameter.  The authors note a merit of this model as being feasible in four spacetime dimensions and thus not requiring complications associated with compactifcation of additional dimensions during the inflationary phase.  While this thesis does not consider a supersymmetric extension of our model, it is noted that tachyon modes would necessarily be eliminated in such a model and we do not consider such modes in this thesis.

\subsection{String Cosmology and Anisotropy}\label{String Cosmology and Anisotropy}
In this thesis we are interested in any fundamental spatial anisotropy, or preferred direction, which may have originated in the evolution of the early universe as hinted by our string model.  Analysis of any such anisotropies are a key motivation behind the study of the cosmic microwave background from the Wilkinson Microwave Anisotropy Probe, WMAP, and other data sources.   The main phenomenological motivation of our string solution involves coupling it to an electromagnetic field and considering the rotational effect on the electric field.  The phenomenon known as  \textit{cosmic birefringence}, \cite{Prasanta}, \cite{Optical_Anisotropy3} refers to additional rotation in the plane of polarisation by a electromagnetic wave traveling through the universe, after Faraday rotation (caused by galactic magnetic fields) has been accounted for.  An optical anisotropic or birefringent substance refers to one in which the refractive index varies with spatial direction, \cite{Optical_Anisotropy7}.  In this context "substance" refers to the intergalactic distances over which electromagnetic radiation travels prior to observation on earth.  Such an affect will cause a phase difference or rotation $\eta$ of the electric field of the radiation along the optic axis.  Such an effect is typically studied by viewing extra-galactic radio waves emitted by pulsars or other distant objects.  This may be evidence of cosmological anisotropy on a vast scale, \cite{Optical_Anisotropy} and the rotation $\theta$ of a polarised wave of wavelength $\lambda$ is defined by:
\be \label{polarised_rotation} \theta (\lambda) = \alpha \lambda^2 + \eta, \ee
where the $\alpha$ term is the Faraday rotational effect and $\eta$ is the wavelength-independent cosmic birefringence effect of interest to this study.  Polarised synchrotron radiation was predicted and observed from distant extra-galactic sources from the 1950s and datasets emerged, \cite{Optical_Anisotropy} with information on the redshift, $z$ and the angular orientation of the source-galaxy optic axis (which sets the source polarisation angle, $\psi$). The fitting parameter $\alpha$ depends on the density of electrons and magnetic field lines in the line of sight of the radiation.  \cite {Optical_Anisotropy2} notes that studies to determine cosmic birefringence originating in large scale anisotropy account for Faraday rotation on a galaxy by galaxy basis in the line of sight of observation, and the residual angle $\eta$ is deemed to be due to cosmic birefringence.  In \cite{Optical_Anisotropy} (1997) a systematic $\eta$ angle effect is reported, correlated with angular positions and source distances, with dependence on redshift $z$ ruling out local effects.  More recent work, \cite{Optical_Anisotropy5} (2009) notes that several studies of the BOOMERanG 2003 flights data (which are devoted to measurement of polorisation of the CMB radiation) may show cosmological birefringence effects of $\eta \sim -4.3^o \pm 4.1^o$.  Meanwhile \cite{Optical_Anisotropy6} (2010) reports on further tests to look at cosmic birefringence from eight radio galaxies at redshift $z > 2$, and reports no $\eta$ rotation within a limit of $\eta \sim -0.8^o \pm 2.2^o$.  While experimental research into cosmic birefringence remains ongoing, we consider theories which predict it with a focus on string related theories.\\\\\cite{Optical_Anisotropy6} notes that the potential for the universe on a large scale to rotate (other than through Faraday rotation) the plane of polarisation of light could have a number of possible origins: the presence of a pseudo scalar condensate, neutrino number asymmetry and the violation of certain fundamental symmetries such as CPT.  \cite{Optical_Anisotropy6} notes that the starting point for modeling cosmic birefringence due to pseudo scalar effects is the interaction between the photon and a scalar field $\phi$ (not the dilaton here).
\be \label{cosmic_bire} \mathcal{L} = -\frac{1}{4} F_{\mu\nu}F^{\mu\nu} - 
\frac{1}{2}
\partial^{\mu}\phi\partial_{\mu}\phi - U(\phi) - \frac{g_{\phi}}{4}\phi F_{\mu\nu}\widetilde{F}^{\mu\nu}, \ee
with $F_{\mu\nu}\widetilde{F}^{\mu\nu}$ defined in a similar manner to the axion interaction Lagrangian in eq (\ref{general theta terms}). (One notes the negative sign in front of the pseudo-scalar kinetic term $\frac{1}{2}
\partial^{\mu}\phi\partial_{\mu}\phi$ is a matter of convention deployed in this paper, and also by Peccei in \cite{Peccei3}.  We deploy the opposite convention in our treatment of the pseudo-scalar axion in (\ref{model})).   \cite{Optical_Anisotropy6} notes that in the lowest order in fluctuations, the photon is coupled to the time derivative of the scalar field $\phi$ and thus different configurations of $\eta(\tau)$, where $\tau$ is conformal time, will lead to differing results for the cosmic birefringence angle.  The angle $\eta$ is related to variation of $\phi(\tau)$:
\be \label{cosmic birefringence_angle} \eta(\tau) = \frac{g_{\phi}}{2}[\phi(\tau_0) - \phi(\tau)]. \ee
\cite{Prasanta} and \cite{Optical_Anisotropy3} consider the cosmic birefringence effects of the  the antisymmetric tensor field $B_{\mu\nu}$ in low energy effective string theories.   In \cite{Optical_Anisotropy3}, the field strength of the antisymmetric tensor is coupled to an electromagnetic gauge field $A_{\mu}$ to produce an "auxiliary" field $T_{\mu\nu\lambda}$:
\be \label{EM couple} T_{\mu\nu\lambda} = \sqrt{G} \partial_{[\mu}B_{\nu\lambda]} + \frac{1}{3}\sqrt{G} A_{[\mu}F_{\nu\lambda]}, \ee
where $F_{\mu\nu}$ is the electromagnetic tensor and $G$ is Newton's constant.  In this it is assumed that $H_{\mu\nu\lambda} = \partial_{[\mu}B_{\nu\lambda]}$ varies only with cosmic time, $\tau$, and 
$H_{\mu\nu\lambda}=\varepsilon_{\mu\nu\lambda 3}\partial^3 h$ where $h$ is the string axion pseudo scalar field.  With a flat spacetime cosmology configuration with $h = h'\tau + h_0$ (where $h'$ and $h_0$ are values of $h$ at times $\tau$ and $\tau = 0$ respectively) the authors compute a cosmic birefringence angle $\eta$ is proportional to $h'\tau$, that is the cosmic time that has elapsed between the values $h'$ and $h_0$.  In another scenario associated with large values of $\tau$, \cite{Optical_Anisotropy3} finds the rotation for the radiation and matter dominated time periods goes as $\tan^{-1}\frac{1}{\tau}$ and $\tan^{-1}\frac{1}{\tau^3}$ respectively.  In a follow up paper, \cite{Prasanta} using similar methods but with a heterotic string theory, similar results are noted for the flat spacetime case.  Our treatment of the bosonic string axion similarly couples to an electromagnetic field and we likewise consider optical anisotropy effects. 

\chapter {Non-perturbative String Cosmology}\label{chapter non-perturbative Approach to String Cosmology}

\section{Our Approach to the Bosonic String}

\subsection{Introduction}
In our treatment of the quantization of the axion, we used an alternative approach to the Wilsonian exact renormalisation group (ERG) as outlined in section (\ref{An Alternative, Exact Approach}) and implemented in section (\ref{Quantization of the Axion}).  Our aim in this section of the thesis is to apply these same techniques to a conformally invariant bosonic string model including the string-axion, motivated by cosmology.  As with our treatment of the the QCD axion, the rationale for applying ERG-based methods resulting in non-perturbative equations is that we are examining theories which approach the high energy limit in terms of using perturbative techniques.   We commence with a worldsheet action based on (\ref{string action backgds}), without tachyon fields.  As with our treatment of the axion, where the scale factor $f$ was used to construct our final exact equation (\ref{final_result_1}), in the treatment of the bosonic string, we utilise $\lambda$ which is a parameter of string theory, (where $\lambda \equiv \frac{1}{\ap}$ in the full quantum theory, and $\ap$ is the Regge slope).  We seek a $\lambda$-independent solution to an evolution equation of the effective bosonic string action in four spacetime dimensions.  We demonstrate that our resulting configuration is conformally invariant by showing the beta functions can vanish to all orders in $\ap$.  This contrasts with most conventional critical string cosmology models which are generally limited to the beta functions vanishing at two loops or less.  We achieve this via a two step process.  Firstly, we show the beta functions of our solution are self-homogeneous to all orders in the time ($X^0$) dimension.  Secondly, utilising established methods which state the beta functions may be transformed, with the physics unaffected, by arbitrary linear transformations of the spacetime coordinates of the string background fields, we show that they can be made to vanish.  It is stressed that both of these steps are required for our demonstration of conformal invariance to all orders.  We then look at a particular area of cosmological implications.  This involves the coupling of the field strength $H_{\rho\mu\nu}$ to an electromagnetic field, solving the resulting equations of motion, and examining any resulting optical activity demonstrated by the configuration, following work done in \cite{Optical_Anisotropy3}.  We initially fix one index of  $H_{\rho\mu\nu}$ (an anisotropic configuration of the antisymmetric tensor) and then we consider the general case with all indices free (isotropic).  It is stressed at this point that the key result of our research in this area is to demonstrate a technique for showing conformal invariance to all orders in $\ap$ for a bosonic string cosmology model.  The predictive powers of the model may be incomplete as the resulting metric does not support an inflationary scenario.  As a note on content, it is acknowledged that a large portion of the derivations in this section were published in \cite{Alexandre4} and \cite{Alexandre5} both of which built on earlier work by Alexandre and Mavromatos in \cite {Alexandre1} and \cite{Alexandre2}. 

\subsection{The Effective String Action}
We commence with the following bare action of the bosonic string in background fields, based on (\ref{string action backgds}), with $\lambda \equiv \frac{1}{\ap}$, $g_{\mu\nu} = \eta_{\mu\nu}$,  $B_{\mu\nu} = a_{\mu\nu}$ and the dilaton field varying in the time $X^0$ dimension only (i.e. an isotropic and locally flat universe).  We consider in Euclidean format and hence the imaginary component of (\ref{string action backgds}) reverts to real:
\be \label{string_action_model} 
\hspace{-0.5cm}S_{\lambda}  = \frac{1}{4\pi} \int d^2\xi \sqrt{\gamma} \,\Bigl\{\lambda\left[\gamma^{ab} \eta_{\mu\nu} +\epsilon^{ab}a_{\mu\nu}\right]\partial_a  X^\mu \partial_b  X^\nu + R^{(2)} \Phi(X^0) \Bigr\}. 
\ee
This is the bare theory.  Here $X^{\mu}$ are the microscopic fields over which the path integral in (\ref{string_partition}) is taken to compute the partition function. We consider flat spacetime $\eta_{\mu\nu}$ and hence we assume no dependence by the worldsheet metric $\gamma^{ab}$ on $X^{\mu}$. We will use techniques outlined in section (\ref{The Effective Action and Potential}) to construct a quantised effective theory and define the Legendre effective action $\Gamma$. We note that these techniques may only be employed if the evolution parameter is multiplied by terms quadratic in the field $X^{\mu}$.  We allow for $X^0$ (i.e. time) dependence in all background fields in an assumed form for the effective action later.  First we define the partition function, $Z$, and connected diagram generator, $W$, for our theory, using (\ref{partitiondefi}) and associated results.
\be \label{string_partition} Z =  \int {\cal D} [X^\mu] \exp -\left(S_{\lambda}
+ S_S \right) = \exp (-W), \ee
where we have added a source term $S_S$ which interacts with the fields $X^{\mu}$ (where, comparing the notation deployed in section (\ref{The Effective Action and Potential}), $J^{\nu} \equiv \sqrt{\gamma}R^{(2)}\eta_{\mu\nu}V^{\mu}$).
\be \label{string_source} S_S = \frac{1}{4\pi} \int d^2\xi \sqrt{\gamma}R^{(2)}\eta_{\mu\nu}V^{\mu}X^{\nu}. \ee
Following from (\ref{classicalfield}) the classical fields $X^\mu_{cl}$ are defined by:
\be \label{string_classicalfield} X^\mu_{cl}(\xi) =\frac{1}{Z} \int {\cal D}[X^\mu]\, X^\mu \exp -\left(S_{\lambda} + S_S \right) = \frac{\delta W}{\delta
J(\xi)}  = \frac{1}{\sqrt{\gamma_{\xi}}R_{\xi}^{(2)}}\frac{\delta W}{\delta
V_{\mu}(\xi)} . \ee
Hence as in (\ref{secondderivative}) taking the second functional derivative gives (where the notation $<.....>$ is as used in (\ref{two_point_correlation}), and $\xi$ and $\zeta$ are worldsheet parameter variables of integration):
\be
\frac{1}{\sqrt{\gamma_\zeta\gamma_\xi}R^{(2)}_\zeta
R^{(2)}_\xi}\frac{\delta^2 W}{\delta V_\mu(\zeta)\delta V_\nu(\xi)}
=X_{cl}^\nu(\xi)X_{cl}^\mu(\zeta)-< X^\nu(\xi) X^\mu(\zeta)>. \ee
As in (\ref{legeffact}) we then introduce the Legendre transform of $W$:
\be
\Gamma=W-\int d^2\xi\sqrt{\gamma}R^{(2)}V^\mu X_\mu. \ee 
Using the result (\ref{legdir}) we take the first derivative of $\Gamma$ with respect to the classical field, and then the second as in (\ref{inverse relationship}).
\bea\label{derivG}
\frac{1}{\sqrt{\gamma_\xi}R^{(2)}_\xi}\frac{\delta\Gamma}{\delta
X_{cl}^\mu(\xi)}&=&-V_\mu(\xi),\nn
\frac{1}{\sqrt{\gamma_\xi\gamma_\zeta}R^{(2)}_\xi R^{(2)}_\zeta}
\frac{\delta^2\Gamma}{\delta X_{cl}^\nu(\zeta)\delta X_{cl}^\mu(\xi)}&=&
-\left(\frac{\delta^2W}{\delta V_\nu(\zeta)\delta
V_\mu(\xi)}\right)^{-1}. \eea  
We now deploy the methods described in section (\ref{An Alternative, Exact Approach}) to derive an exact equation for the evolution of the effective action, $\Gamma$ of our theory with the parameter $\lambda$.  We make several comments on this method first.
\begin{itemize}
\item As with the use of the parameter $f$ in the treatment of the axion, we use the parameter $\lambda$ as a tool in setting up evolution equations for our effective action, following which we can examine solutions with no variation of the action with our chosen parameter, $\lambda$.
\item Considering the bare action (\ref{string_action_model}) the progression $\lambda \rightarrow \infty$ corresponds to $\ap \rightarrow 0$ and thus approaches the classical limit of our theory where the kinetic term dominates and dilaton mediated interactions are negligible.
\item As $\lambda \rightarrow \frac{1}{\ap}$ we approach the full quantum theory as the $\Phi(X^0)$ interactions become important.
\end{itemize}
Where a dot signifies a derivative with respect to $\lambda$ and noting the result $\dot{W}=\dot{\Gamma}$ as derived in (\ref{legeffact_b_derivative}) we have an exact
equation for the evolution of $\Gamma$.
\bea\label{evolGappendix} \dot\Gamma&=&\frac{1}{4\pi}\int
d^2\xi\sqrt{\gamma}\left(\gamma^{ab}\eta_{\mu\nu}+\varepsilon^{ab}a_{\mu\nu}\right)
<\partial_a X^\mu\partial_b X^\nu> \nn
&=& \frac{1}{4\pi}\mbox{Tr}\left[
\left(\gamma^{ab}\eta_{\mu\nu}+\varepsilon^{ab}a_{\mu\nu}\right)
\frac{\partial}{\partial\xi^a}\frac{\partial}{\partial\zeta^b} < X^\mu (\xi) X^\nu (\zeta) >\right]. \eea
In this we define the trace as:
\be \mbox{Tr}[.....\,\,\,] = \int d^2\xi \, d^2\zeta \sqrt{\gamma_{\xi}\gamma_{\zeta}} \,\, \delta^2(\xi - \zeta)[.....\,\,\,]. \ee
Using the results (\ref{string_classicalfield}) and (\ref{derivG}) we have an exact evolution equation.
\bea\label{evolGappendix1} \dot\Gamma&=&\frac{1}{4\pi}\int
d^2\xi\sqrt{\gamma}\left(\gamma^{ab}\eta_{\mu\nu}+\varepsilon^{ab}a_{\mu\nu}\right)
\partial_a X_{cl}^\mu\partial_b X_{cl}^\nu \nn
&&+\frac{1}{4\pi}\mbox{Tr}\left\{\left(\gamma^{ab}\eta_{\mu\nu}+\varepsilon^{ab}a_{\mu\nu}\right)
\frac{\partial}{\partial\xi^a}\frac{\partial}{\partial\zeta^b}
\left(\frac{\delta^2\Gamma}{\delta X_{cl}^\nu(\zeta)\delta X_{cl}^\mu(\xi)}\right)^{-1}\right\}. \eea 
As with our ansatz for the assumed form of the Legendre action of the axion in (\ref{assumed form}) we now also assume a form for the Legendre effective action of the bosonic string so we may equate two separate, $\lambda$-dependent expressions for $\Gamma$.  In the ansatz for the bosonic string's effective action we assume a form for the gravitational tensor which allows for curvature but varies only in the $X^0$ time dimension, with $D$ the total spacetime dimensionality, which we will later assume to be four as it becomes a free parameter of the theory.  In this we have an isotropic universe in terms of curvature.
\be \label{graviton_assumed_form} g_{\mu\nu} =  \mbox {diag} \left(\kappa(X^0), \tau_1(X^0),.....\tau_{D-1}(X^0)\right). \ee
We denote the antisymmetric tensor $B_{\mu\nu}$ again varying only with the $X^0$ time dimension.  Using (\ref{graviton_assumed_form}) the assumed form (as opposed to the exact expression given in eq. (\ref{evolGappendix})) for the effective action is as follows.
\bea\label{gradexp}
\Gamma_\lambda&=&\frac{1}{4\pi} \int
d^2\xi\sqrt{\gamma}\left\{\gamma^{ab}\kappa(X^0)\partial_a
X^0\partial_bX^0 +\gamma^{ab}\sum_{i=1}^{D-1}\tau_i(X^0)\partial_a
X^i\partial_b X^i\right.  \nn &&
~~~~~~~~~~~~~~~+\varepsilon^{ab}B_{\mu\nu}(X^0)\partial_a
X^\mu\partial_b X^\nu+R^{(2)}\phi(X^0)\Bigg\}, \eea where
$\kappa,\tau_i,\phi$ are $\lambda$-dependent functions of $X^0$. We thus have established two expressions for the Legendre effective action $\Gamma$ for our theory in (\ref{gradexp}), an assumed form, and in (\ref{evolGappendix1}) an exact non-perturbative differential equation.  We utilise these, along with the constraints of conformal invariance expressed through (\ref{beta_functions}) to search for meaningful solutions.

\subsection{Detailed Calculations}
We wish to evaluate our exact evolution equation (\ref{evolGappendix1}) by substituting on the assumed form for $\Gamma$ in eq. (\ref{gradexp}).  We take a flat worldsheet metric (the fiducial gauge) $\gamma^{ab}=\delta^{ab}$ and configurations of the classical fields constant in spacetime, $X_{cl}^\mu(\xi)=x^\mu$.  For brevity we now drop the \textit{cl} subscript denoting a classical field so that $X_{cl}^\mu \equiv X^\mu=x^\mu$. The second functional derivatives of $\Gamma$ in (\ref{evolGappendix1}) are ( with no summation on $i$) and $\Delta$ is the worldsheet Laplacian operator.
\bea \label{second_functional_derivatives}
\frac{\delta^2\Gamma}{\delta X^0(\zeta)\delta X^0(\xi)}|_{X^{\mu}=0}&=&
-\frac{\kappa}{2\pi}\Delta\delta^2(\xi-\zeta)+\frac{R^{(2)}\phi^{''}}{4\pi}\delta^2(\xi-\zeta),\nn
\frac{\delta^2\Gamma}{\delta X^i(\zeta)\delta X^i(\xi)}|_{X^{\mu}=0}&=&
-\frac{\tau_i}{2\pi}\Delta\delta^2(\xi-\zeta)\nn
\frac{\delta^2\Gamma}{\delta X^i(\zeta)\delta X^j(\xi)}|_{X^{\mu}=0}&=&0~~~~~~~i\ne j
\eea
where a prime denotes a derivative with respect to $x_0$. We are thus considering a constant field configuration $x^{\mu}$ where the off diagonal spacetime terms play no role.  As such,
the antisymmetric tensor $B_{\mu\nu}$ does not play a role in the evolution equation for $\Gamma_\lambda$, as
its contribution is proportional to:
\be
\varepsilon^{ab}\partial_a\partial_b~\delta^{(2)}(\xi-\zeta)=0.
\ee
With regards to conversion to momentum space on the worldsheet, we note that our closed string model is described by a spherical worldsheet with non zero curvature which we neglect for the purposes of taking Fourier transforms required for momentum space formulation using 
$F(p) = \int d^2 \xi e^{-i \textbf{p.$\xi$ }}\,F(\xi)$ (in eq (\ref{phase_space_derivatives}) below $\zeta, \xi$ become $p, q$). For the purposes of evaluating eqs. (\ref{second_functional_derivatives}) we consider a small but non zero $R^{(2)}$.  In phase space on the worldsheet, (\ref{second_functional_derivatives}) read as:
\bea \label{phase_space_derivatives}
\frac{\delta^2\Gamma}{\delta X^0(p)\delta X^0(q)}&=&
\frac{1}{4\pi}\left(2\kappa p^2+R^{(2)}\phi^{''}\right)\delta^2(p+q);\nn
\frac{\delta^2\Gamma}{\delta X^i(p)\delta X^i(q)}&=&
\frac{\tau_i ~p^2}{2\pi}\delta^2(p+q).
\eea
The area of a sphere with curvature scalar $R^{(2)}$ is $8\pi/R^{(2)}$, so
with the constant configuration $X^\mu=x_\mu$ we have
\be\label{Gammaconfig}
\Gamma=2\phi_\lambda(x_0),
\ee
where the $\lambda$ in $\phi_\lambda$ signifies a $\lambda$ dependence.  The term containing the trace in (\ref{evolGappendix1}) is computed as follows.  The results are in phase space using the results in eqs. (\ref{phase_space_derivatives}) and where $[......]$  in eq. (\ref{TrABA}) represents $\left[\eta_{\mu\nu} \frac{\partial}{\partial p}\frac{\partial}{\partial q}
\left(\frac{\delta^2\Gamma}{\delta X^\nu(p)\delta X^\mu(q)}\right)^{-1}\right]$.
\bea\label{TrABA}
\frac{1}{4\pi}\mbox{Tr} [......]
&=&-\int\frac{d^2p}{(2\pi)^2}\left(\frac{p^2}{2\kappa p^2+R^{(2)}\phi^{''}}
+\frac{1}{2}\sum_{i=1}^{D-1}\frac{1}{\tau_i}\right)\frac{8\pi}{R^{(2)}}\nn
&=&-\frac{\Lambda^2}{R^{(2)}}\left(\frac{1}{\kappa}+\sum_{i=1}^{D-1}\frac{1}{\tau_i}\right)
+\frac{\phi^{''}}{2\kappa^2}\ln\left(1+\frac{2\Lambda^2\kappa}{R^{(2)}\phi^{''}}\right).
\eea
To calculate this trace, we rely on the result:
\be
\delta^2(p=0)=~\mbox{world-sheet area}~=\frac{8\pi}{R^{(2)}}.
\ee
We have used the definition of the trace as $\mbox{Tr}[\,\,\,] = \int d^2 p \, d^2 q \sqrt{\gamma_{p}\gamma_{q}} \,\, \delta^2(p-q)[\,\,\,]$.  In this, $\Lambda$ is the worldsheet high energy cutoff used to regularise the theory.  We note here that $\Lambda$ does not appear in any of our final results, nor do we attempt to evaluate it.  The evolution equation for $\Gamma$, expressed in effect as an evolution of $\phi$
is finally obtained by putting together results
(\ref{evolGappendix}), (\ref{Gammaconfig}) and (\ref{TrABA}):
\be \label{phi_evolution}
\dot\phi=-\frac{\Lambda^2}{2R^{(2)}}\left(\frac{1}{\kappa}+\sum_{i=1}^{D-1}\frac{1}{\tau_i}\right)+
\frac{\phi^{''}}{4\kappa^2}\ln\left(1+\frac{2\Lambda^2\kappa}{R^{(2)}\phi^{''}}\right).
\ee
We would like to find solutions independent of $\lambda$ (i.e. where $\dot\phi = 0$) and configurations which preserve conformal invariance.  A fixed point solution is possible for $\kappa=F\phi^{''}$ where $F$ is a constant: 
\be \label{evolution_phi}
\kappa\dot\phi=-\frac{\Lambda^2}{2R^{(2)}}\left(1+\kappa\sum_{i=1}^{D-1}\frac{1}{\tau_i}\right)
+\frac{1}{4F}\ln\left(1+\frac{2\Lambda^2F}{R^{(2)}}\right),
\ee
if the following condition is also satisfied:
\be
\kappa\sum_{i=1}^{D-1}\frac{1}{\tau_i}=-1+\frac{R^{(2)}}{2\Lambda^2F}
\ln\left(1+\frac{2\Lambda^2F}{R^{(2)}}\right)
=\mbox{constant}.
\ee
For large $\Lambda$, this constant is negative, representing a Minkowski signature in target spacetime, a desired configuration.  We have then the following conditions to satisfy, in order to have an $\alpha^{'}$-independent solution:
\bea\label{conditions}
\kappa(X^0)&\propto&\phi^{''}(X^0)\nn
\sum_{i=1}^{D-1}\frac{1}{\tau_i(X^0)}&\propto&\frac{1}{\kappa(X^0)}.
\eea
We now test for and apply the further constraints of conformal invariance.

\subsection{Conformal Invariance}\label{conformal_anisotropy}
\subsubsection{Introduction}

We will base our demonstration of conformal invariance in the vanishing of the beta functions in (\ref{beta_functions}) for our theory.  A key point of our derivation resulting in (\ref{phi_evolution}) is that it originates from a non-perturbative expression.  We thus want to demonstrate conformal invariance to all orders in $\ap$.  This contrasts with the common approach in string cosmology where the expressions in (\ref{beta_functions}) are usually considered to first or at most second order in $\ap$.  We will do this by demonstrating first and second order $\ap$ conformal invariance and by induction showing this, under certain assumptions, can be applied to all orders.\\\\
To first order in $\alpha^{'}$, the beta functions (\ref{beta_functions}) for the bosonic world-sheet
$\sigma$-model theory in graviton, antisymmetric tensor and dilaton backgrounds are,  \cite{low.ener.eff.} and \cite{metsaev}:
\bea\label{weylconditions}
\beta_{\mu\nu}^{g(1)}&=&R_{\mu\nu}+2\nabla_\mu\nabla_\nu\phi-\frac{1}{4}H_{\mu\rho\sigma}H_\nu^{~\rho\sigma}\nn
\beta_{\mu\nu}^{B(1)}&=&-\frac{1}{2}\nabla^\rho H_{\rho\mu\nu}+\partial^\rho\phi H_{\rho\mu\nu}\nn
\beta^{\phi(1)}&=&\frac{D-26}{6\alpha^{'}}-\frac{1}{2}\nabla^2\phi+\partial^\rho\phi\partial_\rho\phi
-\frac{1}{24}H_{\mu\nu\rho}H^{\mu\nu\rho}
\eea
We note that the terms in (\ref{weylconditions}) are expressed in terms of $R_{\mu\nu}$ and $H_{\rho\mu\nu}$ rather than the tensors $g_{\mu\nu}$ and $B_{\mu\nu}$.  We now consider an ansatz for $H_{\rho\mu\nu}$.
\be
\label {axion} H_{\rho\mu\nu} =
\omega_0\cdot(X^0)^m\varepsilon_{\rho\mu\nu\sigma}(X^0)\partial^\sigma h,
\ee where
\be h=h_1X^1+h_2X^2+h_3X^3 = \left(\sum_{i=1}^3 h_iX^i\right). \ee 
(Note: the dot in $\omega_0\cdot(X^0)^m$ indicates multiplication of $\omega_0$ by $(X^0)^m$ whereas $\varepsilon_{\rho\mu\nu\sigma}(X^0)$ indicates an $(X^0)$ dependence in $\varepsilon_{\rho\mu\nu\sigma}$). We consider the spacetime metric as a function of the $X^0$ coordinate:
\be \label{metric_general_case} g_{\mu\nu}=\mbox{diag}(\kappa(X^0), \tau_i(X^0), \tau_j(X^0),
\tau_k(X^0)). \ee
In (\ref{axion}), $\varepsilon_{\rho\mu\nu\sigma}$ is the Levi-Civita tensor in four dimensions which has some $X^0$ dependence.  In the linear sigma string cosmology model, \cite{Mavromates}, the linear function $h$ plays the role of the string-axion (also noted in (\ref{string_axion})), (with $\omega(X^0)\equiv \exp(2\Phi)$).  We initially limit our choice of $\omega_0 \cdot (X^0)^m \epsilon_{\rho\mu\nu\sigma}\partial^{\sigma}h$ such that:
\bea \label{fixing_h_inx3}
h&=&h_0X^3\nn
\omega_0 \cdot (X^0)^m &=&1,
\eea
where where $h_0$ is a constant and thus the antisymmetric tensor field strength is anisotropic as the $X^3$ spatial direction is preferred. For this by definition $(\varepsilon_{0123})^2 = \det (g_{\mu\nu}(X^0))$ and thus we have for the field strength:
\be \label{fieldstrengthansatz}
H_{0ij}= \sqrt{\kappa\tau_i\tau_j\tau_k}\frac{h_k\epsilon_{ijk}}{\tau_{k}(X^0)}. \ee
We have taken an anisotropic configuration of the antisymmetric tensor field strength. The more general isotropic case with the full form of (\ref{axion}) will be considered later in section (\ref{Generalisation to an Isotropic Universe}).

\subsubsection{Structure of the First Order Beta Functions}
The first order beta functions we will use are noted in (\ref{weylconditions}).  We assume a $X^0$ dependence for $\phi$, $\kappa$ and $\tau_i$ given our constraints on these parameters in (\ref{conditions}), where $\phi_0$ and $\kappa_0$ are constants.  Prime denotes derivative with respect to $X^0$.
\bea\label{ansatz}
\phi^{'}(X^0)&=&\frac{\phi_0}{(X^0)^n},\nn
\kappa(X^0)&=&\frac{\kappa_0}{(X^0)^{n+1}}\nn
\tau_i(X^0)&=&\frac{-\kappa_0}{(X^0)^{n_i}}.
\eea
We now show that $n=1$ to ensure the functions in (\ref{weylconditions}) are homogeneous in powers of $X^0$ given the properties of the spacetime (i.e. the form of the geometric parameters and operators $R_{\mu\nu}$, R, $\nabla_{\mu}$ etc).  This is obviously essential to ensure any meaning to the summations of the terms in the beta functions.  For the spacetime metric $g_{\mu\nu}(X^0)=\mbox{diag}(\kappa(X^0),\tau_1(X^0), ...,\tau_{D-1}(X^0))$,
the non-vanishing components of the Christoffel symbols are (without summation over the space index $i$) 
\be
\Gamma^i_{~0i}=\frac{\tau_i^{'}}{2\tau_i},~~~~~~~
\Gamma^0_{~00}=\frac{\kappa^{'}}{2\kappa},~~~~~~~~
\Gamma^0_{~ii}=-\frac{\tau_i^{'}}{2\kappa},
\ee
so that the non vanishing covariant derivatives of the dilaton are
\bea
\nabla_0\nabla_0\phi&=&\phi^{''}-\frac{\kappa^{'}}{2\kappa}\phi^{'},\nn
\nabla_i\nabla_i\phi&=&\frac{\tau_i^{'}}{2\kappa}\phi^{'}.
\eea
The non-vanishing components of the Ricci and Riemann tensors are
\bea
R_{0i0}^{~~~~i}&=&-\left(\frac{\tau_i^{'}}{2\tau_i}\right)^{'}+\frac{\kappa^{'}\tau_i^{'}}{4\kappa\tau_i}
-\left(\frac{\tau_i^{'}}{2\tau_i}\right)^2  ,\nn
R_{iji}^{~~~~j}&=&-\frac{\tau_i^{'}\tau_j^{'}}{4\kappa\tau_j}\nn
R_{00}&=&\sum_{i=1}^{D-1}\left[-\left(\frac{\tau_i^{'}}{2\tau_i}\right)^{'}+\frac{\kappa^{'}\tau_i^{'}}{4\kappa\tau_i}
-\left(\frac{\tau_i^{'}}{2\tau_i}\right)^2\right]\nn
R_{ii}&=&-\frac{\tau_i^{'}}{4\kappa}\left( \sum_{j\ne i}^{D-1}\frac{\tau_j^{'}}{\tau_j}\right) 
+\frac{\tau_i}{\kappa}\left[ -\left(\frac{\tau_i^{'}}{2\tau_i}\right)^{'}+\frac{\kappa^{'}\tau_i^{'}}{4\kappa\tau_i}
-\left(\frac{\tau_i^{'}}{2\tau_i}\right)^2\right]
\eea
Together with the condition $\kappa\propto\phi^{''}$,  with power laws for the metric components,
$R_{00}$ is homogeneous, in terms of powers of $X^0$, to $\nabla_0\nabla_0\phi$ only if 
\bea
\phi(X^0)&=&\phi_0\ln(X^0)\nn
\kappa(X^0)&=&\frac{\kappa_0}{(X^0)^2},
\eea
\\\\
and thus we have $n=1$ in (\ref{ansatz}).  Given (\ref{fieldstrengthansatz}) the non-vanishing components of $H_{\rho\mu\nu}$ are (such that $\rho\neq\mu\neq\nu$, and listing only one permutation here for brevity):
\be
H_{012}=\varepsilon_{0123}\partial^3 h=-\kappa_0 h_0(X^0)^{(n_3-n_1-n_2-2)/2}.
\ee
From our condition $h=h_0X^3$ we have singled out the $X^3$ direction, thus by symmetry $n_1=n_2$ (although we leave these denoted as $n_1$ and $n_2$ for the time being).  We may now evaluate the first order in $\ap$ beta functions in terms of $n_1$, $n_2$, $n_3$, the time dimension $X^0$ and constants $h_0$, $\kappa_0$ and $\phi_0$ with a view to finding beta functions homogeneous in $X^0$.  The equations in (\ref{weylconditions}) may be evaluated, (without summation on the index $i$). 
\bea
\beta_{00}^{g(1)}&=&\frac{-1}{4(X^0)^2}\left[2h_0^2(X^0)^{n_3}+\sum_{j=1}^3 n_j^2\right] \nn
\beta_{ii}^{g(1)}&=&\frac{n_i}{4(X^0)^{n_i}}\left[2h_0^2(X^0)^{n_3}
+4\phi_0+\sum_{j=1}^3 n_j\right]~~\longrightarrow~~i=1,2 \nn
\beta_{33}^{g(1)}&=&\frac{n_3}{4(X^0)^{n_3}}\left[4\phi_0+\sum_{j=1}^3 n_j\right]\nn
\beta_{12}^{B(1)}&=&\frac{h_0}{4}(X^0)^{(n_3-n_1-n_2)/2}\left[n_3-4\phi_0\right]\nn
\beta^{\phi(1)}&=&-\frac{11}{3\alpha^{'}}+\frac{\phi_0}{4\kappa_0}\left[ 4\phi_0+\sum_{j=1}^3 n_j\right]
-\frac{h_0^2}{4\kappa_0}(X^0)^{n_3}.
\eea
In order for these beta functions to be homogeneous, it is necessary to have $n_3=0$.
The second equation of (\ref{conditions}) can then be satisfied with $n_1=n_2=2$.
The non-vanishing components of $H_{\rho\mu\nu}$ are:
\be
H_{012}=-\frac{\kappa_0 h_0}{(X^0)^3}.
\ee
Meanwhile the non-vanishing components of the antisymmetric tensor $B_{\mu\nu}$ are 
\be
B_{12}=-B_{21}=-\frac{2\kappa_0 h_0}{n_3-n_1-n_2}(X^0)^{(n_3-n_1-n_2)/2}.
\ee
Thus we have a configuration for the beta functions in which they are homogeneous to first order in $\ap$: 
\bea\label{solution}
g_{\mu\nu}(X^0)&=&\mbox{diag}\left(\frac{\kappa_0}{(X^0)^2},\frac{-\kappa_0}{(X^0)^2},
\frac{-\kappa_0}{(X^0)^2},-\kappa_0\right) \nn
B_{\mu\nu}(X^0)&=&\left(\delta_{\mu 1}\delta_{\nu 2}-\delta_{\mu 2}\delta_{\nu 1}\right)\frac{\kappa_0 h_0}{2(X^0)^2}\nn
\phi(X^0)&=&\phi_0\ln(X^0).
\eea
We note here that the components of the matrix $g_{\mu\nu}(X^0)$ are dimensionless, with $\kappa_0$ a constant and the time-variable $X^0$ rescaled so as to be dimensionless.  We now check homogeneity in the beta functions to the next order in $\ap$.

\subsubsection{Structure of the Second Order Beta Functions}\label{structure_second_order}
The expressions in (\ref{weylconditions}) may be extended to the next order in $\ap$.  The precise expressions for the second order beta functions depend on the renormalisation scheme employed, \cite{metsaev}.  However, the $X^0$ dependence will remain unaffected by constants and/or signs in front of the terms.  We consider a representative list of possible terms (expressed by contractions of the Ricci or Riemann tensors, the dilaton field and the antisymmetric tensor field strength).  We rely on \cite{metsaev} for a list of such terms in second order in $\ap$.  We note that while the list of second order beta function terms is not the complete list from \cite{metsaev}, we have taken all possible combinations of the parameters $R_{\mu\nu}$, $H_{\mu\nu\lambda}$, $R_{\mu\nu\sigma\rho}$ and $\phi$ and derivatives thereof, within this list.  Alternative contractions of the combinations of the parameters simply involve the metric $g_{\mu\nu}$ or its inverse and do not change the power of $X^0$ for that term.  For example, a term of the form:
\be \label{tensor_1} R_{0\alpha\beta\gamma} R_0^{~\alpha\beta\gamma} \ee
is of the order $(X^0)^{-2}$.  However this simply a component of the tensor: 
\be \label{tensor_2} R_{\mu\alpha\beta\gamma} R_{\nu}^{~\alpha\beta\gamma}, \ee
and the square of this is:
\be \label{tensor_2} R_{\mu\alpha\beta\gamma} R_{\rho}^{~\alpha\beta\gamma}R^{\rho}_{~\delta\epsilon\lambda} R_{\nu}^{~\delta\epsilon\lambda}
 = R_{\mu\alpha\beta\gamma} R_{\rho}^{~\alpha\beta\gamma} g^{\rho\omega} R_{\omega\delta\epsilon\lambda} R_{\nu}^{~\delta\epsilon\lambda}, \ee
which is of order $(X^0)^{-2+2-2} = (X^0)^{-2}$ as the inverse metric $g^{\rho\omega}$ is of order $(X^0)^{2}$.  Thus in taking higher powers of the terms one must contract the indices carefully with the metric or its inverse in order to preserve covariance.  We thus conclude that all powers of the terms containing indices are of order $(X^0)^{-2}$.  The terms from \cite{metsaev} along with the powers of $X^0$ they imply based on our solution (\ref{solution}) are computed below.

\begin{itemize}
\item for $\beta^g_{00}$ and $\beta^g_{ii}$, $i=1,2$:\\
\bea
R_{0\alpha\beta\gamma}R_0^{~\alpha\beta\gamma}&=&\frac{6}{\kappa_0}(X^0)^{-2}\nn
R_{i\alpha\beta\gamma}R_i^{~\alpha\beta\gamma}&=&-\frac{6}{\kappa_0}(X^0)^{-2}\nn
R^{\alpha\beta\rho\sigma}H_{0\alpha\beta}H_{0\rho\sigma}&=&-\frac{4h_0^2}{\kappa_0}(X^0)^{-2}\nn
R^{\alpha\beta\rho\sigma}H_{i\alpha\beta}H_{i\rho\sigma}&=&\frac{2h_0^2}{\kappa_0}(X^0)^{-2}\nn
H_{\rho\sigma 0}H^{\sigma\alpha\beta}H^\rho_{~\beta\gamma}H^\gamma_{~\alpha 0}
&=&\frac{2h_0^4}{\kappa_0}(X^0)^{-2}\nn
H_{\rho\sigma i}H^{\sigma\alpha\beta}H^\rho_{~\beta\gamma}H^\gamma_{~\alpha i}
&=&-\frac{4h_0^4}{\kappa_0}(X^0)^{-2}\nn
\nabla_0 H_{\alpha\beta\gamma}\nabla_0H^{\alpha\beta\gamma}&=&\frac{54h_0^2}{\kappa_0}(X^0)^{-2}\nn
\eea

\item for $\beta^B_{33}$:
All the contributions vanish.

\item for $\beta^B_{12}$:\\
\bea
R_{1\gamma\alpha\beta}\nabla^\gamma H^{\alpha\beta}_{~~~2}&=&
-\frac{6h_0}{\kappa_0}(X^0)^{-2}\nn
\nabla_\gamma H_{\alpha\beta 1}H_{2\rho}^{~~\alpha}H^{\beta\gamma\rho}&=&-\frac{3h_0^3}{\kappa_0}(X^0)^{-2}\nn
\nabla_\beta \left( H_{\alpha\rho\sigma}H_1^{~~\rho\sigma}\right) H_2^{~~\alpha\beta}&=&\frac{2h_0^3}{\kappa_0}(X^0)^{-2}\nn
H_{\alpha\rho\sigma}H_\beta^{~~\rho\sigma}\nabla^\alpha H^\beta_{~12}&=&-\frac{6h_0^3}{\kappa_0}(X^0)^{-2}\nn
\eea

\item for $\beta^\phi$:\\
\bea
\left( H_{\alpha\beta\gamma}H^{\alpha\beta\gamma}\right)^2&=&\frac{36h_0^4}{\kappa_0^2}\nn
R_{\lambda\mu\nu\rho}R^{\lambda\mu\nu\rho}&=&\frac{6}{\kappa_0^2}\nn
H_{\alpha\beta}^{~~~\mu}H^{\alpha\beta\nu}\nabla_\mu\nabla_\nu\phi&=&\frac{2h_0^2\phi_0}{\kappa_0^2}\nn
R^{\alpha\beta\rho\sigma}H_{\alpha\beta\lambda}H_{\rho\sigma}^{~~~\lambda}&=&-\frac{h_0^2}{\kappa_0^2}\nn
H_{\alpha\beta}^{~~~\mu}H^{\alpha\beta\nu}H_{\gamma\delta\mu}H^{\gamma\delta}_{~~~\nu}&=&\frac{20h_0^4}{\kappa_0^2}\nn
\nabla_\lambda H_{\alpha\beta\gamma}\nabla^\lambda
H^{\alpha\beta\gamma}&=&\frac{54h_0^2}{\kappa_0^2}
\eea

\end{itemize}
Thus to this order in $\ap$ the beta functions are homogeneous within each beta function, in powers of $X^0$.  We are thus are satisfied that we have shown homogeneity of the beta functions in $X^0$ to second order in $\ap$.  While we have not done a similar exercise for third order or higher beta functions, we note we have taken all possible combinations of the parameters $R_{\mu\nu}$, $H_{\mu\nu\lambda}$, $R_{\mu\nu\sigma\rho}$ which produce the correct tensor form of the beta function concerned.  Higher powers of these combinations will need to be contracted with the metric or its inverse which will produce a similar dependence on $X^0$.  The leaves the scalar field $\phi$ and the scalar curvature $R$ as potentially producing terms with differing powers of $X^0$.  \cite{metsaev} notes that the dilaton field only occurs in the beta functions as a derivative (i.e. $\nabla_\nu\phi$).  $R$, which is constant for our configuration, does not appear in any of the first or second order beta terms and we assume it is absent from higher order terms.  Thus, while we do not prove third order or higher terms are homogeneous in $X^0$ we work with this assumption.  

\subsubsection{Field Re-definition and Conformal Invariance}\label{Field Re-definition and Conformal Invariance}
Thus the beta functions using our solution are homogeneous to first and second order in $\ap$.  We extrapolate this based on the fact that to all orders, by design, beta functions are homogeneous in the spacetime indices.  Given our consistent configuration (\ref{solution}), we expect this to be 
true for all orders as we have shown that all possible contractions of the spacetime indices in the beta functions produce terms homogeneous in $X^0$.  We can write down a general power dependence for the beta functions in terms of $X^0$ in terms of powers of $\ap$ and $X^0$.
\bea
\beta^g_{00}&=&\frac{1}{(X^0)^2}\sum_{n=0}^\infty \xi_n(\alpha^{'})^n=\frac{E_0}{(X^0)^2}\nn
\beta^g_{11}&=&\beta^g_{22}=\frac{1}{(X^0)^2}\sum_{n=0}^\infty \zeta_n(\alpha^{'})^n=\frac{E_1}{(X^0)^2}\nn
\beta^g_{33}&=&\sum_{n=0}^\infty \chi_n(\alpha^{'})^n=E_2\nn
\beta^B_{12}&=&\frac{1}{(X^0)^2}\sum_{n=0}^\infty \delta_n(\alpha^{'})^n=\frac{E_3}{(X^0)^2}\nn
\beta^\phi&=&\frac{1}{\alpha^{'}}\sum_{n=0}^\infty \eta_n(\alpha^{'})^n=E_4,
\eea
where the coefficients $\xi_n,\zeta_n,\chi_n,\delta_n,\eta_n$ are independent of $\alpha^{'}$.  The terms $E_{0,1,2,3,4}$ are constants and it is seen that $\beta^\phi$ and $\beta^g_{33}$ are independent of $X^0$.  We need now to make the step from the beta functions being homogeneous to all orders, to conformal invariance. We do this using local field redefinitions.  In (\ref{local transformations}) we noted that the vanishing of the beta functions is not affected by a transformation which consists of a re-definition of the beta functions.
\be \label{local transformations1} \beta^i \rightarrow \tilde\beta_i = \beta^i  + \delta \beta^i, \ee 
where $\delta \beta^i$ is defined as $(\tilde g^j-g^j)\frac{\partial\beta^i}{\partial g^j}
-\beta^j\frac{\partial}{\partial g^j}\left(\tilde g^i-g^i\right)$  in eq. (\ref{redefbeta}) as $\tilde\beta_i - \beta^i$. \cite{Mavromates} notes that this property is due to target space diff invariance, a symmetry essential in general relativity, and denotes the transformed $\hat{\beta}^i$ as the Weyl anomaly coefficients.  This point is developed explicitly for use with a graviton and dilaton background in \cite{Alexandre2} with reference to work conducted related to second order beta functions in \cite{metsaev}.  In \cite{Alexandre4} this is extended to beta functions in the action under consideration in this thesis, i.e. (\ref{gradexp}).
In \cite{Alexandre4} it is noted that the target space physics in the one loop model (that is one order in $\ap$) is unchanged under the following transformations.
\bea\label{reparametrization}
\tilde g_{\mu\nu}&=&g_{\mu\nu}+\alpha^{'}g_{\mu\nu}\left(a_1 R+a_2 \partial^\rho\phi\partial_\rho\phi
+a_3\nabla^2\phi+a_4H_{\rho\mu\nu}H^{\rho\mu\nu}\right)\nn
\tilde B_{\mu\nu}&=&B_{\mu\nu}+\alpha^{'}\left(b_1\nabla^\rho H_{\rho\mu\nu}+b_2\partial^\rho\phi H_{\rho\mu\nu}\right)\nn
\tilde\phi&=&\phi+\alpha^{'}\left(c_1R+c_2\partial^\rho\phi\partial_\rho\phi+c_3\nabla^2\phi
+c_4H_{\rho\mu\nu}H^{\rho\mu\nu}\right), 
\eea
where ($a_1,...,b_1,...,c_1,...$) are dimensionless, unspecified parameters.  These transformations do not change the $X^0$ dependence of the fields in our configuration given in eq. (\ref{solution}) as can be seen by computing the relationship between the initial and transformed fields.
\bea \label{transformed_fields}
\tilde g_{\mu\nu}&=&A_1g_{\mu\nu}\nn
\tilde B_{\mu\nu}&=&A_2B_{\mu\nu} \nn
\tilde\phi&=&\phi+ A_3, 
\eea
where $A_1, A_2, A_3$ are constants.  However as noted in \cite{metsaev}, the transformations (\ref{reparametrization}) do change the beta functions as follows:
\be\label{redefbeta}
\tilde\beta_i=\beta_i+(\tilde g^j-g^j)\frac{\partial\beta^i}{\partial g^j}
-\beta^j\frac{\partial}{\partial g^j}\left(\tilde g^i-g^i\right).
\ee
In the above (\ref{redefbeta}) the factors $g^i$ represent the fields $g_{\mu\nu}$, $B_{\mu\nu}$ and $\phi$ in turn (with $\beta_i$ the associated beta functions), and the $j$ index notation represents integration over $X^0$.  One may then fine tune the parameters ($a_1,...,b_1,...,c_1,...$) (which are arbitrary) to make the transformed beta functions $\tilde\beta_i$ vanish.  We note that the terms $\tilde\beta_i$ remain homogeneous in $X^0$ and homogeneous to the original beta functions $\beta_i$ as no operations raising or lowering powers of $X^0$ have been conducted in (\ref{redefbeta}). \cite{Alexandre4} notes that given the homogeneity of the beta functions to all orders and the ability to arbitrarily rescale the fields, the following conditions are necessary to ensure vanishing of the $\tilde\beta_i$ functions to second order in $\ap$.
\bea\label{tildebeta0}
\tilde\beta^g_{00}&=&\frac{\tilde E_0}{(X^0)^2}=0\nn
\tilde\beta^g_{11}&=&\tilde\beta^g_{22}=\frac{\tilde E_1}{(X^0)^2}=0\nn
\tilde\beta^g_{33}&=&\tilde E_2=0\nn
\tilde\beta^B_{12}&=&\frac{\tilde E_3}{(X^0)^2}=0\nn
\nn\tilde\beta^\phi&=&\tilde E_4=0,
\eea
where the constants $\tilde E_{0,1,2,3,4}$ are linear functions of the parameters ($a_1, ...,b_1,...,c_1,...$).  The equations (\ref{tildebeta0}) consist of five parameters $E_{0,1,2,3,4}$ which are in turn linear functions of the arbitrary constants $a_i$, $b_i$ and $c_i$ in (\ref{reparametrization}).  There are ten such constants and six linear equations and thus solutions may always be found.  By induction, \cite{Alexandre4} argues that this can be extended to all orders and thus we conclude that our non-perturbative evolution equation (\ref{evolution_phi}) with the conditions (\ref{conditions}) can be shown to have non vanishing beta functions to all orders, therefore conformally invariant and of interest in the string cosmology context.

\section{Cosmology}
We have demonstrated that there is a viable (conformally
invariant) solution to the non-perturbative effective action on the
world-sheet.  We can now utilise the results in target space and
suggest cosmological properties.  We first consider what kind of physical universe our solution represents and then couple it to a real-world field, the electromagnetic field.  

\subsection{The Dynamics of the Universe} \label{The Dynamics of the Universe}
In the comments surrounding eq. (\ref{effective_string_action_E}) it was noted that the Einstein frame for a worldsheet effective action and ensuing results is a convenient one in terms of describing the real world picture of the universe's evolution.  The generalised expression for the Einstein frame line element is quoted in eq. (\ref{Einstein_action}).  We have in $D=4$ using, our configuration (\ref{solution}) (with our scale factors $a_i\equiv a(t)$ now generalised to allow variation with spatial direction):
\bea
\label{einstein_frame_metric}
g^E_{\mu\nu}(t)dx^\mu dx^\nu&=&dt^2-a_1^2(t)(dx^1)^2-a_2^2(t)(dx^2)^2-a_3^2(t)(dx^3)^2\nn
&=&\exp\left\{-2\phi(x^0)\right\}g_{\mu\nu}(x_0)dx^\mu dx^\nu\nn
&=&(x^0)^{-2\phi_0}\kappa_0\left[\left(\frac{dx^0}{x^0}\right)^2
-\left(\frac{dx^1}{x^0}\right)^2-\left(\frac{dx^2}{x^0}\right)^2-(dx^3)^2\right].
\eea
Here, the $x^\mu$ are the zero modes of $X^\mu$ (here the classical field, i.e. $X^0=x^0$), and $a_i(t)$ are the scale factors in the different
space directions.  In this, $t$ is the Einstein frame time dimension (or cosmic time).  We have then
\be
\frac{dt}{dx^0}=\sqrt{\kappa_0} (x^0)^{-1-\phi_0},
\ee
such that
\be\label{tx^0}
t=\frac{\sqrt{\kappa_0}}{|\phi_0|}(x^0)^{-\phi_0}.
\ee
(We note here we only consider the solution such that $\frac{dt}{dx^0}>0$ as we require a cosmic time which does not diminish).  The scale factors are:
\bea\label{scalefactors}
a_1(t)=a_2(t)&=&a_0~t^{1+1/\phi_0}\nn
a_3(t)&=&a_0~t
\eea
where $a_0$ is a constant.  The linear sigma string cosmology model for which $B_{\mu\nu} = 0$, (\ref{linear_background}) shows linear expansion in all spatial directions and a flat Minkowski metric. Our model shows linear expansion in the direction for which we fixed the antisymmetric tensor field strength in the $X^3$ direction in eq. (\ref{fixing_h_inx3}) with dilaton ($\phi_0$) dependent expansion in the $x_1$ and $x_2$ directions for which $H_{\rho\mu\nu}$ is not fixed.  We also note than when $\phi_0 = -1$ and when $B_{\mu\nu} = 0$ we revert to this currently observed isotropic, Minkowski universe, (\ref{linear_background}).  We will couple this configuration to an electromagnetic field in the next section.   It was discussed in section (\ref{Tachyon Cosmology}) that our configuration as above gives rise to a power law dependence by the scale factors $a_i$ with cosmic time, in the Einstein frame.  We noted that in the $\sigma$-model frame as described by eq. (\ref{effective_string_action}) a de Sitter, inflationary universe is possible, but the introduction of the $\exp(-2\phi(x^0))$ factor in front of the metric in the Einstein frame removes, following Big Bang, the possibility for an exponential relationship representing inflation, i.e. $a(t) \sim \exp(Ht)$ as our configuration for $\phi$ involves a logarithm ($\phi(X^0) = \phi_0 \ln(X^0)$).  \cite{Mavromates} notes the importance of a non trivial dilaton field in accounting for inflationary and other expanding universes in string cosmology, as seen in our configuration for the scale factors in eq. (\ref{scalefactors}). \cite{Mavromates} also notes Pre Big Bang (PBB) models in which the singularity is replaced by a dilaton-potential barrier such that before and after the dilaton becomes strongly coupled.  The weakly coupled stage prior to big bang has the solution (not derived here), \cite{Mavromates}:
\be \label{PBB_dilaton} \phi = -(1-\Sigma_i a_i)\ln (-t), \,\, \Sigma_i a_i^2 = 1, t<0. \ee
Here $a_i$ are the scale factors, $t$ is cosmic time, and $\phi$ is the dilaton field.  We note this is similar to our configuration of the dilaton field in eq. (\ref{solution}).  \cite{Mavromates} qualitatively argues that at cosmic time $t<0$ in this PBB scenario and after crossing the dilaton-potential barrier, one expects an inflationary phase followed by a graceful and ensuing normal expansion to modern times.  While we will not analyse the PBB scenario in detail, we note that our configuration may encompass an inflationary phase within the context of PBB theory. \\\\It is noted in passing that the treatment of the bosonic string in this thesis including the tachyon is conducted in \cite{Alexandre_tachyon} and the configuration in this study allows for a de Sitter universe with the condition $2 \phi + T = 0$, where $T$ is the tachyon field.  We also acknowledge that there is an issue of how to emerge from anisotropic expanding universe described by the conditions (\ref{scalefactors}) to a isotropic Minkowski model.  While we do not cover this point in detail, it is possible that a time dependence on $h$ in (\ref{fixing_h_inx3}) could address this issue, i.e. the string axion field, which relaxes to zero in cosmic time: $h(t) \rightarrow 0$. 

\subsection{Optical Anisotropy}\label{optical_anisotropy_anisotropic_config}
We described recent research in section (\ref{String Cosmology and Anisotropy}) which dealt with how the coupling of a pseudo-scalar field such as the string axion to the photon may result in optical anisotropy on a universal scale. We wish to test this with our configuration.  First we reformulate our field configurations in (\ref{solution}) by considering the $\phi_0 = -1$ solution such that the universe exhibits linear expansion in the $X^3$ direction but static in the other spatial directions.  We rescale $x_0 \rightarrow \frac{|\phi_0|}{\sqrt{\kappa_0}}(x^0)$ so that $t=x_0$ where $t$ is the Einstein frame cosmic time.  We take $\ap = \lambda = 1$ and the solution becomes:
\bea\label{configg}
g_{\mu\nu}(t)&=&\mbox{diag}(1,-1,-1,-t^2)\nn
H_{012}(t)&=&-\frac{\kappa_0 h_0}{t^3}\nn
\phi(t)&=&-\ln t.
\eea
Motivated by our desire to couple the string axion $h$ to a photon represented by a gauge field $A_{\rho}$ we introduce a modified field strength, (in a similar to manner to that done in \cite{Optical_Anisotropy3}).
$\tilde H_{\rho\mu\nu}$:
\be
\tilde H_{\rho\mu\nu}=H_{\rho\mu\nu}+\frac{1}{M}A_{[\rho}F_{\mu\nu]},
\ee
where $M$ is a parameter with mass dimensions dependent on the string model and $F_{\mu\nu}$ the electromagnetic tensor.  In this solution we assume the string lives in $D=4$ spacetime dimensions and we refer again to our non-perturbative demonstration of conformal invariance which does not limit the value of $D$.  $\ap$ is the only dimensionful parameter in the bare string action and since it has units $\sqrt{l}$ we regard $M \propto \sqrt{\ap}$ and as the string mass scale.  We could like to couple the photon field strength $F_{\mu\nu}$ to the string background fields defined by the solution in (\ref{configg}) and hence identify any anisotropies in $F_{\mu\nu}$ in the low energy effective action.  We note that while our configuration in (\ref{configg}) is based on a non-perturbative effective action equation, we are not able to formulate such a non-perturbative approach for an effective action involving the background fields represented in (\ref{string_action_model}).  We point out that our results have been designed to apply to a late (post inflationary) era in the universe and we thus justify at this stage using a one loop model in the Einstein frame, (zeroth order in $\ap$), \cite{low.ener.eff.} and used in \cite{Alexandre5}.  We note that this action, quoted here without derivation, is a perturbative effective action.  While our model and configuration (\ref{solution}) is based on non-perturbative methods, we assume this configuration to be valid for our purposes in weak field configurations, $R$, $H_{\rho\mu\nu}$ and $F_{\mu\nu}$.
\be \label{one-loop-effective}
S_{eff}=-\int d^4x \sqrt{|g|}\left\lbrace \frac{44}{3 \ap}e^{2\phi} + R-2\partial_\mu\phi\partial^\mu\phi
-\frac{e^{-4\phi}}{12}\tilde H_{\rho\mu\nu}\tilde H^{\rho\mu\nu}
-\frac{e^{-2\phi}}{4}F_{\mu\nu}F^{\mu\nu}\right\rbrace.
\ee
As discussed surrounding our eq. (\ref{effective_string_action_E}) from \cite{Mavromates}, there is a relative change of sign between the curvature and dilaton kinetic term, when transforming from the $\sigma$-model frame to the physical Einstein frame. Inserting our conditions (\ref{configg}) into this and considering only the electromagnetic sector we have:
\be\label{gauge}
S_{EM}=\int dt d\vec x\left\lbrace -\frac{t^2}{4}F_{\mu\nu}F^{\mu\nu}
+\epsilon t A_{[0}F_{12]}\right\rbrace,
\ee
where $\epsilon=\frac{\kappa_0h_0}{M}$.   This part of the action as it is is valid only for a specific gauge transformation $A_\mu \rightarrow A_\mu+\partial_\mu\psi$ as this transformation will also impact the antisymmetric tensor $B_{\mu\nu}$ which must transform as $B_{\mu\nu}\rightarrow B_{\mu\nu}-\frac{\psi}{M} F_{\mu\nu}$ for the action (\ref{one-loop-effective}) to remain invariant.  
The term $\epsilon t A_{[0}F_{12]}$ arises in order to make the action (\ref{gauge}) invariant under the transformation of $B_{\mu\nu}$.  The factor $t^2$ in front of $F_{\mu\nu}F^{\mu\nu}$ arises from our configuration $\phi(t)=-\ln t$.  We consider this effective action in terms of the equations of motion of the electromagnetic field, that is:
\be \label{eom_em} \partial_{\nu} \left( \frac{\partial \mathcal{L}}{\partial(\partial_{\nu}A_{\mu}}\right) - \frac{\partial \mathcal{L}}{\partial A_{\mu}} = 0, \ee
where $\mathcal{L} = -\frac{t^2}{4}F_{\mu\nu}F^{\mu\nu} +\epsilon t A_{[0}F_{12]}$.  
\bea\label{newMax}
{\bf\nabla}\cdot{\bf\tilde B}&=&0\\
{\bf\nabla}\cdot{\bf E}&=&\frac{2\epsilon}{t}B_3\nn
{\bf\nabla}\times{\bf\tilde E}&=&-\partial_t{\bf\tilde B}\nn
\left({\bf\nabla}\times{\bf B}\right)_1&=&\partial_t E_1+\frac{3}{t}E_1
-\frac{2\epsilon}{t} E_2+\frac{2\epsilon}{t^2}A_2,\nn
\left({\bf\nabla}\times{\bf B}\right)_2&=&\partial_t E_2+\frac{3}{t}E_2+
\frac{2\epsilon}{t} E_1-\frac{2\epsilon}{t^2}A_1,\nn
\left({\bf\nabla}\times{\bf B}\right)_3&=&\partial_t E_3+\frac{5}{t}E_3\nonumber
\eea
where we define
\bea
&&{\bf E}=(E_1,E_2,E_3),~~~~~~{\bf B}=(B_1,B_2,B_3)\nn
&&{\bf\tilde E}=(E_1,E_2,t^2E_3),~~~~{\bf\tilde B}=(t^2B_1,t^2B_2,B_3).
\eea 
Here $E_i,B_j$ are the components of the electric and magnetic fields in the "conventional" electromagnetic Lagrangian $\mathcal{L} = -\frac{1}{4}F_{\mu\nu}F^{\mu\nu}$.  We note the presence of the gauge fields $A_1$ and $A_2$ in (\ref{newMax}) is a result of the term $\epsilon t A_{[0}F_{12]}$ in the action (\ref{gauge}).  We notice from (\ref{newMax}) that the the $X^3$ direction is anisotropic, as expected from our choice of pseudo-scalar axion $h = h_0X^3$.  We hence look at the time ($t$) varying (represented by the overdot) propagation of an electromagnetic wave in the $X^3$ direction arising from (\ref{newMax}).
\be\label{evolE}
{\bf\ddot E}+\frac{3}{t}{\bf\dot E}+\frac{2\epsilon}{t}{\bf \dot E_\bot}=0,
\ee
where we have truncated to include only powers of $1/t$, justified by the relatively large cosmological time scales we are concerned with.  Here ${\bf E}=(E_1,E_2,0)$ and ${\bf E_\bot}=(-E_2,E_1,0)$.  
This has a solution, by inspection, of 
\be\label{solE}
{\bf E}=\frac{{\bf K}f(x^3)}{t^2}\exp\left(-2i\epsilon\ln t\right) ,
\ee
where ${\bf K}$ is a constant complex number and $(x^1,x^2)$ are in the complex plane.  $f(x^3)$ is an arbitrary function of the $x^3$ spatial coordinate.  From the ${\bf\nabla}\times{\bf B}$ expressions in (\ref{newMax}) we also have:
\be\label{solB}
{\bf\nabla}\times{\bf B}=\frac{{\bf K}f(x^3)}{t^3}\exp\left(-2i\epsilon\ln t\right) ,
\ee
We take ${\bf B}$ and ${\bf E}$ as representing an electromagnetic wave lying in the $(x_1, x_2)$ plane and propagating in the $x_3$ optic axis.  If we take $f(x^3) = \cos(x^3)$ we have solutions for the ${\bf B}$ and ${\bf E}$ fields:
\bea \label{solE_and_B}
{\bf E} &=& {\bf K} \exp\left(-2i\epsilon\ln t\right) \frac{\cos(x^3)}{t^2} \nn
{\bf B} &=& i{\bf K} \exp\left(-2i\epsilon\ln t\right) \frac{\sin(x^3)}{t^3}. \eea
Using this we obtain for the term $-\frac{e^{-2\phi}}{4}F_{\mu\nu}F^{\mu\nu}$ in the effective action:
\be \label{F_term_in_action} -\frac{e^{-2\phi}}{4}F_{\mu\nu}F^{\mu\nu} =  -\frac{t^2}{2}\left(|{\bf E} \cdot \widetilde{{\bf E}}| + |{\bf B} \cdot \widetilde{{\bf B}}|\right) = -\frac{{\bf K}^2}{2t^2}. \ee
In arguments that will not be detailed here, \cite{Alexandre5}, it is conjectured that the configuration (\ref{configg}) coupled to the gauge field $A_{\mu}$ is conformally invariant.  This is done by applying again the argument of homogeneity combined with the potential to redefine the fields to achieve the vanishing of the beta functions.  As seen from eqs. (\ref{solE}), 
the effect of the dilaton (i.e. the factor $e^{-2\phi} \equiv t^{-2}$ in front of $F_{\mu\nu}F^{\mu\nu}$) is to damp the amplitude of the electric field: 
\be
|{\bf E}|=\frac{|{\bf K}|}{t^2}.
\ee 
Another aspect of eqs. (\ref{solE_and_B}) of interest is the change in polarisation which is generated by the term $\frac{2\epsilon}{t}{\bf \dot E_\bot}$ in the equation of motion of the electric field where the direction of the electric field (in the $x_1, x_2$ plane) rotates with the angle:
\be \label{anisotropic_polarisation}
\Delta(t)= |\arg ({\bf E}) - \arg({\bf E})^*| =  4\epsilon \ln t = \frac{4\kappa_0 h_0}{M}\ln t.
\ee
This contrasts with the relationship postulated by \cite{Optical_Anisotropy6} in (\ref{cosmic birefringence_angle}) which shows a simple proportionality between the change in polarisation and cosmic time (in this source denoted by $\eta$ and $\tau$ respectively).\\\\At this stage we make a comment on the term $\frac{44}{3 \ap}e^{2\phi}$ present in our effective spacetime action in the Einstein frame, eq. (\ref{one-loop-effective}) which derives from the $D=4$ case in the general expression in (\ref{effective_string_action_E}).  We note this implies a negative cosmological constant relative to the Einstein curvature term.  We do not explore a solution for this and note that the phenomenological value of our model may be limited if this negative term cannot be compensated for by other terms in the effective action.

\section{Generalisation of the Antisymmetric Tensor}\label{Generalisation to an Isotropic Universe}
\subsection{Introduction}
In the last section we chose an anisotropic configuration of the antisymmetric tensor field strength, $H_{\rho\mu\nu}$, using constraints (\ref{fixing_h_inx3}).  We now return to the generalised form of (\ref{axion}):
\be
\label {axion1} H_{\rho\mu\nu} =
\omega_0\cdot (X^0)^m\varepsilon_{\rho\mu\nu\sigma}(X^0)\partial^\sigma h,
\ee 
Previously we found that $R_{00}$ is homogeneous to $\nabla_0\nabla_0\phi$ only if 
\bea \label{config_isotropic}
\phi(X^0)&=&\phi_0\ln(X^0)\nn
\kappa(X^0)&=&\frac{\kappa_0}{(X^0)^2},
\eea
and thus a requirement for homogeneous beta functions $\beta^g_{00}$ and $\beta^{\phi}$ as denoted in (\ref{weylconditions}).  As in (\ref{ansatz}) we assume a form for the components of the metric $\tau_i(X^0)$.
\be \tau_i(X^0)=\frac{-\kappa_0}{(X^0)^{n_i}}. \ee
We now set up a more general ansatz for $H_{\rho\mu\nu}$, using similar notation as in (\ref{fieldstrengthansatz}).
\be \label{general_antis_tensor}
H_{0ij}= -\kappa_0 h_k(X^0)^{(m-1 + (n_i + n_j + n_k)/2)}\epsilon_{ijk}. \ee
where we have taken $\omega_0\cdot (X^0)^m = (X^0)^m$.  We again consider the conformal invariance of this configuration. 

\subsection{Conformal Invariance}\label{conformal_invariance_2}
As in the previous section, we subject our configuration and model to the two-step test of conformal invariance.  First we ensure that the configuration can be homogeneous to all orders in $\ap$ in the beta functions in the variable $X^0$.  Then we redefine the fields as before to make the beta functions vanish.  We consider the first order beta functions.  We use the configuration of $H_{0ij}$ in (\ref{general_antis_tensor}) in the beta functions in (\ref{weylconditions}), (without summation on the index $i$,
where $i,j,k$ all run between 1 to 3). 
\bea \label{isotropic_first_Wely}
\beta_{00}^{g(1)}&=&\frac{-1}{4(X^0)^2} \sum_{i=1}^3 n_i^2 -
\frac{1}{2}\omega_0^2\sum_{i=1}^3 h_i^2 (X^0)^{2m+n_i-2} \nn
\beta_{ii}^{g(1)}&=&\frac{n_i}{4(X^0)^{n_i}}\left[4\phi_0+\sum_{i=1}^3
n_i \right]+\frac{1}{2}\omega_0^2 \sum_{j=1}^3 h_j^2
(X^0)^{2m+n_j-n_i} \nn
\beta_{ij}^{g(1)}&=&-\beta_{ji}^{g(1)}=-\frac{\omega_0^2 (h_i
h_j)(X^0)^{2m}}{2}~~\longrightarrow~~i\neq j=1,2,3 \nn
\beta_{ij}^{B(1)}&=&-\beta_{ji}^{B(1)}=
\omega_0h_k(X^0)^{-(\frac{n_i}{2}+\frac{n_j}{2}-\frac{n_k}{2}-m)}\left[\frac{n_i}{4}+\frac{n_j}{4}-\frac{n_k}{4}-\frac{m}{2}
+ \phi_0\right]\nn
\beta^{\phi(1)}&=&-\frac{11}{3\alpha^{'}}+\frac{\phi_0}{4\kappa_0}\left[
4\phi_0+\sum_{i=1}^3 n_i\right] - \sum_{i=1}^3
\frac{h_i^2\omega_0^2}{4 \kappa_0}(X^0)^{2m+n_i} \eea
For these to be homogeneous, the constraint must apply: $2m+n_i=0$ for those $i$ which $h_i \neq 0$.  We are interested in looking at the most general, isotropic configuration so we take $n_1=n_2=n_3$ by symmetry, but for the time being leave the notation separate as $n_{i,j,k}$.  We now look at representative terms of the next order beta functions (again quoted from \cite{metsaev}) for homogeneity (here $i\neq j\neq k
=1,2,3$).

\begin{itemize}

\item for $\beta^g_{00}$:
\bea \label{betasecond}
    R_{0\alpha\beta\gamma}R_0^{~\alpha\beta\gamma}&=&\frac{1}{16\kappa_0}(X^0)^{-2} \sum_{i=1}^3n_i^4\nn
    R^{\alpha\beta\rho\sigma}H_{0\alpha\beta}H_{0\rho\sigma}&=&-\frac{1}{4\kappa_0}\sum_{i,j,k=1}^3
    h_k^2n_in_j(X^0)^{2m-2+n_k}\nn H_{\rho\sigma0}H^{\sigma\alpha\beta}H^\rho_{~\beta\gamma}H^\gamma_{~\alpha
    0} &=&\frac{2}{\kappa_0}(X^0)^{4m-2}\sum_{i,j,k=1}^3h_i^2 (X^0)^{n_i}\sum_{l=1}^3 h_l^2(X^0)^{n_l}\nn
    \nabla_0H_{\alpha\beta\gamma}\nabla_0H^{\alpha\beta\gamma}&=&\frac{1}{2\kappa_0}
    \sum_{i,j,k=1}^3h_k^2\left(n_i+n_j-n_k-2m\right)^2(X^0)^{2m+n_k-2}\nonumber
\eea

\item for $\beta^g_{ii}$ (no summation on $i$): \bea
    R_{i\alpha\beta\gamma}R_i^{~\alpha\beta\gamma}=-\frac{1}{8\kappa_0}(X^0)^{-n_i}n_i^2\sum_{k=1}^3n_k^2\nonumber
    \eea

\item for $\beta^g_{ij}$, with $i\ne j$:
\bea
R^{\alpha\beta\rho\sigma}H_{i\alpha\beta}H_{j\rho\sigma}&=&
    \frac{n_k^2}{2\kappa_0}h_ih_j(X^0)^{2m}\nn H_{\rho\sigma
    i}H^{\sigma\alpha\beta}H^\rho_{~\beta\gamma}H^\gamma_{~\alpha j} &=&-\frac{8}{\kappa_0}(X^0)^{4m}
    h_ih_j\sum_{l=1}^3 h_l^2(X^0)^{n_l}\nonumber
\eea

\item for $\beta^B_{ij}$, with $i\ne j\ne k$:
\bea R_{i\gamma\alpha\beta}\nabla^\gamma
    H^{\alpha\beta}_{~~~j}&=& -\frac{n_i^2h_k}{8\kappa_0}(n_i+n_j-n_k-2m) (X^0)^{-(n_i+n_j-n_k-2m)/2}\nn
    \nabla_\gamma H_{\alpha\beta i}H_{j\rho}^{~~\alpha}H^{\beta\gamma\rho}&=&
    \frac{1}{2\kappa_0}(n_i+n_j-n_k-2m) \Big[h_k^3(X^0)^{(-n_i-n_j+3n_k+6m)/2}\nn
    &&+h_kh_i^2(X^0)^{(n_i-n_j+n_k+6m)/2}\Big]\nn &&+\frac{1}{2}(n_i-n_j+n_k-2m)
    h_kh_j^2(X^0)^{(-n_i+n_j+n_k+6m)/2}\nonumber
\eea

\item for $\beta^\phi$:
\bea \left(H_{\alpha\beta\gamma}H^{\alpha\beta\gamma}\right)^2&=&
    \frac{36}{\kappa_0^2}\sum_{i=1}^3h_i^4(X^0)^{4m+2n_i}\nn
    R_{\lambda\mu\nu\rho}R^{\lambda\mu\nu\rho}&=&\frac{1}{16\kappa_0^2} \sum_{i,j=1}^3 n_i^2(2n_i^2+n_j^2)\nn
    H_{\alpha\beta}^{~~~\mu}H^{\alpha\beta\nu}\nabla_\mu\nabla_\nu\phi&=&
    \frac{\phi_0}{\kappa_0^2}\left[\sum_{i,j=1}^3 n_ih_j^2(X^0)^{2m+n_j}\right ]\nn
    R^{\alpha\beta\rho\sigma}H_{\alpha\beta\lambda}H_{\rho\sigma}^{~~~\lambda}&=&
    -\frac{1}{2\kappa_0^2}\left[\sum_{i,j,k=1}^3 n_i(n_i+n_j)h_k^2(X^0)^{2m+n_k}\right ]\nn
    H_{\alpha\beta}^{~~~\mu}H^{\alpha\beta\nu}H_{\gamma\delta\mu}H^{\gamma\delta}_{~~~\nu}&=&
    \frac{4}{\kappa_0^2}(X^0)^{4m}\Bigg[\sum_{i,j=1}^3h_i^2h_j^2(X^0)^{n_i+n_j} \nn &~~~& +
    \sum_{i,j=1}^{3,\,i\neq j}h_i^2[3h_i^2(X^0)^{2n_i} + 4h_j^2(X^0)^{n_i+n_j}]\Bigg] \nn \nabla_\lambda
    H_{\alpha\beta\gamma}\nabla^\lambda H^{\alpha\beta\gamma}&=&\frac{3}{2\kappa_0^2}
    \sum_{i,j,k=1}^3h_k^2\left(n_i+n_j-n_k-2m\right)^2(X^0)^{2m+n_k}\nonumber
\eea

\end{itemize} 
In terms of homogeneity in $X^0$ the above do not provide any new constraints compared with those in the first order in (\ref{isotropic_first_Wely}).  We thus assume, as with the anti-symmetric tensor-anisotropic case presented in section (\ref{conformal_anisotropy}) that our configuration is homogeneous to all orders and as in the anisotropic case, we may rescale the fields to make the beta functions vanish and hence demonstrate conformal invariance to all orders in $\ap$.  We utilise similar methods and arguments to section (\ref{structure_second_order}) in assuming conformal invariance to all orders following examination of the second order terms.

\subsection{Cosmology Discussion}\label{optical_anisotropy_isotropic_config}
As in the eq. (\ref{einstein_frame_metric}) the spacetime metric in the Einstein frame $g^E_{\mu\nu}$ is related to our metric $g_{\mu\nu}$:
\bea\label{einstein_frame_metric2}
ds^2 = g^E_{\mu\nu}(t)dx^\mu dx^\nu&=&dt^2-a_1^2(t)(dx^1)^2-a_2^2(t)(dx^2)^2-a_3^2(t)(dx^3)^2\nn
&=&\exp\left\{-2\phi(x^0)\right\}g_{\mu\nu}(x_0)dx^\mu dx^\nu. \eea
We consider the representative case where $h_1=h_2=0$ and $h_3 = h \neq 0$.  For brevity we denote $n_1 = n_2 = n$ and $n_3 = -2m$.  Using our configuration in eq. (\ref{config_isotropic}) we have:
\be\label{einsteinframe}
dt^2-\sum_{i=1}^3
a_i^2(t)(dx^i)^2=(x^0)^{-2\phi_0}\kappa_0 \left[\left(\frac{dx^0}{x^0}\right)^2-\sum_{i=1}^3
\frac{(dx^i)^2}{(x^0)^{n_i}}\right],
\ee
and as in (\ref{tx^0}) we have:
\be\label{tx0} 
t=\frac{\sqrt{\kappa_0}}{|\phi_0|}(x^0)^{-\phi_0}. 
\ee 
This configuration (analogous to (\ref{solution})) is in the string frame:
\bea\label{solution1}
g_{\mu\nu}(X^0)&=&\kappa_0~\mbox{diag}((X^0)^{-2},-(X^0)^{-n},-(X^0)^{-n},-(X^0)^{2m})\nn
H_{012}(X^0)&=&-\kappa_0
h(X^0)^{-1-n}\nn \phi(X^0)&=&\phi_0\ln(X^0).
\eea
From eqs.(\ref{einsteinframe}) and (\ref{tx0}), the scale
factors in the Einstein frame read then
\bea\label{scalefactors1} a_1(t)=a_2(t)&=&a_0~t^{1+n/(2\phi_0)}\nn
a_3(t)&=&\tilde a_0~t^{1-m/\phi_0}
\eea
where $a_0,\tilde a_0$ are constants, and the corresponding target space
is a power-law expanding Universe.  For $H_{012}$ we have set  $h_1=h_2=0$, $h_3=h\ne 0$ and take $n_1=n_2=n$, but note that similar configurations (which by symmetry will be physically identical) exist for $H_{013}$ and $H_{023}$.  We now note that the (\ref{scalefactors1}) scale factors no longer necessarily represent a linearly expanding universe, when $\phi_0 = -1$, as did the scale factors in (\ref{scalefactors}).  In that case, the specific $m=0$ condition was imposed.  Here we consider the more general isotropic configuration of $H_{\rho\mu\nu}$ with dependence on $n$, $m$ and $\phi_0$.  We again couple to the electromagnetic field, and focus on, for (\ref{solution1}) a configuration leading to a Minkowski geometry, i.e. $m=-1$, $\phi_0 = -1$ and $n=0$.  We note that since $n_3$ does not appear in the physical configurations below, the constraint $2m+n_i=0$ (for those $i$ which $h_i \neq 0$) does not impact the result and $m$ becomes somewhat arbitrary.
\bea\label{config_isotropic_1}
g^E_{\mu\nu}&=&\eta_{\mu\nu} \nn H_{012}(t)&=& = H_{013}=H_{023} = -\frac{\kappa_0 h}{t^3}\nn
\phi(t)&=&-\ln t.
\eea
We now couple the configuration (\ref{config_isotropic_1}) to the photon field via the effective action in (\ref{one-loop-effective}) and the gauge field action in (\ref{gauge}), and the following modified Maxwell's equations are arrived at.
\bea\label{newMax1}
\vec\nabla\cdot\vec B&=&0\\
\vec\nabla\cdot\vec E&=&\frac{2\epsilon}{t}B_3\nn
\vec\nabla\times\vec E&=&-\partial_t\vec B\nn
\vec\nabla\times\vec B&=&\partial_t\vec E+\frac{2}{t}\vec E +\frac{\epsilon}{t}\vec E_\bot-,\nonumber
\eea
where $\vec
E_\bot=(-E_2,E_1,0)$ and $\vec A_\bot=(-A_2,A_1,0)$ and as in eqs. (\ref{newMax}) we have ignored terms of order $1/t^2$ or higher.
As previously, motivated by cosmic birefringence, we consider a plane wave in the $(x_1, x_2)$ plane moving in the $x_3$ direction. 
Using the complex notation
${\bf E}=E_1+iE_2$, we obtain the following equation of motion
\be\label{evoleq}
\ddot{\bf E}-{\bf E}^{''}+\frac{2+i\epsilon}{t}\dot{\bf E}=0,
\ee
where overdot is $t$ derivative and prime is spatial derivative.  By inspection, ${\bf E}={\bf E_0}t^a\cos(t-x^3)$, (where ${\bf E_0}$ and $a$ are complex constants), satisfies equation (\ref{evoleq}) for $a=-1-i\varepsilon/2$, neglecting terms of  $1/t^2$ or higher.  The solution is then:
\be\label{solE}
{\bf E}=\frac{{\bf E_0}}{t}\exp\left( -i\frac{\epsilon}{2}\ln t\right)\cos(t-x^3).
\ee
As in our previous solution, the dilaton provides for a damping effect (the $1/t$ term in (\ref{evoleq}) while the antisymmetric tensor, via the field strength, provides a varying polarisation of the electric field in the $(x_1, x_2)$ plane, in a similar manner to the result (\ref{anisotropic_polarisation}) we have for the isotropic $H_{\rho\mu\nu}$ case:
\be\label{oa}
\Delta (t) =|\mbox{arg}({\bf E})-\mbox{arg}({\bf E^\star})|=\frac{\kappa_0h}{M} \ln t,
\ee
where $t$ can be seen as the time interval between emission and observation of the electromagnetic waves, which we note provides the same optical anisotropy effect as in the case where the field strength tensor was not generalised as in (\ref{anisotropic_polarisation}).

\section{Summary and Discussion}
The primary result of this section of the thesis is to derive a non-perturbative formulation for the effective action of the bosonic string, on the worldsheet.  We then showed that a viable solution of this was conformally invariant, again, non-perturbatively, and therefore suitable for consideration in string cosmology.  To achieve this we commenced with a bare string action on the worldsheet with background fields $\eta_{\mu\nu}$, $a_{\mu\nu}$ and $\Phi(X^0)$ representing the graviton, antisymmetric tensor and dilaton fields respectively.  Here $\lambda$ is a varying parameter which equals $1/\ap$ in the full quantum theory of (\ref{string_action_model1}), where $\ap$ is the Regge slope.
\be \label{string_action_model1} 
\hspace{-0.5cm}S_{\lambda}  = \frac{1}{4\pi} \int d^2\xi \sqrt{\gamma} \,\Bigl\{\lambda\left[\gamma^{ab} \eta_{\mu\nu} +\epsilon^{ab}a_{\mu\nu}\right]\partial_a  X^\mu \partial_b  X^\nu + R^{(2)} \Phi(X^0) \Bigr\}. \ee
We then applied techniques with roots in the exact renormalisation group (ERG) methods, outlined in section (\ref{An Alternative, Exact Approach}) of this thesis, to derive a form for the Legendre effective action associated with (\ref{An Alternative, Exact Approach}):
\bea\label{evolGappendix2} \dot\Gamma&=&\frac{1}{4\pi}\int
d^2\xi\sqrt{\gamma}\left(\gamma^{ab}\eta_{\mu\nu}+\varepsilon^{ab}a_{\mu\nu}\right)
\partial_a X_{cl}^\mu\partial_b X_{cl}^\nu \nn
&&+\frac{1}{4\pi}\mbox{Tr}\left\{\left(\gamma^{ab}\eta_{\mu\nu}+\varepsilon^{ab}a_{\mu\nu}\right)
\frac{\partial}{\partial\xi^a}\frac{\partial}{\partial\zeta^b}
\left(\frac{\delta^2\Gamma}{\delta X_{cl}^\nu(\zeta)\delta X_{cl}^\mu(\xi)}\right)^{-1}\right\}. \eea 
Here the overdot on $\Gamma$ indicates a derivative with respect to $\lambda$, which we use in an analogous manner to which the high energy cutoff is used in the exact renormalisation group evolution equations such as that derived in Appendix (\ref{appendix Exact renormalisation})  We then derive fixed point or $\dot\Gamma=0$ solutions.  We then plug into (\ref{evolGappendix2}) an assumed form for the effective action $\Gamma$ in which the three background fields vary only with the $X^0$ component of spacetime and for which the spacetime metric is given by: $g_{\mu\nu} =  \mbox {diag} \left(\kappa(X^0), \tau_1(X^0),.....\tau_{D-1}(X^0)\right)$, where $D$ is the spacetime dimensionality.  At this point we note that the method employed thus far is designed to show that a conformally invariant configuration is possible for our non-perturbative expression (\ref{evolGappendix2}).  We acknowledge that the configuration in (\ref{ansatz1}) is intended as a demonstrative solution only and for the reasons below, its applications in string cosmology may be limited.  Given this, we find an acceptable solution to (\ref{evolGappendix2}) is given by the following constraints where prime denotes a derivative with respect to $X^0$.  Here $\phi_0$ and $\kappa_0$ are constants and $n, n_i$ are parameters to be determined.
\bea\label{ansatz1}
\phi^{'}(X^0)&=&\frac{\phi_0}{(X^0)^n},\nn
\kappa(X^0)&=&\frac{\kappa_0}{(X^0)^{n+1}}\nn
\tau_i(X^0)&=&\frac{-\kappa_0}{(X^0)^{n_i}}.
\eea
We then set out to show through our assumptions that the configuration (\ref{ansatz1}) is conformally invariant by showing the beta functions representing our effective action vanish to all orders in $\ap$.  The beta functions are given in (\ref{weylconditions}).  We do this by first showing that the beta functions with (\ref{ansatz1}) are homogeneous in $X^0$ to first order in $\ap$.  We then show the same for selected terms from a complete list of the second order beta functions which make up all possible combinations of the field parameters.  Using arguments outlined in section (\ref{structure_second_order}) we deduce, but do not comprehensively prove, homogeneity in $X^0$ to be true to all orders in $\ap$.  We find that such conformally invariant configuration of (\ref{ansatz1}) is given by:
\bea\label{solution1}
g_{\mu\nu}(X^0)&=&\mbox{diag}\left(\frac{\kappa_0}{(X^0)^2},\frac{-\kappa_0}{(X^0)^2},
\frac{-\kappa_0}{(X^0)^2},-\kappa_0\right) \nn
B_{\mu\nu}(X^0)&=&\left(\delta_{\mu 1}\delta_{\nu 2}-\delta_{\mu 2}\delta_{\nu 1}\right)\frac{\kappa_0 h_0}{2(X^0)^2}\nn
\phi(X^0)&=&\phi_0\ln(X^0).
\eea
In this we have assumed a form for the antisymmetric field strength such that $H_{012}=-\frac{\kappa_0 h_0}{(X^0)^3}$ and thereby limiting this parameter to one particular spacetime configuration (hence anisotropic for the time being).  We then make use of a property of the beta functions under which the following transformation leaves the underlying physics invariant, including the fact that they may be found to vanish.
\bea\label{reparametrization1}
\tilde g_{\mu\nu}&=&g_{\mu\nu}+\alpha^{'}g_{\mu\nu}\left(a_1 R+a_2 \partial^\rho\phi\partial_\rho\phi
+a_3\nabla^2\phi+a_4H_{\rho\mu\nu}H^{\rho\mu\nu}\right)\nn
\tilde B_{\mu\nu}&=&B_{\mu\nu}+\alpha^{'}\left(b_1\nabla^\rho H_{\rho\mu\nu}+b_2\partial^\rho\phi H_{\rho\mu\nu}\right)\nn
\tilde\phi&=&\phi+\alpha^{'}\left(c_1R+c_2\partial^\rho\phi\partial_\rho\phi+c_3\nabla^2\phi
+c_4H_{\rho\mu\nu}H^{\rho\mu\nu}\right), 
\eea
where ($a_1,...,b_1,...,c_1,...$) are dimensionless, arbitrary parameters.  Thus we can choose these parameters in such as way that the beta functions in (\ref{reparametrization1}) vanish and hence we have shown conformal invariance to all orders in $\ap$.  This proof relies on the fact that we have also shown them to be self homogeneous (and therefore the sums of terms within each beta function may cancel) in $X^0$ for our non-perturbative solution to (\ref{ansatz1}).\\\\In terms of cosmology in the Einstein frame, (\ref{solution1}) represents a metric:
\be g^E_{\mu\nu}(t)dx^\mu dx^\nu=dt^2-a_1^2(t)(dx^1)^2-a_2^2(t)(dx^2)^2-a_3^2(t)(dx^3)^2, \ee 
where
\bea\label{scalefactors1}
a_1(t)=a_2(t)&=&a_0~t^{1+1/\phi_0}\nn
a_3(t)&=&a_0~t
\eea
The solution shows linear expansion in the direction for which we fixed the antisymmetric tensor field strength in the $X^3$ direction in eq. (\ref{fixing_h_inx3}) with dilaton ($\phi_0$) dependent expansion in the $x_1$ and $x_2$ directions for which $H_{\rho\mu\nu}$ is not fixed.  When $\phi_0 = -1$ and when $B_{\mu\nu} = 0$ we revert to the currently observed isotropic, Minkowski universe.
Firstly, we acknowledge that our model does not take into account (nor allow for) an inflationary scenario (i.e. $a(t) \sim \exp(Ht)$) and that our results are concerned with post inflationary phenomena.  We do however note that our configuration can allow for inflation in Pre Big Bang theories. Secondly, we acknowledge that there is an issue of how to emerge from anisotropic expanding universe described by the conditions (\ref{scalefactors}) to a isotropic Minkowski model.  While we do not cover this point in detail, it is possible that a time dependence on $h$ in (\ref{fixing_h_inx3}) could address this issue, i.e. the string axion field, which relaxes to zero in cosmic time: $h(t) \rightarrow 0$. \\\\ We then reformulate (\ref{solution1}) to read in the Einstein frame:
\bea\label{configg1}
g_{\mu\nu}(t)&=&\mbox{diag}(1,-1,-1,-t^2)\nn
H_{012}(t)&=&-\frac{\kappa_0 h_0}{t^3}\nn
\phi(t)&=&-\ln t.
\eea
Motivated by optical anisotropy, we insert this configuration (\ref{configg1})into a one loop effective action in target space where the antisymmetric tensor has been coupled to the electromagnetism via a modified field strength such that $\tilde H_{\rho\mu\nu}=H_{\rho\mu\nu}+\frac{1}{M}A_{[\rho}F_{\mu\nu]}$ where $M$ is the string mass scale and $F_{\mu\nu}$ is the electromagnetic tensor.  This results in modified Maxwell's equations which admit as a solution:
\bea \label{solE_and_B1}
{\bf E} &=& {\bf K} \exp\left(-2i\epsilon\ln t\right) \frac{\cos(x^3)}{t^2} \nn
{\bf B} &=& i{\bf K} \exp\left(-2i\epsilon\ln t\right) \frac{\sin(x^3)}{t^3}. \eea
Here ${\bf E}$ and ${\bf B}$ represent an electromagnetic wave lying in the $(x_1, x_2)$ plane and propagating in the $x_3$ optic axis whose angle of polarisation changes with cosmic time $t$.
\be \label{anisotropic_polarisation1}
\Delta(t)= |\arg ({\bf E}) - \arg({\bf E})^*| =  4\epsilon \ln t = \frac{4\kappa_0 h_0}{M}\ln t.
\ee
We then generalise the configuration for $H_{\rho\mu\nu}$ from $H_{012}=-\frac{\kappa_0 h_0}{(X^0)^3}$ to:
\be
\label {axion11} H_{\rho\mu\nu} =
(X^0)^m\varepsilon_{\rho\mu\nu\sigma}(X^0)\partial^\sigma h,
\ee 
and repeat the process of demonstrating conformal invariance, devising the Einstein metric and coupling to the electromagnetic field in a spacetime effective action.  The result is a similarly logarithmic dependence by the polarisation on cosmic time as in (\ref{anisotropic_polarisation1}).  This phenomena is also known as cosmic birefringence as described in section (\ref{String Cosmology and Anisotropy}) and efforts are underway to examine evidence for this in the cosmic microwave background and from studies of radiation from distant extra-galactic objects which emit polarised radiation, \cite{Optical_Anisotropy}, \cite {Optical_Anisotropy2}, \cite{Optical_Anisotropy5}.  Theoretical treatments similar to that conducted in this thesis result in a cosmic time dependence by the polarisation proportional to $h'\tau$, \cite{Optical_Anisotropy3} and $\tan^{-1}\frac{1}{\tau}$ and $\tan^{-1}\frac{1}{\tau^3}$ with the variation due to the configuration of the Einstein metric considered.  The logarithmic dependence of our solution arises as a result of the power law constraints we established in (\ref{ansatz1}), which in turn were motivated by the need to prove conformal invariance in a non-perturbative manner. 

\chapter{Conclusion}

\section{Full Quantization of the Axion}
In section (\ref{Flattening of the Axion Potential}) we described how the spinodal instability in the QCD axion's potential flattened it when interactions are not considered.  We derived a non-perturbative expression for the evolution of the effective potential with the scale factor $f$ as in (\ref{final_result_2}) and found a flat-potential solution consistent with the well known property of convexity outlined in eq. (\ref{convex suppression}).  We then assigned a boundary condition to this expression such that the potential when $f = \Lambda$ is some value $U_{\Lambda}$ leading to:
\be \label{general_solution1.8} U_{eff}(f) = \frac{\Lambda^4}{32\pi^2} \ln \left(\frac{f}{\Lambda}\right) + U_{\Lambda}. \ee
Physically, at energy level $\Lambda$, which represents the upper limit cutoff of our theory, we consider that the evolution of the axion field is in its earliest describable form.  In this sense, for very small $\theta$ (the dynamical axion parameter), the commonly used cosine form of the axion potential (\ref{axion effective potential form}) may be equated to a double well scalar potential as in (\ref{approximation_cosine}).  We consider that such a double well potential represents the origin of the spontaneously broken $U(1)_{PQ}$ symmetry responsible for the axion field's development.  We consider that our result (\ref{general_solution1.8}) describes the system on the verge of spontaneous symmetry breaking.  At this point the axion has yet to evolve fully and the effects of the spinodal instability serve to flatten the potential, prior to any interactions.  We consider that the term $U_{\Lambda}$ in (\ref{general_solution1.8}) is a one loop correction to this.  We identify it with the one loop correction to the quantized form of (\ref{SSB_axion}) at  $\mu = 0$, which we compute using an established result quoted in \cite{Zee}.  For $\Lambda = f$, the result which can be expressed in terms of the axion mass $m_a$ and the ratio of $\frac{m^4}{f^4}$ where $m$ is an arbitrary mass scale less than $\Lambda$, and $f$ is the scale factor which we here set equal to $\Lambda$.
\be \label{final result 4.8} U_{eff, f=\Lambda} \sim   4\times 10^{-2}\, \frac{m^4}{f^4}m_a^4,  \ee
where $\frac{m^4}{f^4}$ is a ratio less than one.  We interpret this as an energy density associated with the axion at the earliest phase in its development, prior to significant interactions, and influenced by spinodal instability effects. (At later stages the axion acquires mass and resolves the CP problem as well as providing a candidate for cold dark matter).  With the caveat that the ratio $\frac{m^4}{f^4}$ is not determined by our theory and using $10^{-4}eV < m_a < 10^{-1}eV$, \cite{axion dark matter} it could be that the result in eq. (\ref{final result 4.8}) it is not orders of magnitude away from required values for dark energy ($U_{DE} \sim (10^{-3} eV)^4$, \cite {Spinodal Instabilities and the Dark Energy Problem}).\\\\We acknowledge the limitations in this result and treatment as follows.  Firstly, our model and assumption of the flattening of the potential does not take into account interactions of the axion scalar field with other particles, in particular with gravity.  Section (\ref{Axion Interactions_section}) shows how interactions will impact this flattening.  Secondly, our result contains indeterminate energy scales (e.g. $\frac{m^4}{f^4}$) which lessen the usefulness in phenomenological comparisons.  Further work in this area would involve an extension of section (\ref{Axion Interactions_section}) to devise and quantize a more realistic model of the axion scalar field in the early universe including coupling to gravity, including perhaps, spacetime curvature which may be relevant at the energies and epochs we are considering.  This could then lead to a more detailed investigation of the flattening effect of the axion's effective potential at high energy scales and lead to more justifiable allocation of values to the at-present arbitrary energy scales in the derivation and results.  

\section{The Bosonic String Axion}
The primary result of this section of the thesis is to derive a non-perturbative formulation for the effective action of the bosonic string including the string axion, on the worldsheet, (\ref{evolGappendix2}).   We showed that a viable solution of this (\ref{ansatz1})) is conformally invariant and therefore suitable for consideration in string cosmology.  We achieved this demonstration of conformal invariance in a non-perturbative manner, by demonstrating homogeneity of the beta functions with time; combined with field rescaling to allow for cancellation of all terms.  This contrasts with the conventional method in string cosmology of canceling a perturbative expression for the beta functions up to some order in $\ap$.  We find that a configuration for (\ref{ansatz1}) is given by:
\bea\label{solution8}
g_{\mu\nu}(X^0)&=&\mbox{diag}\left(\frac{\kappa_0}{(X^0)^2},\frac{-\kappa_0}{(X^0)^2},
\frac{-\kappa_0}{(X^0)^2},-\kappa_0\right) \nn
B_{\mu\nu}(X^0)&=&\left(\delta_{\mu 1}\delta_{\nu 2}-\delta_{\mu 2}\delta_{\nu 1}\right)\frac{\kappa_0 h_0}{2(X^0)^2}\nn
\phi(X^0)&=&\phi_0\ln(X^0).
\eea
In this we have assumed a form for the antisymmetric field strength such that $H_{012}=-\frac{\kappa_0 h_0}{(X^0)^3}$ and thereby limiting this parameter to one particular spacetime configuration (hence anisotropic). In this $h_0$ is the constant field strength of the bosonic string axion.  Motivated by optical anisotropy, we insert this configuration (\ref{solution8})into a one loop effective action in target space where the antisymmetric tensor has been coupled to the electromagnetism via a modified field strength such that $\tilde H_{\rho\mu\nu}=H_{\rho\mu\nu}+\frac{1}{M}A_{[\rho}F_{\mu\nu]}$ where $M$ is the string mass scale and $F_{\mu\nu}$ is the electromagnetic tensor.  The result is a logarithmic dependence by the polarisation on cosmic time as in (\ref{anisotropic_polarisation1}).  This phenomena is also known as cosmic birefringence as described in section (\ref{String Cosmology and Anisotropy}).  The logarithmic dependence of our solution arises as a result of the power law constraints we established in (\ref{ansatz1}), which in turn were motivated by the need to prove conformal invariance in a non-perturbative manner.  We make several comments on this section of the thesis with a view to future work.
\begin{itemize}

\item Section (\ref{Tachyon Cosmology}) noted that the potential for tachyon modes arising in the theory can be removed by applying supersymmetry.  This was not done in our work and is an obvious extension of the model.

\item It was discussed in section (\ref{Tachyon Cosmology}) that our configuration as above gives rise to a power law dependence by the scale factors $a_i$ with cosmic time, in the Einstein frame. The predictive powers of the model may be incomplete as the resulting metric does not support an inflationary scenario other than in a Pre Big Bang scenario.  Further work could involve relaxing our constrain of considering $D=4$ spacetime dimensions and/or considering tachyon modes in our non-sypersymmetric string model as was done in  \cite{Alexandre_tachyon}.

\item We note the $\frac{44}{3 \ap}e^{2\phi}$ present in our effective spacetime action in the Einstein frame, eq. (\ref{one-loop-effective}) which derives from the $D=4$ case in the general expression in (\ref{effective_string_action_E}).  We note this implies a negative cosmological constant relative to the Einstein curvature term.  We do not explore a solution for this and note that the phenomenological value of our model may be limited if this negative term cannot be compensated for by other terms in the effective action, the form of  which we have assumed.

\item  We acknowledge that there is an issue of how to emerge from anisotropic expanding universe described by the conditions (\ref{scalefactors}) to a isotropic Minkowski model.  While we do not cover this point in detail, it is possible that a time dependence on $h$ in (\ref{fixing_h_inx3}) could address this issue, i.e. the string axion field, which relaxes to zero in cosmic time: $h(t) \rightarrow 0$.

\item The next steps in this part of the thesis would be as follows.  We would develop, as far as possible, a more rigorous demonstration of conformal invariance of our configuration via closer examination of the higher order beta functions.  We could further generalise the bosonic string action to incorporate supersymmetry.  

\end{itemize}

\chapter{Appendix}

\section{Derivation of Exact renormalisation Group Equations}\label{appendix Exact renormalisation}

The Wilsonian effective action in  (\ref {Wilson_effaction}) can be written as:
\be\label{05} \exp (- S_{k-\delta k}(\Phi))= \int D\varphi \exp ({- S_k(\Phi +
\varphi)}). \ee
The field $\phi(p)=\Phi(p)+\varphi(p)$ is split into two sectors, with $p$ the momentum variable:  
\bea \label{sharpcutoff}
 \phi(p) &= \varphi(p)&\text {for}\qquad {k-\delta k<|p|\leq k}\,\,\,\text {(the UV modes)}\nn
\phi(p) &= \Phi(p)&\text {for}\qquad{|p|\leq k-\delta k}\,\,\,\text
{(the IR modes),} \eea
where $\phi(p)$ is zero outside the regions specified in (\ref{sharpcutoff}). Expanding $S_k(\Phi + \varphi)$ around $\Phi$ gives, (where the phase space volume within $p\rightarrow \Lambda$ is $V = \int \frac{d^Dp}{(2\pi)^D})$:
\be \label{taylor} S_k(\Phi + \varphi)= S_k(\Phi) + \frac{1}{V}\int _k
\frac{\delta S_k(\Phi)}{\delta \Phi (p)}\varphi(p)
\,+\,\frac{1}{2V^2}\int _k\int _k \varphi
(p)\frac{\delta^2S_k(\Phi)}{\delta \Phi (p)\delta \Phi
(q)}\varphi (q)  +  {\cal O} (\delta k)^2\ee
In (\ref{taylor}), the term $\frac{\delta S_k(\Phi)}{\delta \Phi
(p)}$ is the derivative of the
functional $\delta S_k(\Phi)$ with respect to the function $\Phi
(p)$. The integrals $\int_k$ are defined in D dimensional Euclidean space, and where $d\Omega_D$ is the solid angle in $D$ dimensions.
\be \label{dp}\int_{k}\equiv \int_{k-\delta
k}^{k}\frac{d^Dp}{(2\pi)^D} \,=\,\int_{k-\delta
k}^{k}\frac{d\Omega_D
p^{D-1}dp}{(2\pi)^D}\ee
If there is an expression $f(p^2)$ under the integral in (\ref{dp})
such that the integral only depends on the magnitude of the momentum
($p^2$), we can perform the $d\Omega_D$ integral to
give:
\be \label{dp2}\ \int_{k-\delta k}^{k}\frac{d\Omega_D
p^{D-1}dp}{(2\pi)^D}f(p^2)\,=\,\Omega_D k^{D-1} \delta k \frac{1}{(2\pi)^D}f(k^2).\ee
Thus with these conditions each $\int_k$ contributes a factor proportional to $\delta k$.
In (\ref{taylor}) the terms higher than
$\frac{1}{2V^2}\int _k\int _k \varphi
(p)\frac{\delta^2S_k(\Phi)}{\delta \Phi (p)\delta \Phi (q)}\varphi
(q)$ are at least quadratic in $\varphi$ and therefore the $\int _k$
integrals can be evaluated to give a factor of at least $(\delta
k)^2$ in the terms.  Hence we use the notation ${\cal O} (\delta k)^2$ in (\ref{taylor}). By substituting the expansion in
(\ref{taylor}) into (\ref {05}), and expanding $\exp ({\cal O} (\delta
k)^2$ as $(1 + {\cal O} (\delta k)^2)$
we obtain the following.
\begin{multline} \label{006} \exp (S_{k}(\Phi)-S_{k-\delta k}(\Phi))= \\
\int D\varphi \exp - \left (\frac{1}{V}\int _k \frac{\delta S_k(\Phi)}{\delta
\Phi (p)}\varphi(p) +  \frac{1}{2V^2}\int _k\int _k \varphi
(p)\frac{\delta^2S_k(\Phi)}{\delta \Phi (p)\delta \Phi
(q)}\varphi (q) \right) \\ (1 + {\cal O}(\delta k)^2) \end{multline} 
The right-hand-side of (\ref{006}) can be evaluated by considering it as a Gaussian integral with $B_{ij}$ and $A_i$ as matrices, and using the following, where $K$ and $K'$ are constants.
\begin{multline} (\Pi_k\int d\xi_k)\exp -\left( \frac{1}{2}\xi_iB_{ij}\xi_j + A_i\xi_i\right) \\= K \times \sqrt{\frac{1}{det [B]}} \exp \frac{1}{2}(A_i B_{ij}^{-1}A_j) 
\\ = K' \times \exp \left( {-\frac{1}{2}}\mbox{Tr}(\ln B_{ij}) + \frac{1}{2} A_i B_{ij}^{-1}A_j\right).\end{multline}
We use the following equivalences:
\bea B_{ij} &\equiv & \frac{\delta^2S_k(\Phi)}{\delta \Phi (p)\delta \Phi (q)} \nn
A_i &\equiv & \frac{\delta S_k(\Phi)}{\delta \Phi (p)} \nn
\xi_k &\equiv & \varphi(p) \nn
\Pi_k &\equiv & \frac{1}{V^2}\int_k\int_k, \eea
we compute (\ref {006}):
\begin{multline} \label{007}\exp(S_{k}(\Phi)-S_{k-\delta k}(\Phi))=  
\exp(-\frac{1}{2}\mbox{Tr}\left(\ln \frac{\delta^2S_k(\Phi)}{\delta \Phi
(p)\delta \Phi(q)}\right) + \\ \mbox{Tr}\left(\frac{\delta S_k(\Phi)}{\delta \Phi (p)}\left(\frac{\delta^2S_k(\Phi)}{\delta \Phi (p)\delta \Phi (q)}\right)^{-1}\frac{\delta S_k(\Phi)}{\delta \Phi (q)}\right)) + \int D\varphi {\cal O} (\delta k)^2 \end{multline}
In the right-hand-side of (\ref{007}), we assume that the factor $\frac{\delta^2S_k(\Phi)}{\delta \Phi
(p)\delta \Phi(q)}$ in the logarithm has been multiplied by a unitary constant of units $k^2$ to avoid taking the logarithm of a dimensionful parameter. In (\ref{007}), the  trace taken over the products of the matrices $B_{ij}$ and $A_i$ is defined as $\mbox{Tr} [...] = \int_k dp \int_k dq (2\pi)^D\delta^D (p+q)[....]$.\\\\We now evaluate the factor $\frac{\delta S_k(\Phi)}{\delta \Phi(p)}$ in (\ref{007}). The effective action can be expressed based on the gradient expansion approximation:
\be \label{phi4i} S(\Phi)_{eff}=\int
d^Dx\,\left(\frac{1}{2}Z(\Phi)\partial_\mu\Phi\partial^\mu\Phi +
U_{eff}(\Phi)\right),\ee
where $U_{eff}(\Phi)$ is the effective potential containing the mass and
interaction terms and $Z(\Phi)$ contains all fluctuations involving derivatives of the field $\Phi$.  Considering $\frac{\delta S(\Phi)}{\delta \Phi
(y)}$:
\be \label{firstderivative} \frac{\delta S_k(\Phi)}{\delta \Phi (y)} =
\int d^Dx [U_{eff}'(\Phi(x)) +
\frac{1}{2}Z'(\Phi(x))\partial_\mu\Phi\partial^\mu\Phi -
Z(\Phi(x))\partial_\mu \partial^\mu\Phi] \delta ^D(x-y).\ee
where $U_{eff}'(\Phi)$ is the partial derivative with respect to
$\Phi(y)$.  Evaluating $\delta ^D(x-y)$ under the integral, using integration by parts on the second term on the right-hand-side
and taking $\int d^Dx\, \partial_\mu \,(\delta
(x-y)\partial^\mu \Phi)\,\,\rightarrow 0$ in the integration limits,
gives: 
\be \label{integrationparts} \frac{\delta
S_k(\Phi)}{\delta \Phi (y)} = U_{eff}'(\Phi(y)) +
\frac{1}{2}Z'(\Phi(y))\partial_\mu\Phi\partial^\mu\Phi -
Z(\Phi(y))\partial_\mu \partial^\mu\Phi, \ee
In a constant field configuration $\Phi_0$, the factor $\partial^\mu\partial_\mu\Phi$
vanishes and the expression becomes, in momentum space:
\bea \label{phi4ii} \frac{\delta S(\Phi)}{\delta \Phi (p)}|_{\Phi_0} &=& \frac{1}{(2\pi)^D}\int d^Dy \exp(-i\textbf{p}.\textbf{y})  U_{eff}'(\Phi_0) \nn &=&   
\delta^D(p) U_{eff}'(\Phi_0). \eea
The integral in (\ref{006}) is between the values of
$p$ = $k-\delta k$ and $k$.  The $\delta^D(p)$
factor in (\ref{phi4ii}) indicates evaluation only at $p=0$ and
hence the entire expression for $\frac{\delta S(\Phi)}{\delta \Phi
(p)}|_{\phi_0}$ vanishes when integrated in the shell for $\Phi = $
constant as does the entire term $\left(\frac{\delta
S_k(\Phi)}{\delta \Phi (p)}\left(\frac{\delta^2S_k(\Phi)}{\delta
\Phi (p)\delta \Phi (q)}\right)^{-1}\frac{\delta S_k(\Phi)}{\delta
\Phi (q)}\right)$ in (\ref{007}), resulting in:
\begin{multline} \label {008}\exp (S_{k}(\Phi)-S_{k-\delta k}(\Phi))= 
\exp \left(-\frac{1}{2}\mbox{Tr}(\ln \frac{\delta^2S_k(\Phi)}{\delta
\Phi (p)\delta \Phi
(q)})\right)+\int D\varphi \, {\cal O} (\delta k)^2. \end{multline}
Evaluating the trace in (\ref {008}) as:
\be \label {trace evaluation}\mbox{Tr}(f(p,q))\equiv \int_k dp \int_k dq (2\pi)^D\delta^D (p+q)(f(p,q)), \ee
and the delta function in (\ref{trace evaluation}):
\be \label {trace evaluation 1}\mbox{Tr}(f)=\int_p
\delta^D (0)f(p,-p), \ee
we use the identity $\delta^D(p)=\int d^Dx \exp (ipx)$ and that $\delta^D(0)=\int
d^Dx$ = $V_D$ where $V_D$ is the volume integration over space-time in
the $D$ dimensions (assumed to be finite):
\be \label {trace
evaluation
2}\mbox{Tr}(f)= V_D \int_p f(p,-p). \ee
Using $f(p,q)= \ln \frac{\delta^2S_k(\Phi)}{\delta \Phi (p)\delta
\Phi (q)}$, (\ref {008}) becomes:
\begin{multline} \label {009i}\exp (S_{k}(\Phi)-S_{k-\delta k}(\Phi))=
\exp \left(-\frac{V_D}{2}\int_p\ln \frac{\delta^2S_k(\Phi)}{\delta
\Phi (p)\delta \Phi (-p)}\right)+\int D\varphi \, {\cal O} (\delta k)^2.
\end{multline}
The $\int_p$ in (\ref {009i}) can be evaluated as
in (\ref {dp2}) as the field is quadratic in $p$ and substituting into
(\ref{008}) gives: 
\begin{multline} \label {009iii}\exp (S_{k}(\Phi)-S_{k-\delta k}(\Phi))= 
\exp \left(-\frac{1}{2}\Omega_D k^{D-1} \delta k
\frac{V_D}{(2\pi)^{D}}\ln \frac{\delta^2S_k(\Phi)}{\delta \Phi
(p)\delta \Phi
(-p)}\right)+ \\ \int D\varphi \, {\cal O} (\delta k)^2 \end{multline}
We now take the logarithm of both sides, using on the right-hand-side $\ln (x+\epsilon) = \ln x + \epsilon$
given $x\gg\epsilon$ (or in this case $\delta k \gg
(\delta k)^2$).
We then divide both sides by a factor of $\delta k$ and finally take
the limit as $\delta k \rightarrow 0$.  This limit produces the
derivative with respect to $k$ on the LHS of (\ref{008}), that is:
\be \partial_k \, S_k (\Phi_0)=\lim _{\delta
k \rightarrow 0}\, \frac{(S_{k}(\Phi_0)-S_{k-\delta
k}(\Phi_0))}{\delta k}.\ee
On the right-hand-side of (\ref{009iii}), the ${\cal O}(\delta k)^2$ factor vanishes in
this limit to give an exact form for the variation of the action
with cutoff momentum or energy scale:
\be \label {010}\partial_k S_k (\Phi_0) =
-\frac{1}{2}\Omega_D k^{D-1} \frac{V_D}{(2\pi)^{D}}\ln
\left[\frac{\delta^2S_k(\Phi_0)}{\delta \Phi (p)\delta \Phi (-p)} \right],
\ee
where here $k^2=p^2$.  This is
known as the Wegner-Houghton equation, derived in the path integral
formalism, \cite{non-perturbative renormalisation Flow}.\\\\ Considering the effective action in (\ref{phi4i}), we differentiate (\ref{integrationparts}) with respect to $\Phi(z)$ with the objective of evaluating $\frac{\delta^2S_k(\Phi_0)}{\delta \Phi (p)\delta \Phi (-p)}$ in terms of the $U_{eff}$ and $Z(\Phi)$. 
\begin {multline} \label{second derivative} \frac{\delta^2S_k(\Phi)}{\delta
\Phi (y)\delta \Phi (z)} = [U_k''(\Phi) +
\frac{1}{2}Z''(\Phi(y))\partial_\mu\Phi\partial^\mu\Phi -
Z'(\Phi(y))\partial_\mu \partial^\mu\Phi \\-
Z(\Phi(y))\partial_\mu\partial^\mu ]\delta ^D(y-z) +
Z'(\Phi(y))\partial_\mu (\delta ^D(y-z))\partial^\mu\Phi. \end
{multline}
Evaluating $\frac{\delta^2S_k(\Phi)}{\delta
\Phi (y)\delta \Phi (z)}$ at $\Phi_0$, a constant field value,
results in the terms containing $\partial^\mu\Phi$ vanishing and:
\be \label{second derivative1} \frac{\delta^2S_k(\Phi)}{\delta \Phi
(y)\delta \Phi (z)}|_{\Phi_0} = [U_k''(\Phi_0) -
Z(\Phi_0(y))\partial_\mu\partial^\mu]\delta ^D(y-z). \ee 
This can be expressed in momentum space as, using $-\partial_\mu\partial^\mu \equiv k^2$:
\be \label{second derivative2} \frac{\delta^2S_k(\Phi)}{\delta \Phi
(p)\delta \Phi (q)}|_{\Phi_0} = [U_k''(\Phi_0) + Z(\Phi_0)k^2]\delta
^D(p+q).
\ee
Thus in a constant field configuration $\Phi=\Phi_0$, (\ref{010}) reduces to, with a static field potential $U_{eff}(\Phi_0)$  and a configuration $U_{k}(0)$.
\be \label {final3}\partial_{k}U_{eff}(\Phi_0) -
\partial_{k}U_{eff}(0)=-\frac{\Omega_D k^{D-1}}{2(2\pi)^{D}}
\ln\left(\frac{U_{eff}''(\Phi_0) +
Z(\Phi_0)k^2}{U_{eff}''(0) + Z(0)k^2}\right). \ee

\end{document}